# Кинетика последовательных реакций первого порядка в условиях хроматографического разделения

С.В. Ермолаев, А.К. Скасырская

Последовательными реакциями называют процессы, в которых в которых продукт одного превращения является исходным веществом для следующего. Последовательные реакции первого порядка записывают в виде:

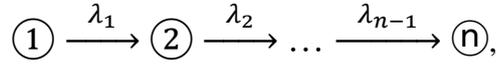

где ⓘ – $i$-ое вещество, $\lambda_i$ – это константа скорости $i$-ой реакции, $i = 1 \div n$. Первый порядок означает, что скорости изменения количества веществ зависят от самих количеств в первой степени. Они определяются системой дифференциальных уравнений $\frac{dN_i}{dt} = \lambda_{i-1}N_{i-1} - \lambda_i N_i$, решение которой сводится к уравнению Бейтмана [1, 2]:

$$N_i(t) = \sum_{j=1}^{i} \left[ N_j^0 \cdot \left( \prod_{k=j}^{i-1} \lambda_k \right) \cdot \left( \sum_{k=j}^{i} \frac{e^{-\lambda_k t}}{\prod_{j \neq k}^{i}(\lambda_j - \lambda_k)} \right) \right] \qquad (1),$$

где $N_j^0$ – количества веществ в начальный момент времени.

Цепочка радиоактивных превращений является последовательностью реакций 1-го порядка, в которых константа скорости $i$-ой реакции называется постоянной распада $i$-ого радионуклида. Одним из интересных и практически важных является случай, когда $\lambda_1 \ll \lambda_i, i = 2 \div n - 1$. В этом случае спустя определенное время достигается состояние:

$$N_i = N_1^0 \cdot \left( \prod_{j=1}^{i-1} \lambda_j \right) \cdot \frac{e^{-\lambda_1 t}}{\prod_{j=2}^{i}(\lambda_j - \lambda_1)} = N_1 \frac{\prod_{j=1}^{i-1} \lambda_j}{\prod_{j=2}^{i}(\lambda_j - \lambda_1)} \qquad (2),$$

называемое подвижным равновесием. При этом, поскольку активность радионуклида связана с его количеством $A_i = \lambda_i N_i$, активности всех радионуклидов становятся близки:

$$A_i = A_1 \frac{\prod_{j=2}^{i} \lambda_j}{\prod_{j=2}^{i}(\lambda_j - \lambda_1)} \approx A_1 \qquad (2a)$$

На использовании подвижного равновесия строится принцип работы радионуклидных генераторов (Рис. 1), цепочка которых обычно ограничивается двумя участниками.

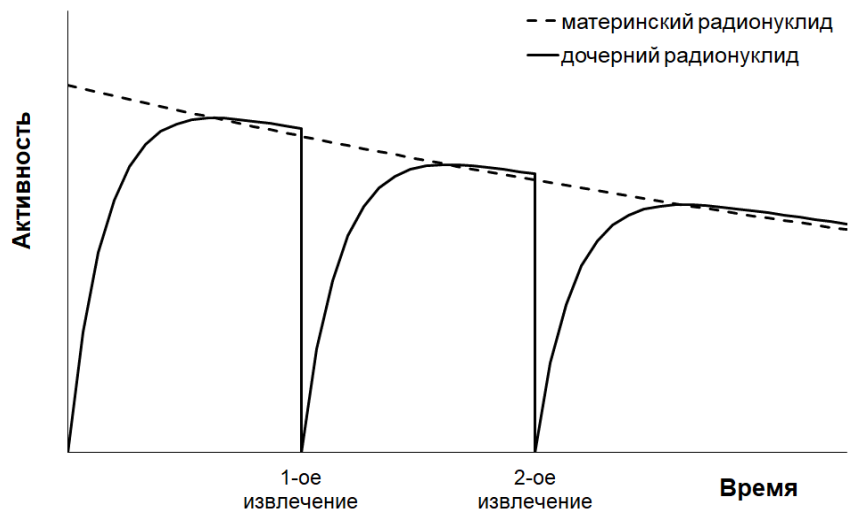

**Рис. 1**. Принцип работы радионуклидного генератора.

По достижении подвижного равновесия дочерний радионуклид отделяют от материнского, т.е. извлекают из генератора, и используют. А в генераторе снова начинается накопление дочернего радионуклида до наступления равновесия, после чего цикл повторяется.

Генераторы радионуклидов успешно применяются в ядерной медицине [3]: при длительном сроке службы, определяемом, главным образом, периодом полураспада материнского радионуклида, они обеспечивают многократность использования дочернего. Разделение радионуклидов в распространенных сейчас генераторах (например, $^{99}$Mo/$^{99m}$Tc, $^{68}$Ge/$^{68}$Ga, $^{82}$Sr/$^{82}$Rb, $^{44}$Ti/$^{44}$Sc) происходит хроматографически по прямой схеме: материнский радионуклид удерживается сорбентом, заключенным в хроматографическую колонку, а пришедший в равновесие дочерний радионуклид вымывают из генератора в небольшом объеме подвижной фазы – элюата. При несомненных достоинствах у такого подхода есть ограничения. В случае, например, $^{82}$Sr/$^{82}$Rb генератора период полураспада $^{82}$Rb составляет всего 75 секунд. Для снижения потерь $^{82}$Rb из-за распада, стремятся уменьшить длительность его выделения и введения пациенту, это обычно достигается за счет высокой скорости (20-100 мл/мин) потока элюента [4, 5] и влечет за собой повышенные требования к конструкции генератора.

Недавние успехи в создании мощных ускорителей заряженных частиц и развитии радиохимических технологий расширяют возможности производства и дальнейшего применения радиоизотопов, в том числе в виде радионуклидных генераторов. Становятся доступными генераторные пары, перспективные для диагностики и терапии различных заболеваний (например, $^{118}$Te/$^{118g}$Sb [6]), в которых длительность отделения дочернего радионуклида сопоставима с периодом его полураспада. Другими словами, равновесное количество дочернего радионуклида, накапливаемое к началу процесса, сопоставимо с количеством, образующимся в ходе самого отделения. Появляются генераторные системы, в которых целевой дочерний радионуклид получают посредством непрерывного отделения и распада более короткоживущего промежуточного предшественника: $^{223}$Ra/$^{219}$Rn/$^{211}$Pb [7, 8], $^{224}$Ra/$^{220}$Rn/$^{212}$Pb [9, 10], $^{225}$Ac/$^{221}$Fr/$^{213}$Bi [11, 12]. Целевой радионуклид накапливают отдельно, например, на второй хроматографической колонке, со временем система приходит в состояние, в котором целевой радионуклид находится в подвижном равновесии с материнским, но пространственно от него отделен. Затем его извлекают из генератора. Стадия накопления может быть организована как замкнутый контур, в котором циркулирует подвижная фаза, отделяющая промежуточный радионуклид от материнского [9, 12]. Такие генераторы сложнее, чем "прямые" генераторы, но имеют преимущества, в частности, обеспечивают более высокую радионуклидную чистоту продукта [13].

Для создания и оптимизации генератора, в котором период полураспада дочернего радионуклида соизмерим со временем его разделения, надо знать не только количества веществ в разные моменты времени, но и их распределение в генераторной системе. Для этого недостаточно уравнения (1), описывающего $N_i$ только как функцию времени. Целью данной работы является развитие кинетики последовательных реакций первого порядка для условий хроматографического разделения. Предложен подход, описывающий движение в сорбенте веществ-участников реакций с различными скоростями и позволяющий определять концентрацию веществ, движущихся в потоке подвижной фазы, в зависимости от времени протекания реакций и положения в хроматографической системе.

# Описание движения веществ-участников
# последовательных реакций 1-го порядка в хроматографической среде

Движение в сорбенте веществ, одновременно превращающихся друг в друга в цепочке реакций 1-го порядка, приводит к возникновению распределений или профилей концентрации этих веществ, зависящих от времени. В предыдущей работе [11] описан частный случай цепочки из трех радионуклидов, когда материнский зафиксирован в начальном слое сорбента. Теперь будет предложен более общий подход к нахождению концентрации вещества как функции времени и координаты в хроматографической среде и показано влияние на концентрацию таких факторов хроматографического разделения, как:

- изменение скорости пропускания или состава подвижной фазы;
- изменение неподвижной фазы, т.е. выход вещества из колонки с сорбентом в раствор (элюат) или комбинация колонок с разными сорбентами;
- циркуляция подвижной фазы.

Затем на примере цепочки радиоактивных превращений $^{225}Ac \rightarrow \, ^{221}Fr \rightarrow \, ^{213}Bi \rightarrow$ будет построена модель радионуклидного $^{225}Ac/^{213}Bi$ генератора, в котором $^{213}Bi$ получают с помощью непрерывного отделения и распада промежуточного $^{221}Fr$.

## 1. Движение веществ в хроматографической колонке бесконечной длины (одномерная модель)

### 1.1. Цепочка из двух последовательных реакций: ① → ② →

Рассмотрим систему, состоящую из сосуда и достаточно длинной хроматографической колонки постоянного сечения, заполненной сорбентом (Рис. 2). В сосуде находится раствор объемом $V_0$, в котором протекают последовательные реакции 1-го порядка. Ограничимся вначале цепочкой из двух реакций и условием $\lambda_1 \ll \lambda_2$. Со временем вещества в растворе приходят в подвижное равновесие (уравнение (2)), а их концентрации равны $c_1 = \frac{N_1}{V_0}$ и $c_2 = \frac{\lambda_1}{\lambda_2 - \lambda_1} c_1$. Раствор в сосуде граничит с сорбентом, примем сечение их контакта за начало оси $V$.

В исходный момент времени начинается движение раствора, он поступает в колонку с объёмной скоростью $Q$. Предположим, что движение происходит в режиме идеального вытеснения. Вещества ① и ② с начальными концентрациями $c_1^0$ и $c_2^0$ входят в сорбент, где движутся со скоростями $q_1, q_2 \leq Q$, т.е. при входе в сорбент ($V = 0$) их концентрации равны $\frac{dN_i^0}{dV} = c_i^0 \frac{Q}{q_i}, i = 1,2$. Скорость $q_1$ может быть как больше, так и меньше $q_2$, рассмотрим сначала случай $q_1 \geq q_2$.

Спустя некоторое время $t_e$ элюирования фронт раствора в сорбенте пройдет $V_e = Q t_e$, а фронты веществ ① и ② пройдут: $V_{ei} = q_i t_e$. К этому моменту вещества

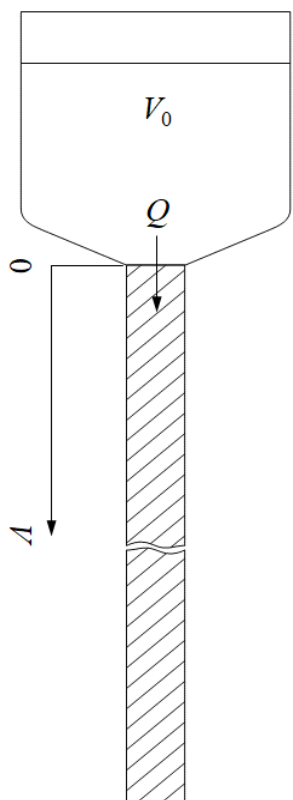

**Рис. 2**. Модель "сосуд и хроматографическая колонка" для изучения движения веществ-участников последовательных реакций 1-го порядка.

① и ② находятся в растворе объемом $V_0 - Qt_e$, оставшемся в сосуде, и в объеме сорбента равном $V_{e1}$. Общее количество вещества ① равно: $N_1(t_e) = \left(c_1^0(V_0 - Qt_e) + \frac{dN_1^0}{dV}V_{e1}\right)e^{-\lambda_1 t_e} = N_1^0 e^{-\lambda_1 t_e}$. Для нахождения общего количества вещества ② надо определить его концентрацию $\frac{dN_2}{dV}$ в сорбенте. Для этого мы будем использовать диаграмму в координатах "$V$-$t$" (Рис. 3), поскольку концентрация движущегося вещества является в общем случае функцией этих переменных.

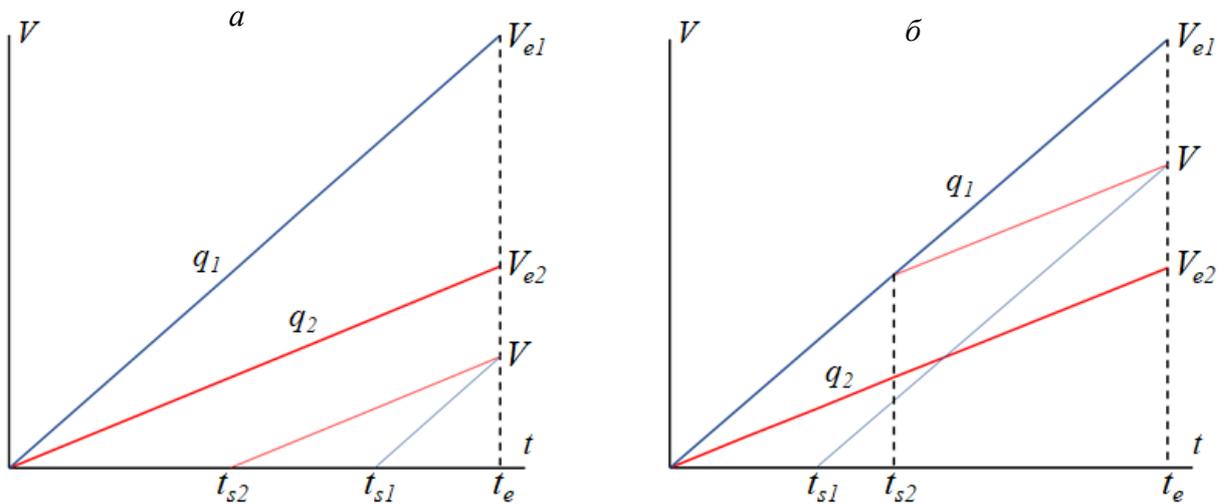

**Рис. 3**. $V$-$t$ диаграммы движения веществ ① и ② в хроматографической колонке для случая $q_1 \geq q_2$: *а* – в диапазоне $0 \div V_{e2}$; *б* – в диапазоне $V_{e2} \div V_{e1}$. Синие линии соответствуют движению фронта $V_{e1}$ и дифференциального элемента $dV$ вещества ① ($dV$①), красные линии – движению фронта $V_{e2}$ и дифференциального элемента $dV$ вещества ② ($dV$②).

Изменение концентрации $\frac{dN_2}{dV}$ в произвольной точке $V$ ( $0 \leq V \leq V_{e1}$ ) хроматографической колонки определяется уравнением материального баланса в виде:
$$\frac{d^2 N_2}{dV dt} = \lambda_1 \frac{dN_1}{dV} - \lambda_2 \frac{dN_2}{dV} \tag{3}$$
Получим решение этого уравнения для диапазонов существования вещества ② в сорбенте $0 \div V_{e2}$ и $V_{e2} \div V_{e1}$.

**Диапазон $0 \div V_{e2}$ вещества ②** (Рис. 3а). Рассмотрим вещество ② в дифференциальном элементе $dV$, движущемся со скоростью $q_2$ ($dV$②) и оказывающимся в точке $V$ в момент $t_e$. Он стартует в начальной точке хроматографической колонки ($V=0$) в момент $t_{s2}$ и движется в сорбенте в течение $t_2 = t_e - t_{s2} = \frac{V}{q_2}$. Решаем ур-ие (3) относительно $t_2$, используя в качестве граничного условия концентрацию вещества ② в элементе $dV$② при старте: $t_2 = 0$: $\left(\frac{dN_2}{dV}\right)_{t_{s2}} = c_2 \frac{Q}{q_2} = c_2^0 \frac{Q}{q_2} e^{-\lambda_1 t_{s2}}$, и учитывая, что концентрация вещества ① в дифференциальном элементе $dV$, прибывающем со скоростью $q_1$ в точку $V$ в момент $t_e$ ($dV$①) равна $\frac{dN_1}{dV} = c_1^0 \frac{Q}{q_1} e^{-\lambda_1 t_e} = c_1^0 \frac{Q}{q_1} e^{-\lambda_1(t_{s2}+t_2)}$. Получаем:
$$\frac{dN_2}{dV}(t_{s2}, t_2) = \left[\frac{dN_1^0}{dV}\frac{\lambda_1}{(\lambda_2-\lambda_1)}\left(e^{-\lambda_1 t_2} - e^{-\lambda_2 t_2}\right) + \frac{dN_2^0}{dV}e^{-\lambda_2 t_2}\right]e^{-\lambda_1 t_{s2}} \tag{4}$$
Поскольку $t_2 = \frac{V}{q_2}$ (Рис. 3а), преобразовываем ур-ие (4) к виду:

$$\frac{dN_2}{dV}(t_e, V) = \left[\frac{dN_1^0}{dV}\frac{\lambda_1}{(\lambda_2-\lambda_1)}\left(1 - e^{-(\lambda_2-\lambda_1)\frac{V}{q_2}}\right) + \frac{dN_2^0}{dV}e^{-(\lambda_2-\lambda_1)\frac{V}{q_2}}\right]e^{-\lambda_1 t_e} \quad (5)$$

Сравнивая с уравнением Бейтмана (1) для $i = 2$, замечаем, что вместо $e^{-\lambda_2 t_e}$ появился член $e^{-\lambda_1 t_e - (\lambda_2-\lambda_1)\frac{V}{q_2}}$. Введем обозначение $\Delta_0^2 = \lambda_1 t_e + (\lambda_2 - \lambda_1)\frac{V}{q_2}$, имея в виду, что в этом диапазоне вещество ② (верхний индекс) начинает движение в сорбенте с начальной точки $V=0$ (нижний индекс), и перепишем ур-ие (5):

$$\frac{dN_2}{dV}(t_e, V) = \frac{dN_1^0}{dV}\frac{\lambda_1}{(\lambda_2-\lambda_1)}\left(e^{-\lambda_1 t_e} - e^{-\Delta_0^2}\right) + \frac{dN_2^0}{dV}e^{-\Delta_0^2} \quad (6)$$

**Диапазон $V_{e2} \div V_{e1}$ вещества ②** (Рис. 3б). Вещество ② образуется из вещества ①, движущегося с фронтом $V_{e1}$. Аналогично схеме на Рис. 3а, дифференциальный элемент $dV$② отделяется от фронта $V_{e1}$ в момент $t_{s2}$, движется в сорбенте со скоростью $q_2$ и спустя $t_2 = t_e - t_{s2}$ оказывается в точке $V$ ($V_{e2} \le V \le V_{e1}$). Решение ур-ия (3) с граничным условием $t_2 = 0$: $\left(\frac{dN_2}{dV}\right)_{t_{s2}} = 0$ приводит к

$$\frac{dN_2}{dV}(t_{s2}, t_2) = \frac{dN_1^0}{dV}\frac{\lambda_1}{(\lambda_2-\lambda_1)}\left(e^{-\lambda_1 t_2} - e^{-\lambda_2 t_2}\right)e^{-\lambda_1 t_{s2}} \quad (7)$$

Теперь $t_2 = \frac{V_{e1}-V}{q_1-q_2}$ (Рис. 3б). Введем обозначение $\Delta_1^2 = \lambda_1 t_e + (\lambda_2 - \lambda_1)\frac{(V_{e1}-V)}{(q_1-q_2)}$ (верхний индекс – вещество ②, нижний индекс – пересекается с фронтом $V_{e1}$, в данном случае, стартует с него) и приведем ур-ие (7) к виду:

$$\frac{dN_2}{dV}(t_e, V) = \frac{dN_1^0}{dV}\frac{\lambda_1}{(\lambda_2-\lambda_1)}\left(e^{-\lambda_1 t_e} - e^{-\Delta_1^2}\right) \quad (8)$$

Складывая количества вещества ② в растворе объемом $V_0 - Qt_e$, оставшемся в сосуде, и в объеме сорбента равном $V_{e1}$, находим общее количество вещества ② к моменту $t_e$:

$$N_2(t_e) = c_2^0(V_0 - Qt_e)e^{-\lambda_1 t_e} + \int_0^{V_{e1}}\frac{dN_2}{dV}dV = N_2^0 e^{-\lambda_1 t_e} \quad (9)$$

Интегрирование в ур-ии (9) проводим, используя ур-ие (6) в пределах $0 \div V_{e2}$ и ур-ие (8) в пределах $V_{e2} \div V_{e1}$. Рассматривая всю область существования вещества ②, мы видим, что оно осталось в подвижном равновесии с веществом ①, в тоже время его концентрация в хроматографической колонке является функцией времени и координаты.

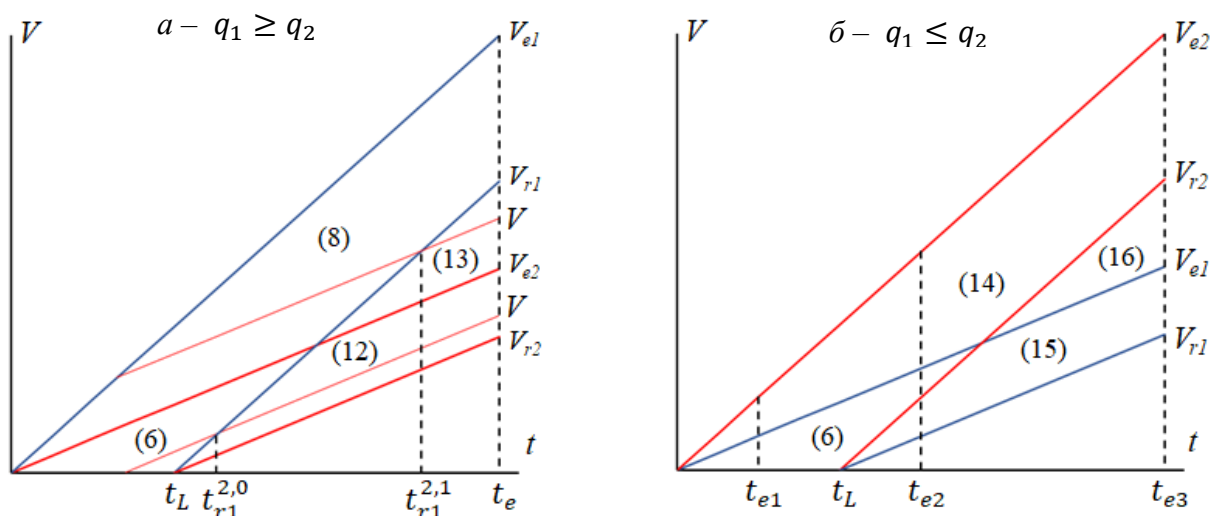

**Рис. 4**. $V$-$t$ диаграммы движения веществ ① и ② в хроматографической колонке после загрузки исходного раствора: $а$ – для случая $q_1 \ge q_2$; $б$ – для случая $q_1 \le q_2$. Синие линии соответствуют движению вещества ①, красные линии – движению вещества ②. В скобках указаны номера уравнений, описывающих концентрацию $\frac{dN_2}{dV} = f(t_e, V)$ в соответствующих областях.

Поступление раствора с веществами ① и ② в колонку заканчивается в момент $t_L = \frac{V_0}{Q}$ ($L$ – loading). Для продолжения их движения пропускаем чистый элюент с той же скоростью $Q$, при этом тылы веществ $V_{r1}$ и $V_{r2}$ отдаляются от начала хроматографической колонки со скоростями $q_1$ и $q_2$. Образуются коридоры веществ ① и ②, ограниченные линиями фронта $V_{ei}$ и тыла $V_{ri}$ и осью $t$. На Рис. 4а показан момент элюирования $t_e > t_L \frac{q_1}{q_1 - q_2}$, когда тыл $V_{r1}$ вещества ① обогнал фронт $V_{e2}$ вещества ②, и коридоры веществ разошлись.

Прослеживая движение дифференциального элемента $dV$②, начало которого показано на Рис. 3а, мы видим, что он пересекает тыл $V_{r1}$ вещества ① в момент $t_{r1}^{2,0} = \frac{t_L q_1 - t_{s2} q_2}{q_1 - q_2}$ (верхний индекс – вещество ② начинает движение в сорбенте с начальной точки $V=0$, нижний индекс – пересекается с тылом $V_{r1}$). До этого момента концентрация $\frac{dN_2}{dV}$ подчиняется ур-ию (4), после – вещество ② теряет связь с веществом ① и распадается с постоянной распада $\lambda_2$:

$$\frac{dN_2}{dV} = \left[\frac{dN_1^0}{dV}\frac{\lambda_1}{(\lambda_2 - \lambda_1)}\left(e^{-\lambda_1(t_{r1}^{2,0}-t_{s2})} - e^{-\lambda_2(t_{r1}^{2,0}-t_{s2})}\right) + \frac{dN_2^0}{dV}e^{-\lambda_2(t_{r1}^{2,0}-t_{s2})}\right]e^{-\lambda_1 t_{s2}-\lambda_2(t_e-t_{r1}^{2,0})} \qquad (10)$$

Выражая с помощью $V$-$t$ диаграммы (Рис. 4а) $t_{s2}$ и $t_{r1}^{2,0}$ через $t_e$ и $V$, находим концентрацию $\frac{dN_2}{dV} = f(t_e, V)$ в диапазоне $V_{r2} \div V_{e2}$ в момент $t_e$:

$$\frac{dN_2}{dV} = \frac{dN_1^0}{dV}\frac{\lambda_1}{(\lambda_2 - \lambda_1)}\left(e^{-\lambda_1 t_e - (\lambda_2 - \lambda_1)\frac{(V_{r1}-V)}{(q_1-q_2)}} - e^{-\Delta_0^2}\right) + \frac{dN_2^0}{dV}e^{-\Delta_0^2} \qquad (11)$$

Введем обозначение постоянной $r_1^2 = -(\lambda_2 - \lambda_1)\frac{t_L q_1}{(q_1 - q_2)}$, которая появляется при пересечении пути дифференциального элемента $dV$② с тылом $V_{r1}$ вещества ①. Поскольку $V_{e1} = V_{r1} + t_L q_1$, ур-ие (11) приходит к виду:

$$\frac{dN_2}{dV} = \frac{dN_1^0}{dV}\frac{\lambda_1}{(\lambda_2 - \lambda_1)}\left(e^{-\Delta_1^2 - r_1^2} - e^{-\Delta_0^2}\right) + \frac{dN_2^0}{dV}e^{-\Delta_0^2} \qquad (12)$$

Так же проследим по $V$-$t$ диаграмме (Рис. 4а) движение дифференциального элемента $dV$②, начало которого показано на Рис. 3б. В момент $t_{r1}^{2,1} = t_{s2} + \frac{t_L q_1}{q_1 - q_2}$ он пересекает тыл $V_{r1}$ вещества ①, до этого концентрация $\frac{dN_2}{dV}$ подчиняется ур-ию (7), после – вещество ② распадается с постоянной распада $\lambda_2$. Преобразуя, получаем концентрацию $\frac{dN_2}{dV} = f(t_e, V)$ в диапазоне $V_{e2} \div V_{r1}$ в момент $t_e$:

$$\frac{dN_2}{dV} = \frac{dN_1^0}{dV}\frac{\lambda_1}{(\lambda_2 - \lambda_1)}\left(e^{-\Delta_1^2 - r_1^2} - e^{-\Delta_1^2}\right) = \frac{dN_1^0}{dV}\frac{\lambda_1}{(\lambda_2 - \lambda_1)}e^{-\Delta_1^2}\left(e^{-r_1^2} - 1\right) \qquad (13)$$

Рассуждая таким же образом, получим с помощью $V$-$t$ диаграммы (Рис. 4б) распределение вещества ② в хроматографической колонке в зависимости от времени для соотношения скоростей $q_1 \leq q_2$. В первом временном интервале $0 \leq t_{e1} \leq t_L$ вещество ② находится в сорбенте в двух диапазонах: в диапазоне $0 \div V_{e1}$ для концентрации $\frac{dN_2}{dV}$ действует ур-ие (6), а в диапазоне $V_{e1} \div V_{e2}$ - ур-ие (14):

$$\frac{dN_2}{dV} = \frac{dN_1^0}{dV}\frac{\lambda_1}{(\lambda_2 - \lambda_1)}\left(e^{-\Delta_1^2} - e^{-\Delta_0^2}\right) + \frac{dN_2^0}{dV}e^{-\Delta_0^2} \qquad (14)$$

Во втором временном интервале $t_L \leq t_{e2} \leq t_L \frac{q_2}{q_2 - q_1}$ вещество ② находится в трех диапазонах, в которых $\frac{dN_2}{dV}$ описывается уравнениями: диапазон $V_{r1} \div V_{r2}$ - ур-ие (15); диапазон $V_{r2} \div V_{e1}$ - ур-ие (6) и диапазон $V_{e1} \div V_{e2}$ - ур-ие (14):

$$\frac{dN_2}{dV} = \frac{dN_1^0}{dV}\frac{\lambda_1}{(\lambda_2-\lambda_1)}\left(e^{-\lambda_1 t_e} - e^{-\Delta_1^2 - r_1^2}\right) \qquad (15)$$

Наконец, в третьем интервале $t_{e3} \geq t_L \frac{q_2}{q_2-q_1}$ распределение вещества ② , существующего в трех диапазонах, подчиняется уравнениям: диапазон $V_{r1} \div V_{e1}$ - ур-ие (15); диапазон $V_{e1} \div V_{r2}$ - ур-ие (16) и диапазон $V_{r2} \div V_{e2}$ - ур-ие (14):

$$\frac{dN_2}{dV} = \frac{dN_1^0}{dV}\frac{\lambda_1}{(\lambda_2-\lambda_1)}\left(e^{-\Delta_1^2} - e^{-\Delta_1^2 - r_1^2}\right) = \frac{dN_1^0}{dV}\frac{\lambda_1}{(\lambda_2-\lambda_1)}e^{-\Delta_1^2}\left(1 - e^{-r_1^2}\right) \qquad (16)$$

На Рис. 5 показаны профили концентрации вещества ② , возникающие в трех временных интервалах, для случаев $q_1 \geq q_2$ и $q_1 \leq q_2$.

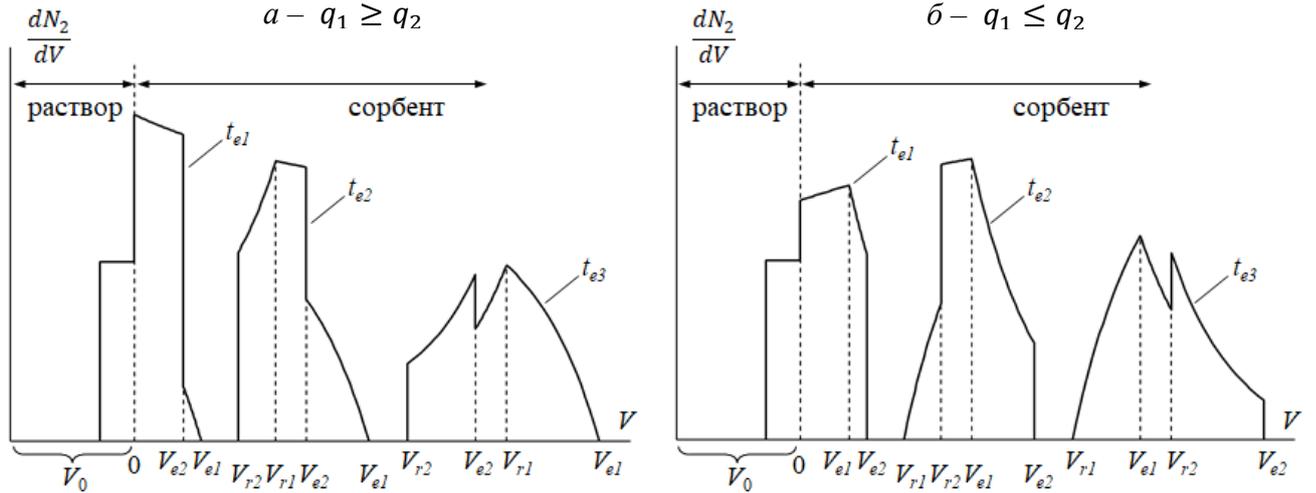

**Рис. 5**. Профили концентрации вещества ② , возникающие в разные моменты $t_e$ элюирования: $a - q_1 \geq q_2$, $0 \leq t_{e1} \leq t_L$, $t_L \leq t_{e2} \leq t_L\frac{q_1}{q_1-q_2}$ и $t_{e3} \geq t_L\frac{q_1}{q_1-q_2}$; $б - q_1 \leq q_2$, $0 \leq t_{e1} \leq t_L$, $t_L \leq t_{e2} \leq t_L\frac{q_2}{q_2-q_1}$ и $t_{e3} \geq t_L\frac{q_2}{q_2-q_1}$. В расчете использованы значения: $V_0$ = 10 мл, $t_L$ = 2.5 мин, $t_{e1}$ = 1.8 мин, $t_{e2}$ = 6.3 мин, $t_{e3}$ = 12.5 мин, $\lambda_1 = 7.7 \cdot 10^{-6}$ с$^{-1}$, $\lambda_2 = 2.4 \cdot 10^{-3}$ с$^{-1}$; $a - q_1$ = 3 мл/мин, $q_2$ = 2.2 мл/мин; $б - q_1$ = 2.2 мл/мин, $q_2$ = 3 мл/мин.

Общее количество вещества ② в любой момент $t_e$ равно: $N_2(t_e) = N_2^0 e^{-\lambda_1 t_e}$, т.е. сохраняется подвижное равновесие.

**Общий вид уравнения для $\frac{dN_2}{dV}$**. Анализируя уравнения (6, 8, 12-16), описывающие профили концентрации вещества ② (Рис. 5), замечаем, что член $e^{-\lambda_1 t_e}$ появляется, когда путь дифференциального элемента $dV$① лежит внутри коридора вещества ① ($(t_e, V) \in 0(V_{r1}) \div V_{e1}$). Подобным образом, член $e^{-\Delta_0^2}$ присутствует в уравнениях, когда путь $dV$② лежит внутри коридора вещества ② ($(t_e, V) \in 0(V_{r2}) \div V_{e2}$). Члены $e^{-\Delta_1^2}$ и $e^{-\Delta_1^2 - r_1^2}$ появляются, когда путь $dV$② имеет точку пересечения с фронтом $V_{e1}$ и тылом $V_{r1}$ вещества ① (path $dV$② ∩ $V_{e1}$ и path $dV$② ∩ $V_{r1}$, соответственно).

Для сведения уравнений (6, 8, 12-16) в общую формулу, воспользуемся структурой уравнения Бейтмана (1), в котором обозначим $\Pi(\lambda_{k,j}) = \prod_{j\neq k}^{i}(\lambda_j - \lambda_k)$ и $B_k = e^{-\lambda_k t}$. По аналогии, для общего выражения концентрации $\frac{dN_i}{dV}$ ($i = 1,2$) напишем:

$$\frac{dN_i}{dV}(t_e, V) = \sum_{j=1}^{i}\left[\frac{dN_j^0}{dV}\cdot\left(\prod_{k=j}^{i-1}\lambda_k\right)\cdot\left(\sum_{k=j}^{i}\frac{b_k^j}{\Pi(\lambda_{k,j})}\right)\right] \qquad (17),$$

где $b_k^j(i=1) = e^{-\lambda_1 t_e}$ в области существования вещества ① в сорбенте ($(t_e, V) \in 0(V_{r1}) \div V_{e1}$), а смысл $b_k^j(i=2)$ раскрывается в Таблице 1.

**Таблица 1**. Члены уравнения (17) для $i = 2$, представленные в виде таблицы.

| $j$ $(1 \div i)$ | $\dfrac{dN_j^0}{dV}\left(\prod\limits_{k=j}^{i-1}\lambda_k\right)$ | $k$ $(j \div i)$ | $\Pi(\lambda_{k,j})$ | $b_k^j$ | наличие в ур-ии (17) (occurrence condition) |
|---|---|---|---|---|---|
| 1 | $\dfrac{dN_1^0}{dV}\lambda_1$ | 1 | $\lambda_2 - \lambda_1$ | $e^{-\lambda_1 t_e}$ | $(t_e, V) \in 0(V_{r1}) \div V_{e1}$ |
|   |   | 2 | $\lambda_1 - \lambda_2$ | $e^{-\Delta_0^2}$ | $(t_e, V) \in 0(V_{r2}) \div V_{e2}$ |
|   |   |   |   | $\mathrm{sgn}(q_1 - q_2) e^{-\Delta_1^2}$ | path $dV② \cap V_{e1}$ |
|   |   |   |   | $-\mathrm{sgn}(q_1 - q_2) e^{-\Delta_1^2 - r_1^2}$ | path $dV② \cap V_{r1}$ |
| 2 | $\dfrac{dN_2^0}{dV}$ | 2 | 1 | $e^{-\Delta_0^2}$ | $(t_e, V) \in 0(V_{r2}) \div V_{e2}$ |

Покажем на примере, как работает ур-ие (17) вкупе с Таблицей 1, для случая $q_1 \leq q_2$, временного интервала $t_{e3} \geq t_L \dfrac{q_2}{q_2 - q_1}$ и диапазона $V_{e1} \div V_{r2}$ (Рис. 4б). Путь любого дифференциального элемента $dV②$, оказывающегося в этом диапазоне в момент $t_{e3}$, пересекает фронт $V_{e1}$ и тыл $V_{r1}$ вещества ① ($dV②$ стартует с тыла $V_{r1}$), поэтому в уравнении для концентрации $\dfrac{dN_2}{dV} = f(t_e, V)$ будут присутствовать, с учетом знаков: $-e^{-\Delta_1^2} + e^{-\Delta_1^2 - r_1^2}$. Путь $dV②$ не попадает в коридор вещества ② так же, как и путь $dV①$ – в коридор вещества ①, поэтому остальные члены в уравнении отсутствуют. В результате, мы приходим к ур-ию (16).

Рассмотрим практически важный случай, когда скорость движения вещества ① в сорбенте крайне мала ($q_1 \to 0$). После загрузки исходного раствора в колонку ($t_e \geq t_L$) вещество ② существует в двух диапазонах $0 \div V_{r2}$ и $V_{r2} \div V_{e2}$, так как фронт $V_{e1}$ и тыл $V_{r1}$ вещества ① почти не движутся: $V_{e1} \approx V_{r1} \to 0$. Ур-ие (16) упрощается следующим образом:

1) $\lim\limits_{q_1 \to 0} e^{-\Delta_1^2} = e^{-\lambda_1 t_e - (\lambda_2 - \lambda_1)\frac{V}{q_2}} = e^{-\Delta_0^2}$; 2) раскладывая в ряд Тейлора и ограничиваясь двумя первыми членами, преобразуем $\lim\limits_{q_1 \to 0}(1 - e^{-r_1^2}) = (\lambda_2 - \lambda_1)\dfrac{q_1}{q_2} t_L$. Учитывая, что $\dfrac{dN_1^0}{dV} = c_1^0 \dfrac{Q}{q_1} = \dfrac{N_1^0}{t_L q_1}$, упрощенное ур-ие (16) принимает вид:

$$\frac{dN_2}{dV} = \frac{\lambda_1}{q_2} N_1^0 e^{-\lambda_1 t_e - (\lambda_2 - \lambda_1)\frac{V}{q_2}} \qquad (16a)$$

Оно совпадает с ур-ием (3а), полученным в нашей предыдущей работе [11] для случая, когда материнское вещество удерживается в начальном слое сорбента.

### 1.2. Цепочка из трех последовательных реакций: ① → ② → ③ →

Применим предложенный подход к цепочке из трех последовательных реакций первого порядка, протекающих в растворе объемом $V_0$ (Рис. 2). При условии $\lambda_1 \ll \lambda_2, \lambda_3$ в растворе со временем устанавливается подвижное равновесие (ур-ие (2)), в котором концентрации веществ равны $c_1 = \dfrac{N_1}{V_0}$, $c_2 = \dfrac{\lambda_1}{\lambda_2 - \lambda_1} c_1$ и $c_3 = \dfrac{\lambda_1 \lambda_2}{(\lambda_2 - \lambda_1)(\lambda_3 - \lambda_1)} c_1$. Когда начинается движение раствора с объёмной скоростью $Q$, вещества с начальными концентрациями $c_1^0$, $c_2^0$ и $c_3^0$ поступают в сорбент, где движутся со скоростями $q_1, q_2, q_3 \leq Q$. В этот момент при входе в хроматографическую колонку ($t_e = 0, V = 0$) их концентрации равны $\frac{dN_i^0}{dV} = c_i^0 \frac{Q}{q_i}, i = 1 \div 3$. Число соотношений скоростей равно 3! = 6.

В предыдущем разделе были введены обозначения величин $\Delta_0^2, \Delta_1^2$ и $r_1^2$, присутствующих при выполнении определенных условий в уравнениях для концентрации вещества ② в сорбенте (см. Таблицу 1). Теперь обозначим их в общем виде.

$$\Delta_0^i = \lambda_1 t_e + (\lambda_i - \lambda_1)\frac{V}{q_i}, \; i = 1 \div 3 \tag{18}$$

$$\Delta_j^i = \lambda_j t_e + (\lambda_i - \lambda_j)\frac{V_{ej}-V}{q_j-q_i}, \; i = 2,3, j = 1 \div i-1 \tag{19}$$

$$r_j^i = -\left((\lambda_j - \lambda_1) + (\lambda_i - \lambda_j)\frac{q_j}{q_j-q_i}\right)t_L = \left(\lambda_1 + \frac{\lambda_j q_i - \lambda_i q_j}{q_j-q_i}\right)t_L, \; i = 2,3, j = 1 \div i-1 \tag{20}$$

Нахождение концентрации $\frac{dN_3}{dV}(t_e, V)$ вещества ③ сводится к решению уравнения материального баланса

$$\frac{d^2 N_3}{dVdt} = \lambda_2 \frac{dN_2}{dV} - \lambda_3 \frac{dN_3}{dV} \tag{21}$$

с использованием известного граничного условия. Участвующая в этом уравнении концентрация $\frac{dN_2}{dV}$ уже определена в предыдущем разделе. Поясним процедуру нахождения $\frac{dN_3}{dV}$ на двух примерах. Перед этим введем для удобства константы, которые будут часто встречаться:

$$k_0 = \frac{q_3}{q_2}, \; \Lambda_0 = \lambda_1 + k_0(\lambda_2 - \lambda_1), \; k_1 = \frac{q_1-q_3}{q_1-q_2}, \; \Lambda_1 = \lambda_1 + k_1(\lambda_2 - \lambda_1)$$

**Пример 1**. На Рис. 6а изображены фронты веществ для соотношения скоростей $q_1 \geq q_2 \geq q_3$ и пути дифференциальных элементов $dV$② и $dV$③, стартующих с фронта $V_{e1}$ вещества ① в моменты $t_{s2}$ и $t_{s3}$ и оказывающихся в момент $t_e$ в некоторой точке $V$ диапазона $V_{e2} \div V_{e1}$.

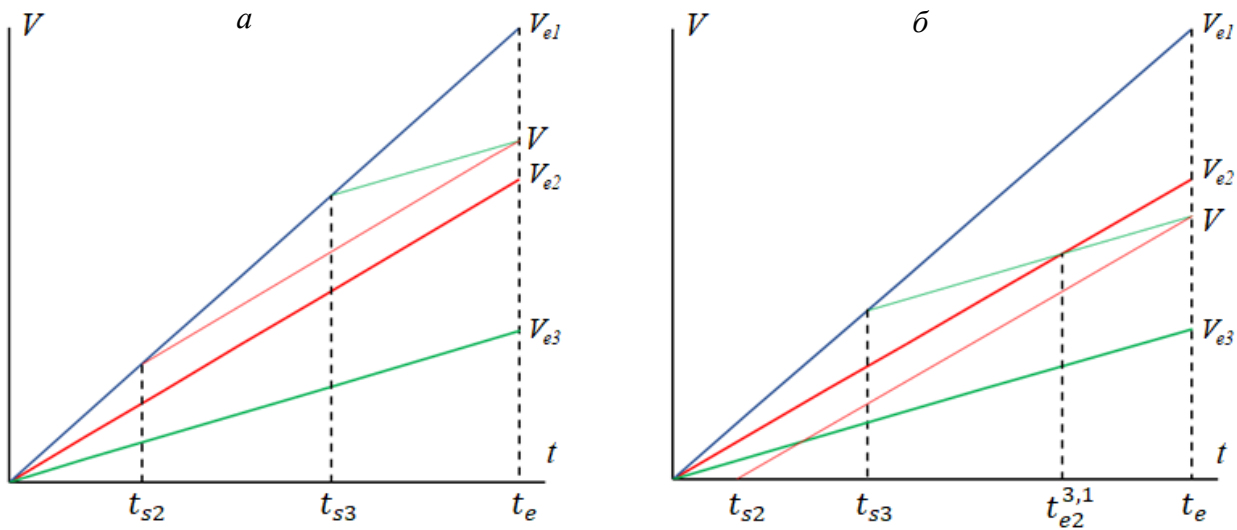

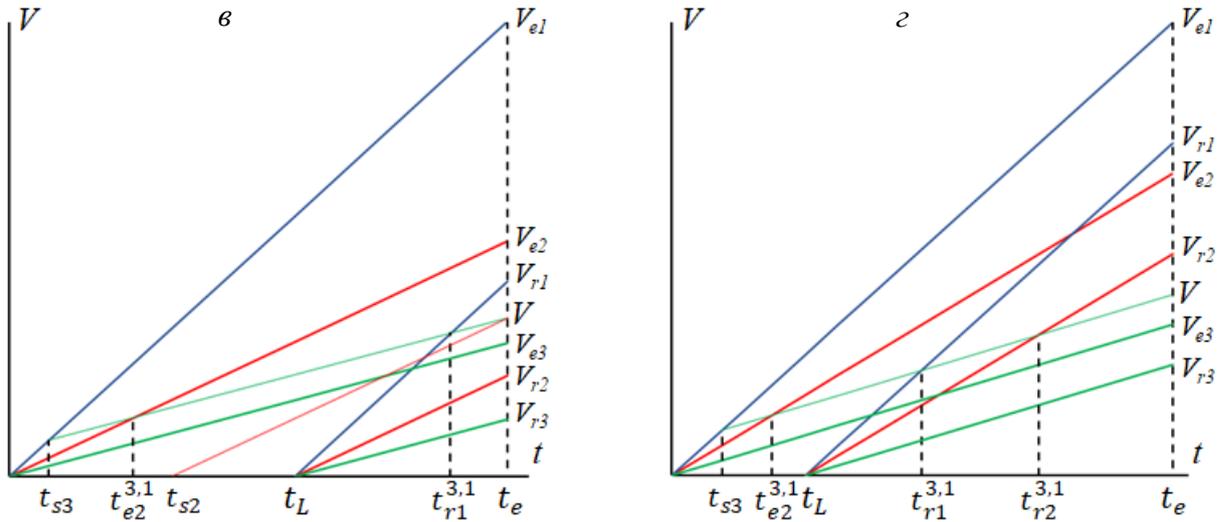

**Рис. 6**. *V-t* диаграммы движения дифференциального элемента $dV$③ в хроматографической колонке для случая $q_1 \geq q_2 \geq q_3$:
*а* – от старта с фронта $V_{e1}$ вещества ① до пересечения с фронтом $V_{e2}$ вещества ②;
*б* – от пересечения с фронтом $V_{e2}$ вещества ② до пересечения с тылом $V_{r1}$ вещества ①;
*в* – от пересечения с тылом $V_{r1}$ вещества ① до пересечения с тылом $V_{r2}$ вещества ②;
*г* – после пересечения с тылом $V_{r1}$ вещества ①.
Синие, красные и зеленые линии соответствуют движению веществ ①, ② и ③.

Концентрация $\frac{dN_2}{dV}$ в этом диапазоне определяется ур-иями (7, 8). Для решения ур-ия (21) относительно времени нахождения $dV$③ в сорбенте $t_3 = t_e - t_{s3}$ воспользуемся ур-ием (7) для $\frac{dN_2}{dV}(t_{s2}, t_2)$, выразив $t_{s2}$ и $t_2$ через $t_{s3}$ и $t_3$ (Рис. 6а): $t_{s2} = t_{s3} + t_3(1 - k_1)$; $t_2 = t_3 k_1$. Тогда

$$\frac{dN_2}{dV}(t_{s3}, t_3) = \frac{dN_1^0}{dV}\frac{\lambda_1}{\lambda_2-\lambda_1}\left(e^{-\lambda_1 t_3} - e^{-\Lambda_1 t_3}\right)e^{-\lambda_1 t_{s3}} \tag{7а}$$

Решение ур-ия (21) с граничным условием $t_3 = 0$: $\left(\frac{dN_3}{dV}\right)_{t_{s3}} = 0$ приводит к

$$\frac{dN_3}{dV}(t_{s3}, t_3) = \lambda_1 \lambda_2 \frac{dN_1^0}{dV}\left(\frac{e^{-\lambda_1 t_3}}{(\lambda_2-\lambda_1)(\lambda_3-\lambda_1)} + \frac{e^{-\Lambda_1 t_3}}{(\lambda_1-\lambda_2)(\lambda_3-\Lambda_1)} + \frac{k_1 e^{-\lambda_3 t_3}}{(\lambda_1-\lambda_3)(\Lambda_1-\lambda_3)}\right)e^{-\lambda_1 t_{s3}} \tag{22}$$

Выражая $t_{s3}$ и $t_3$ через $t_e$ и $V$ (Рис. 6а), получаем:

$$\frac{dN_3}{dV}(t_e, V) = \lambda_1 \lambda_2 \frac{dN_1^0}{dV}\left(\frac{e^{-\Delta_0^1}}{(\lambda_2-\lambda_1)(\lambda_3-\lambda_1)} + \frac{e^{-\Delta_1^2}}{(\lambda_1-\lambda_2)(\lambda_3-\Lambda_1)} + \frac{k_1 e^{-\Delta_1^3}}{(\lambda_1-\lambda_3)(\Lambda_1-\lambda_3)}\right) \tag{23}$$

Концентрация $\frac{dN_3}{dV}$ описывается ур-иями (22, 23) вплоть до момента $t_{e2}^{3,1}$, когда путь $dV$③ пересекает фронт $V_{e2}$ вещества ② (Рис. 6б). В этот момент ур-ие (22) принимает вид:

$$\left(\frac{dN_3}{dV}\right)_{t_{e2}^{3,1}} = \lambda_1 \lambda_2 \frac{dN_1^0}{dV}\left(\frac{e^{-\lambda_1(t_{e2}^{3,1}-t_{s3})}}{(\lambda_2-\lambda_1)(\lambda_3-\lambda_1)} + \frac{e^{-\Lambda_1(t_{e2}^{3,1}-t_{s3})}}{(\lambda_1-\lambda_2)(\lambda_3-\Lambda_1)} + \frac{k_1 e^{-\lambda_3(t_{e2}^{3,1}-t_{s3})}}{(\lambda_1-\lambda_3)(\Lambda_1-\lambda_3)}\right)e^{-\lambda_1 t_{s3}} \tag{22гу}$$

Теперь $dV$③ движется в диапазоне $V_{e3} \div V_{e2}$. Этот диапазон принадлежит диапазону $0 \div V_{e2}$, в котором концентрация $\frac{dN_2}{dV}$ определяется ур-иями (4-6). Подставляем в ур-ие (4) выражения $t_{s2}$ и $t_2$ через $t_{e2}^{3,1}$ и $t_3$ (Рис. 6б) и решаем ур-ие (21) относительно $t_3 = t_e - t_{e2}^{3,1}$, используя ур-ие (22гу) в качестве граничного условия. Получаем уравнение для концентрации вещества ③ в интервале $t_{e2}^{3,1} \div t_{r1}^{3,1}$:

$$\frac{dN_3}{dV} = \lambda_1 \lambda_2 \frac{dN_1^0}{dV}\left(\frac{e^{-\lambda_1 t_e}}{(\lambda_2-\lambda_1)(\lambda_3-\lambda_1)} + \frac{e^{-\lambda_2 t_{e2}^{3,1}-\Lambda_0 t_3}}{(\lambda_1-\lambda_2)(\lambda_3-\Lambda_0)} + \frac{k_1 e^{-\lambda_1 t_{s3}-\lambda_3(t_e-t_{s3})}}{(\lambda_1-\lambda_3)(\Lambda_1-\lambda_3)} + \frac{(k_0-k_1)e^{-\lambda_2 t_{e2}^{3,1}-\lambda_3 t_3}}{(\Lambda_0-\lambda_3)(\Lambda_1-\lambda_3)}\right) +$$

$$+\lambda_2 \frac{dN_2^0}{dV}\left(\frac{e^{-\lambda_2 t_{e2}^{3,1}-\Lambda_0 t_3}}{(\lambda_3-\Lambda_0)} + \frac{e^{-\lambda_2 t_{e2}^{3,1}-\lambda_3 t_3}}{(\Lambda_0-\lambda_3)}\right) \tag{24}$$

Выражая $t_{s3}$, $t_{e2}^{3,1}$ и $t_3$ через $t_e$ и $V$ (Рис. 6б), приводим ур-ие (24) к виду:

$$\frac{dN_3}{dV}(t_e,V) = \lambda_1\lambda_2\frac{dN_1^0}{dV}\left(\frac{e^{-\Delta_0^1}}{(\lambda_2-\lambda_1)(\lambda_3-\lambda_1)} + \frac{e^{-\Delta_0^2}}{(\lambda_1-\lambda_2)(\lambda_3-\Lambda_0)} + \frac{k_1 e^{-\Delta_1^3}}{(\lambda_1-\lambda_3)(\Lambda_1-\lambda_3)} + \frac{(k_0-k_1)e^{-\Delta_2^3}}{(\Lambda_0-\lambda_3)(\Lambda_1-\lambda_3)}\right) +$$
$$+\lambda_2\frac{dN_2^0}{dV}\left(\frac{e^{-\Delta_0^2}}{(\lambda_3-\Lambda_0)} + \frac{e^{-\Delta_2^3}}{(\Lambda_0-\lambda_3)}\right) \qquad (25)$$

В момент $t_{r1}^{3,1}$ путь $dV③$ пересекает тыл $V_{r1}$ вещества ① (Рис. 6в), и ур-ие (24), в котором $t_e = t_{r1}^{3,1}$ и $t_3 = t_{r1}^{3,1} - t_{e2}^{3,1}$, становится граничным условием для следующего участка движения $dV③$ - ур-ием (24гу). Дифференциальный элемент $dV②$, прибывающий в то же время и место $(t_e, V)$, что и $dV③$, также пересекает тыл $V_{r1}$ в момент $t_{r1}^{2,0}$ (Рис. 4а), после чего концентрация $\frac{dN_2}{dV}$ определяется ур-иями (10-12). Выражаем в ур-ии (10) $t_{s2}$, $t_{r1}^{2,0}$ и $t_e - t_{r1}^{2,0}$ через $t_{r1}^{3,1}$ и $t_3$ и решаем ур-ие (21) относительно $t_3 = t_e - t_{r1}^{3,1}$ с граничным условием - ур-ием (24гу):

$$\frac{dN_3}{dV} = \lambda_1\lambda_2\frac{dN_1^0}{dV}\left(\frac{e^{-\lambda_1 t_e - (\lambda_2-\lambda_1)\left(\frac{q_1}{q_2}(t_{r1}^{3,1}-t_L)+k_0 t_3\right)}}{(\lambda_1-\lambda_2)(\lambda_3-\Lambda_0)} - \frac{e^{-\lambda_1 t_{r1}^{3,1} - \Lambda_1 t_3}}{(\lambda_1-\lambda_2)(\lambda_3-\Lambda_1)} + \frac{k_1\left(e^{-\lambda_1 t_{s3} - \lambda_3(t_e-t_{s3})} - e^{-\lambda_1 t_{r1}^{3,1} - \lambda_3 t_3}\right)}{(\lambda_1-\lambda_3)(\Lambda_1-\lambda_3)} +$$
$$+\frac{(k_0-k_1)e^{-\lambda_2 t_{e2}^{3,1} - \lambda_3(t_e-t_{e2}^{3,1})}}{(\Lambda_0-\lambda_3)(\Lambda_1-\lambda_3)}\right) + \lambda_2\frac{dN_2^0}{dV}\left(\frac{e^{-\lambda_1 t_e - (\lambda_2-\lambda_1)\left(\frac{q_1}{q_2}(t_{r1}^{3,1}-t_L)+k_0 t_3\right)}}{(\lambda_3-\Lambda_0)} + \frac{e^{-\lambda_2 t_{e2}^{3,1} - \lambda_3(t_e-t_{e2}^{3,1})}}{(\Lambda_0-\lambda_3)}\right) \quad (26)$$

Выражая $t_{s3}$, $t_{e2}^{3,1}$, $t_{r1}^{3,1}$ и $t_3$ через $t_e$ и $V$ (Рис. 6в), получаем решение в интервале $t_{r1}^{3,1} \div t_{r2}^{3,1}$:

$$\frac{dN_3}{dV}(t_e,V) = \lambda_1\lambda_2\frac{dN_1^0}{dV}\left(\frac{e^{-\Delta_0^2}}{(\lambda_1-\lambda_2)(\lambda_3-\Lambda_0)} - \frac{e^{-\Delta_1^2 - r_1^2}}{(\lambda_1-\lambda_2)(\lambda_3-\Lambda_1)} + \frac{k_1 e^{-\Delta_1^3}\left(1-e^{-r_1^3}\right)}{(\lambda_1-\lambda_3)(\Lambda_1-\lambda_3)} + \frac{(k_0-k_1)e^{-\Delta_2^3}}{(\Lambda_0-\lambda_3)(\Lambda_1-\lambda_3)}\right) +$$
$$+\lambda_2\frac{dN_2^0}{dV}\left(\frac{e^{-\Delta_0^2}}{(\lambda_3-\Lambda_0)} + \frac{e^{-\Delta_2^3}}{(\Lambda_0-\lambda_3)}\right) \qquad (27)$$

Наконец, в момент $t_{r2}^{3,1}$ путь $dV③$ пересекает тыл $V_{r2}$ вещества ②. Как видно из V-t диаграммы на Рис. 6г, после этого на пути $dV③$ больше не будет никаких пересечений. Концентрация $\left(\frac{dN_3}{dV}\right)_{t_{r2}^{3,1}}$ в этот момент определяется ур-ием (26), в котором $t_e = t_{r2}^{3,1}$ и $t_3 = t_{r2}^{3,1} - t_{r1}^{3,1}$. Далее вещество ③ в движущемся дифференциальном элементе $dV③$ распадается с постоянной распада $\lambda_3$, и его концентрация, выраженная через $t_e$ и $V$, равна:

$$\frac{dN_3}{dV}(t_e,V) = \left(\frac{dN_3}{dV}\right)_{t_{r2}^{3,1}} e^{-\lambda_3(t_e - t_{r2}^{3,1})} = \lambda_1\lambda_2\frac{dN_1^0}{dV}\left(\frac{k_1 e^{-\Delta_1^3}\left(1-e^{-r_1^3}\right)}{(\lambda_1-\lambda_3)(\Lambda_1-\lambda_3)} + \frac{(k_0-k_1)e^{-\Delta_2^3}\left(1-e^{-r_2^3}\right)}{(\Lambda_0-\lambda_3)(\Lambda_1-\lambda_3)}\right) +$$
$$+\lambda_2\frac{dN_2^0}{dV}\frac{e^{-\Delta_2^3}\left(1-e^{-r_2^3}\right)}{(\Lambda_0-\lambda_3)} \qquad (28)$$

Таким образом, процедура нахождения концентрации $\frac{dN_3}{dV}(t_e, V)$ состоит из следующих этапов:

1) <u>Построение *V-t* диаграммы</u>

Сначала проводим линии фронтов и тылов веществ, определяя области их существования. Затем в области существования вещества ③ выбираем время $t_e$ элюирования и координату $V$ в хроматографической колонке, для которых надо найти $\frac{dN_3}{dV}$, и проводим через точку $(t_e, V)$ линию с тангенсом угла наклона $q_3$. Эта линия определяет путь дифференциального элемента $dV③$, его пересечения с фронтами и тылами веществ и интервалы (участки) пути между пересечениями. Число $N$ интервалов пути $dV③$ равняется числу итераций нахождения концентрации $\frac{dN_3}{dV}$.

2) <u>Итерации $p = 1 \div N$</u>

i) Обозначаем точку на $p$-ом участке пути $dV③$, проводим к ней линию с тангенсом угла наклона $q_2$ и определяем путь $dV②$ и уравнение, описывающее концентрацию $\frac{dN_2}{dV}$. Путь $dV②$ может начинаться либо с координаты $V=0$ в интервале $0 \div t_L$, либо с фронта или тыла вещества ①. Если к обозначенной точке нельзя провести путь $dV②$, то переходим к пункту *Decay*.

ii) Используя *V-t* диаграмму, преобразуем уравнение для $\frac{dN_2}{dV}$ так, чтобы оно содержало временные интервалы, выраженные через время начала *p*-ого участка пути $dV$③ (например, на первом участке - $t_{s3}$) и время $t_3$ движения $dV$③ от начала *p*-ого участка до обозначенной точки.

iii) Определяем граничное условие, т.е. концентрацию $\frac{dN_3}{dV}$ в начале участка пути $dV$③. На первом участке – это концентрация в момент начала движения $dV$③. На остальных участках – это концентрация $\frac{dN_3}{dV}$ в момент пересечения пути $dV$③ с фронтом или тылом вещества ① или ②, описываемая уравнением предыдущего участка.

iv) Решаем уравнение материального баланса (21) относительно $t_3$ с определенным выше граничным условием. Если искомая точка ($t_e, V$) находится на *p*-ом участке, то преобразуем полученное уравнение для $\frac{dN_3}{dV}$ в виде функции от ($t_e, V$), и на этом процедура завершена. Если нет, то переходим к следующему участку пути $dV$③.

3) *Decay*

Если к точке, обозначенной на пути $dV$③, нельзя провести путь $dV$②, то определяем концентрацию $\frac{dN_3}{dV}$ в момент последнего пересечения. После этого количество вещества ③ в движущемся $dV$③ уменьшается по экспоненциальному закону с постоянной распада $\lambda_3$. Преобразуем уравнение для $\frac{dN_3}{dV}$ в виде функции от ($t_e, V$), процедура завершена.

Рассмотренный в данном примере путь дифференциального элемента $dV$③ закончился в точке ($t_e, V$) в диапазоне между коридорами веществ ② и ③ (Рис. 6г). Ур-ие (28) описывает профиль концентрации $\frac{dN_3}{dV}$ в этом диапазоне в момент $t_e$. Для построения полного профиля концентрации во всей области существования вещества ③ необходимо проследить путь $dV$③, приводящий в каждый диапазон области его существования.

В **Примере 2** предложенная процедура применена для другого соотношения скоростей $q_3 \geq q_1 \geq q_2$, и построены профили концентрации вещества ③ в зависимости от времени $t_e$ движения в сорбенте.

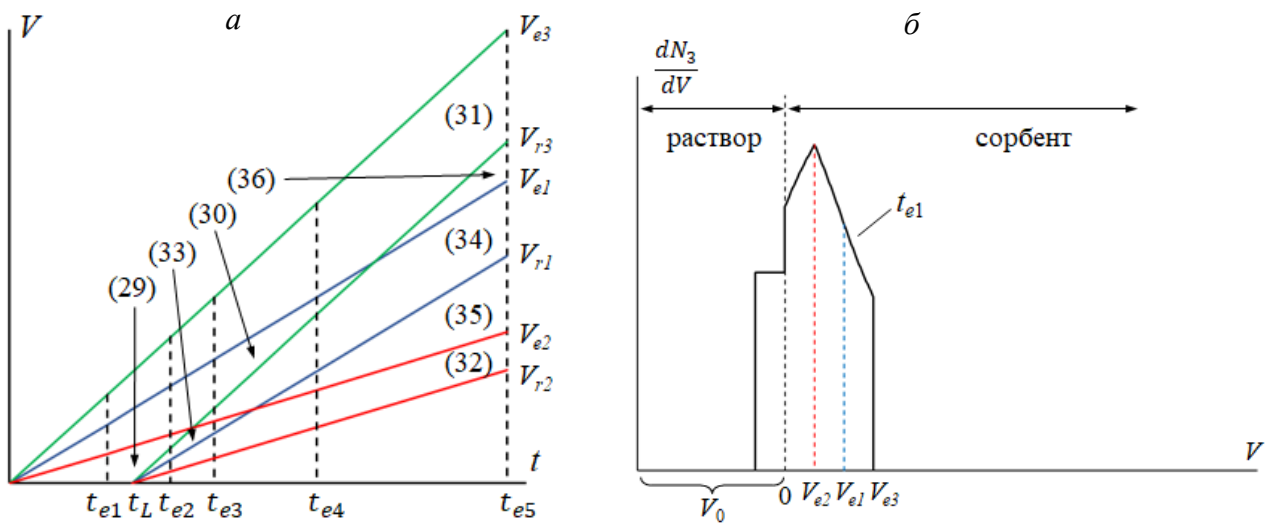

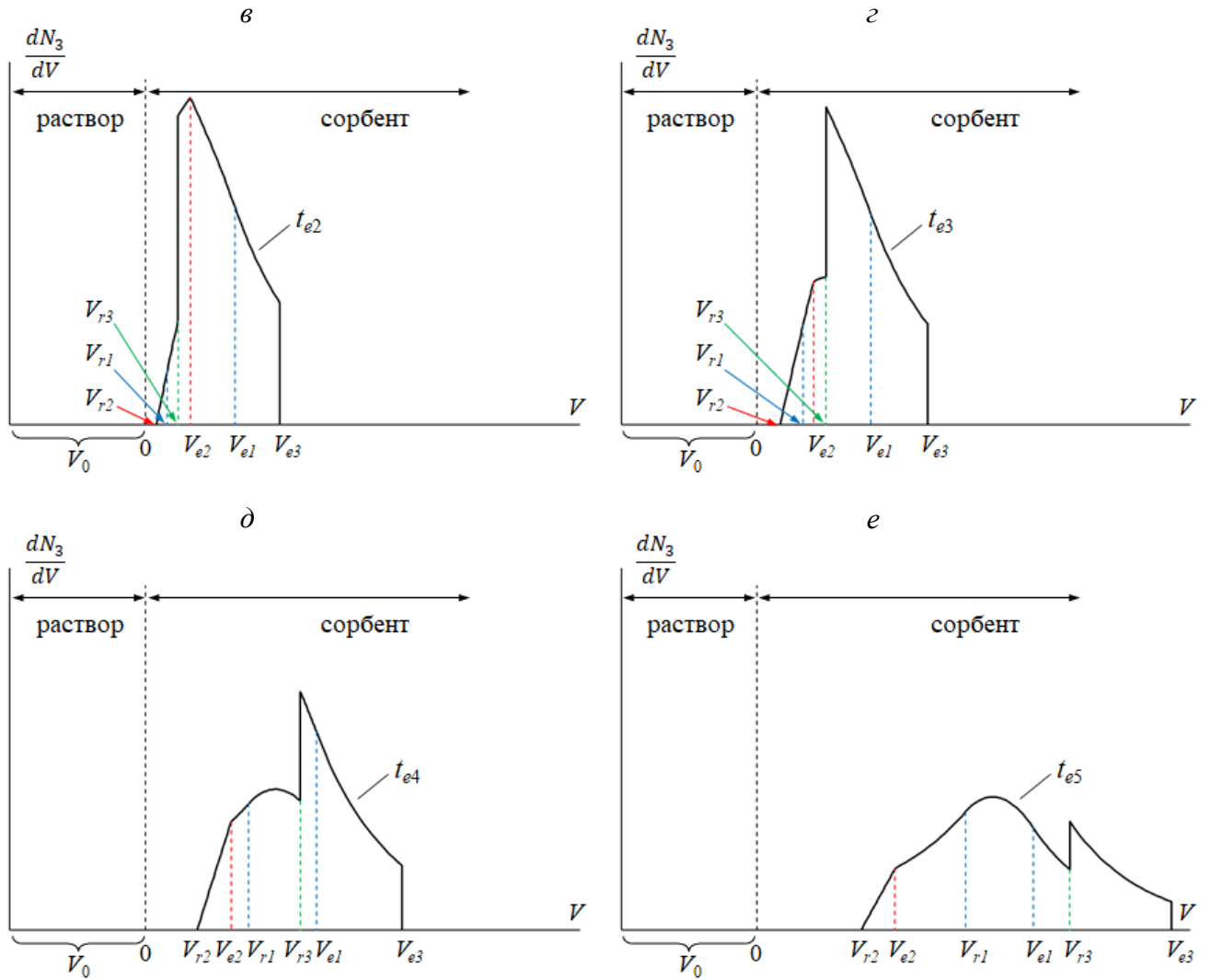

**Рис. 7**. *V-t* диаграмма движения веществ в сорбенте со скоростями $q_3 \geq q_1 \geq q_2$ (*а*, в скобках указаны номера уравнений, описывающих концентрацию $\frac{dN_3}{dV} = f(t_e, V)$ в соответствующих областях) и профили концентрации вещества ③ (*б – е*), возникающие в разные моменты $t_e$ элюирования: *б* – $0 \leq t_{e1} \leq t_L$; *в* – $t_L \leq t_{e2} \leq t_L \frac{q_3}{q_3-q_2}$; *г* – $t_L \frac{q_3}{q_3-q_2} \leq t_{e3} \leq t_L \frac{q_1}{q_1-q_2}$; *д* – $t_L \frac{q_1}{q_1-q_2} \leq t_{e4} \leq t_L \frac{q_3}{q_3-q_1}$; *е* – $t_{e5} \geq t_L \frac{q_3}{q_3-q_1}$. Синие, красные и зеленые линии соответствуют движению веществ ①, ② и ③. В расчете использованы значения: $V_0$ = 10 мл, $t_L$ = 2.5 мин, $q_1$ = 2 мл/мин, $q_2$ = 1 мл/мин, $q_3$ = 3 мл/мин, $\lambda_1 = 7.7 \cdot 10^{-6}$ с$^{-1}$, $\lambda_2 = 2.4 \cdot 10^{-3}$ с$^{-1}$, $\lambda_3 = 3.5 \cdot 10^{-3}$ с$^{-1}$, $t_{e1}$ = 2 мин, $t_{e2}$ = 3.3 мин, $t_{e3}$ = 4.2 мин, $t_{e4}$ = 6.3 мин, $t_{e5}$ = 10.2 мин.

На *V-t* диаграмме (Рис. 7а) показано движение веществ в хроматографической колонке и области, образованные пересечением тылов веществ, движущихся быстрее, с фронтами более медленных веществ. Каждой области соответствует свое уравнение $\frac{dN_3}{dV} = f(t_e, V)$, номер которого указан на Рис. 7а. Эти уравнения, полученные с помощью предложенной процедуры, приведены ниже.

$$\frac{dN_3}{dV}(t_e, V) = \lambda_1 \lambda_2 \frac{dN_1^0}{dV}\left(\frac{e^{-\Delta_0^1}}{(\lambda_2-\lambda_1)(\lambda_3-\lambda_1)} + \frac{e^{-\Delta_0^2}}{(\lambda_1-\lambda_2)(\lambda_3-\Lambda_0)} + \frac{k_0 e^{-\Delta_0^3}}{(\lambda_1-\lambda_3)(\Lambda_0-\lambda_3)}\right) +$$
$$+\lambda_2 \frac{dN_2^0}{dV}\left(\frac{e^{-\Delta_0^2}}{(\lambda_3-\Lambda_0)} + \frac{e^{-\Delta_0^3}}{(\Lambda_0-\lambda_3)}\right) + \frac{dN_3^0}{dV} e^{-\Delta_0^3} \qquad (29)$$

$$\frac{dN_3}{dV}(t_e, V) = \lambda_1 \lambda_2 \frac{dN_1^0}{dV}\left(\frac{e^{-\Delta_0^1}}{(\lambda_2-\lambda_1)(\lambda_3-\lambda_1)} + \frac{e^{-\Delta_1^2}}{(\lambda_1-\lambda_2)(\lambda_3-\lambda_1)} + \frac{k_0 e^{-\Delta_0^3}}{(\lambda_1-\lambda_3)(\Lambda_0-\lambda_3)} - \frac{(k_0-k_1)e^{-\Delta_2^3}}{(\Lambda_0-\lambda_3)(\Lambda_1-\lambda_3)}\right) +$$
$$+\lambda_2 \frac{dN_2^0}{dV} \frac{\left(e^{-\Delta_0^3}-e^{-\Delta_2^3}\right)}{(\Lambda_0-\lambda_3)} + \frac{dN_3^0}{dV} e^{-\Delta_0^3} \qquad (30)$$

$$\frac{dN_3}{dV}(t_e,V) = \lambda_1\lambda_2\frac{dN_1^0}{dV}\left(\frac{k_0 e^{-\Delta_0^3}}{(\lambda_1-\lambda_3)(\Lambda_0-\lambda_3)} - \frac{k_1 e^{-\Delta_1^3}}{(\lambda_1-\lambda_3)(\Lambda_1-\lambda_3)} - \frac{(k_0-k_1)e^{-\Delta_2^3}}{(\Lambda_0-\lambda_3)(\Lambda_1-\lambda_3)}\right) +$$
$$+\lambda_2\frac{dN_2^0}{dV}\frac{\left(e^{-\Delta_0^3}-e^{-\Delta_2^3}\right)}{(\Lambda_0-\lambda_3)} + \frac{dN_3^0}{dV}e^{-\Delta_0^3} \quad (31)$$

$$\frac{dN_3}{dV}(t_e,V) = \lambda_1\lambda_2\frac{dN_1^0}{dV}\left(\frac{e^{-\Delta_0^2}}{(\lambda_1-\lambda_2)(\lambda_3-\Lambda_0)} - \frac{e^{-\Delta_1^2-r_1^2}}{(\lambda_1-\lambda_2)(\lambda_3-\Lambda_1)} + \frac{(k_0-k_1)e^{-\Delta_2^3-r_2^3}}{(\Lambda_0-\lambda_3)(\Lambda_1-\lambda_3)}\right) +$$
$$+\lambda_2\frac{dN_2^0}{dV}\left(\frac{e^{-\Delta_0^2}}{(\lambda_3-\Lambda_0)} + \frac{e^{-\Delta_2^3-r_2^3}}{(\Lambda_0-\lambda_3)}\right) \quad (32)$$

$$\frac{dN_3}{dV}(t_e,V) = \lambda_1\lambda_2\frac{dN_1^0}{dV}\left(\frac{e^{-\Delta_0^1}}{(\lambda_2-\lambda_1)(\lambda_3-\lambda_1)} + \frac{e^{-\Delta_0^2}}{(\lambda_1-\lambda_2)(\lambda_3-\Lambda_0)} + \frac{k_1 e^{-\Delta_1^3-r_1^3}}{(\lambda_1-\lambda_3)(\Lambda_1-\lambda_3)} + \frac{(k_0-k_1)e^{-\Delta_2^3-r_2^3}}{(\Lambda_0-\lambda_3)(\Lambda_1-\lambda_3)}\right) +$$
$$+\lambda_2\frac{dN_2^0}{dV}\left(\frac{e^{-\Delta_0^2}}{(\lambda_3-\Lambda_0)} + \frac{e^{-\Delta_2^3-r_2^3}}{(\Lambda_0-\lambda_3)}\right) \quad (33)$$

$$\frac{dN_3}{dV}(t_e,V) = \lambda_1\lambda_2\frac{dN_1^0}{dV}\left(\frac{e^{-\Delta_0^1}}{(\lambda_2-\lambda_1)(\lambda_3-\lambda_1)} + \frac{e^{-\Delta_1^2}}{(\lambda_1-\lambda_2)(\lambda_3-\Lambda_1)} + \frac{k_1 e^{-\Delta_1^3-r_1^3}}{(\lambda_1-\lambda_3)(\Lambda_1-\lambda_3)} - \frac{(k_0-k_1)e^{-\Delta_2^3}\left(1-e^{-r_2^3}\right)}{(\Lambda_0-\lambda_3)(\Lambda_1-\lambda_3)}\right) +$$
$$+\lambda_2\frac{dN_2^0}{dV}\left(-\frac{e^{-\Delta_2^3}\left(1-e^{-r_2^3}\right)}{(\Lambda_0-\lambda_3)}\right) \quad (34)$$

$$\frac{dN_3}{dV}(t_e,V) = \lambda_1\lambda_2\frac{dN_1^0}{dV}\left(\frac{e^{-\Delta_1^2}\left(1-e^{-r_1^2}\right)}{(\lambda_1-\lambda_2)(\lambda_3-\Lambda_1)} - \frac{(k_0-k_1)e^{-\Delta_2^3}\left(1-e^{-r_2^3}\right)}{(\Lambda_0-\lambda_3)(\Lambda_1-\lambda_3)}\right) + \lambda_2\frac{dN_2^0}{dV}\left(-\frac{e^{-\Delta_2^3}\left(1-e^{-r_2^3}\right)}{(\Lambda_0-\lambda_3)}\right) \quad (35)$$

$$\frac{dN_3}{dV}(t_e,V) = \lambda_1\lambda_2\frac{dN_1^0}{dV}\left(-\frac{k_1 e^{-\Delta_1^3}\left(1-e^{-r_1^3}\right)}{(\lambda_1-\lambda_3)(\Lambda_1-\lambda_3)} - \frac{(k_0-k_1)e^{-\Delta_2^3}\left(1-e^{-r_2^3}\right)}{(\Lambda_0-\lambda_3)(\Lambda_1-\lambda_3)}\right) + \lambda_2\frac{dN_2^0}{dV}\left(-\frac{e^{-\Delta_2^3}\left(1-e^{-r_2^3}\right)}{(\Lambda_0-\lambda_3)}\right) \quad (36)$$

*V-t* диаграмма (Рис. 7а) позволяет выделить пять временных интервалов, в которых образуются различные профили концентрации вещества ③. В первом интервале $0 \leq t_{e1} \leq t_L$ (Рис. 7а, б) одна часть вещества ③ содержится в исходном растворе, которого к этому моменту осталось $V_0 - Qt_{e1}$, а другая часть находится в сорбенте в диапазонах $0 \div V_{e2}$, $V_{e2} \div V_{e1}$ и $V_{e1} \div V_{e3}$. С момента $t_L$ загрузки исходного раствора все вещество ③ находится в сорбенте. В Таблице 2 представлены диапазоны существования вещества ③ в хроматографической колонке, возникающие в различные временные интервалы.

**Таблица 2**. Диапазоны распределения вещества ③ в хроматографической колонке, возникающие в различные временные интервалы элюирования.

| Временной интервал | Рис. | Диапазоны вещества ③ в хроматографической колонке (номер уравнения $\frac{dN_3}{dV} = f(t_e,V)$) | | | | |
|---|---|---|---|---|---|---|
| $0 \leq t_{e1} \leq t_L$ | 7б | $0 \div V_{e2}$ (29) | $V_{e2} \div V_{e1}$ (30) | $V_{e1} \div V_{e3}$ (31) | | |
| $t_L \leq t_{e2} \leq t_L\frac{q_3}{q_3-q_2}$ | 7в | $V_{r2} \div V_{r1}$ (32) | $V_{r1} \div V_{r3}$ (33) | $V_{r3} \div V_{e2}$ (29) | $V_{e2} \div V_{e1}$ (30) | $V_{e1} \div V_{e3}$ (31) |
| $t_L\frac{q_3}{q_3-q_2} \leq t_{e3} \leq t_L\frac{q_1}{q_1-q_2}$ | 7г | $V_{r2} \div V_{r1}$ (32) | $V_{r1} \div V_{e2}$ (33) | $V_{e2} \div V_{r3}$ (34) | $V_{r3} \div V_{e1}$ (30) | $V_{e1} \div V_{e3}$ (31) |
| $t_L\frac{q_1}{q_1-q_2} \leq t_{e4} \leq t_L\frac{q_3}{q_3-q_1}$ | 7д | $V_{r2} \div V_{e2}$ (32) | $V_{e2} \div V_{r1}$ (35) | $V_{r1} \div V_{r3}$ (34) | $V_{r3} \div V_{e1}$ (30) | $V_{e1} \div V_{e3}$ (31) |
| $t_{e5} \geq t_L\frac{q_3}{q_3-q_1}$ | 7е | $V_{r2} \div V_{e2}$ (32) | $V_{e2} \div V_{r1}$ (35) | $V_{r1} \div V_{e1}$ (34) | $V_{e1} \div V_{r3}$ (36) | $V_{r3} \div V_{e3}$ (31) |

Общее количество вещества ③ в любой момент $t_e$, определяемое интегрированием профиля концентрации во всей области его существования, равно: $N_3(t_e) = N_3^0 e^{-\lambda_1 t_e}$, т.е. сохраняется подвижное равновесие.

**Общий вид уравнения для $\frac{dN_3}{dV}$.** Несмотря на громоздкость, уравнения, описывающие концентрацию вещества ③ в сорбенте, строятся по тем же правилам, что и уравнения для $\frac{dN_2}{dV}(t_e, V)$. Сформулируем их в общем виде.

1. Величина $\Delta_0^i$ (ур-ие (18)) появляется в уравнении $\frac{dN_3}{dV} = f(t_e, V)$, когда путь дифференциального элемента $dV$①, прибывающего в точку $(t_e, V)$, начинается с координаты $V=0$, или другими словами, когда путь $dV$① лежит внутри коридора вещества ① ($(t_e, V) \in 0(V_{ri}) \div V_{ei}$).

2. Величина $\Delta_j^i$ (ур-ие (19)) появляется в уравнении $\frac{dN_3}{dV} = f(t_e, V)$ при пересечении пути $dV$① с фронтом $V_{ej}$ вещества ⓙ (path $dV$① $\cap V_{ej}$).

3. Величина $\Delta_j^i + r_j^i$ (ур-ие (20)) появляется в уравнении $\frac{dN_3}{dV} = f(t_e, V)$ при пересечении пути $dV$① с тылом $V_{rj}$ вещества ⓙ (path $dV$① $\cap V_{rj}$).

С помощью приведенной выше процедуры нахождения $\frac{dN_3}{dV}$ можно показать, что эти правила действуют в любой точке $(t_e, V)$ области существования вещества в хроматографической колонке для любого соотношения скоростей $q_i (i = 1 \div 3)$.

Распространим теперь общее уравнение (17), полученное для концентрации $\frac{dN_i}{dV}$ ($i = 1,2$), до $i = 3$. Выражения для входящих в ур-ие (17) членов $\Pi(\lambda_{k,j})(i = 3)$ и $b_k^j(i = 3)$ вместе с условиями их присутствия в уравнении поясняются Таблицей 3.

**Таблица 3**. Члены ур-ия (17) для $i = 3, j = 1 \div i, k = j \div i$, представленные в виде таблицы.

| $j$ | $\frac{dN_j^0}{dV}\left(\prod_{k=j}^{i-1}\lambda_k\right)$ | $k$ | $\Pi(\lambda_{k,j})$ | $b_k^j$ | наличие в ур-ии (17) occurrence condition |
|---|---|---|---|---|---|
| 1 | $\frac{dN_1^0}{dV}\lambda_1\lambda_2$ | 1 | $(\lambda_2 - \lambda_1)(\lambda_3 - \lambda_1)$ | $e^{-\Delta_0^1}$ | $(t_e, V) \in 0(V_{r1}) \div V_{e1}$ |
| | | 2 | $(\lambda_1 - \lambda_2)(\lambda_3 - \Lambda_0)$ | $e^{-\Delta_0^2}$ | $(t_e, V) \in 0(V_{r2}) \div V_{e2}$ |
| | | | $(\lambda_1 - \lambda_2)(\lambda_3 - \Lambda_1)$ | $\text{sgn}(q_1 - q_2)e^{-\Delta_1^2}$ | path $dV$② $\cap V_{e1}$ |
| | | | | $-\text{sgn}(q_1 - q_2)e^{-\Delta_1^2 - r_1^2}$ | path $dV$② $\cap V_{r1}$ |
| | | 3 | $(\lambda_1 - \lambda_3)(\Lambda_0 - \lambda_3)$ | $k_0 e^{-\Delta_0^3}$ | $(t_e, V) \in 0(V_{r3}) \div V_{e3}$ |
| | | | $(\lambda_1 - \lambda_3)(\Lambda_1 - \lambda_3)$ | $\text{sgn}(q_1 - q_3)k_1 e^{-\Delta_1^3}$ | path $dV$③ $\cap V_{e1}$ |
| | | | | $-\text{sgn}(q_1 - q_3)k_1 e^{-\Delta_1^3 - r_1^3}$ | path $dV$③ $\cap V_{r1}$ |
| | | | $(\Lambda_0 - \lambda_3)(\Lambda_1 - \lambda_3)$ | $\text{sgn}(q_2 - q_3)(k_0 - k_1)e^{-\Delta_2^3}$ | path $dV$③ $\cap V_{e2}$ |
| | | | | $-\text{sgn}(q_2 - q_3)(k_0 - k_1)e^{-\Delta_2^3 - r_2^3}$ | path $dV$③ $\cap V_{r2}$ |
| 2 | $\frac{dN_2^0}{dV}\lambda_2$ | 2 | $\lambda_3 - \Lambda_0$ | $e^{-\Delta_0^2}$ | $(t_e, V) \in 0(V_{r2}) \div V_{e2}$ |
| | | 3 | $\Lambda_0 - \lambda_3$ | $e^{-\Delta_0^3}$ | $(t_e, V) \in 0(V_{r3}) \div V_{e3}$ |
| | | | | $\text{sgn}(q_2 - q_3)e^{-\Delta_2^3}$ | path $dV$③ $\cap V_{e2}$ |
| | | | | $-\text{sgn}(q_2 - q_3)e^{-\Delta_2^3 - r_2^3}$ | path $dV$③ $\cap V_{r2}$ |
| 3 | $\frac{dN_3^0}{dV}$ | 3 | 1 | $e^{-\Delta_0^3}$ | $(t_e, V) \in 0(V_{r3}) \div V_{e3}$ |

Константы $\Lambda_0$ и $\Lambda_1$ можно считать заменителями постоянной распада $\lambda_2$ при $q_2 \neq q_3$. Действительно, если $q_2 = q_3$, то $k_0 = k_1 = 1$ и $\Lambda_0 = \Lambda_1 = \lambda_2$. Рассмотрим часть Таблицы 2, начинающуюся с $j = 2$. Понизив индексы $j$ и $k$ на единицу, мы обнаружим, что с точностью до $\Lambda_0 \sim \lambda_2$ она совпадает с Таблицей 1.

Анализируя уравнения $\frac{dN_3}{dV} = f(t_e, V)$, полученные в Примерах 1 и 2, можно убедиться, что они подчиняются общему выражению для $\frac{dN_i}{dV}$ ($i = 1 \div 3$), представленному в виде ур-ия (17) и Таблицы 3. Рассмотрим простой графический способ использования общего выражения для нахождения $\frac{dN_3}{dV}$. На Рис. 8 представлена *V-t* диаграмма движения веществ в хроматографической колонке с соотношением скоростей $q_2 \geq q_3 \geq q_1$ для момента $t_e$ элюирования, когда коридоры веществ разошлись.

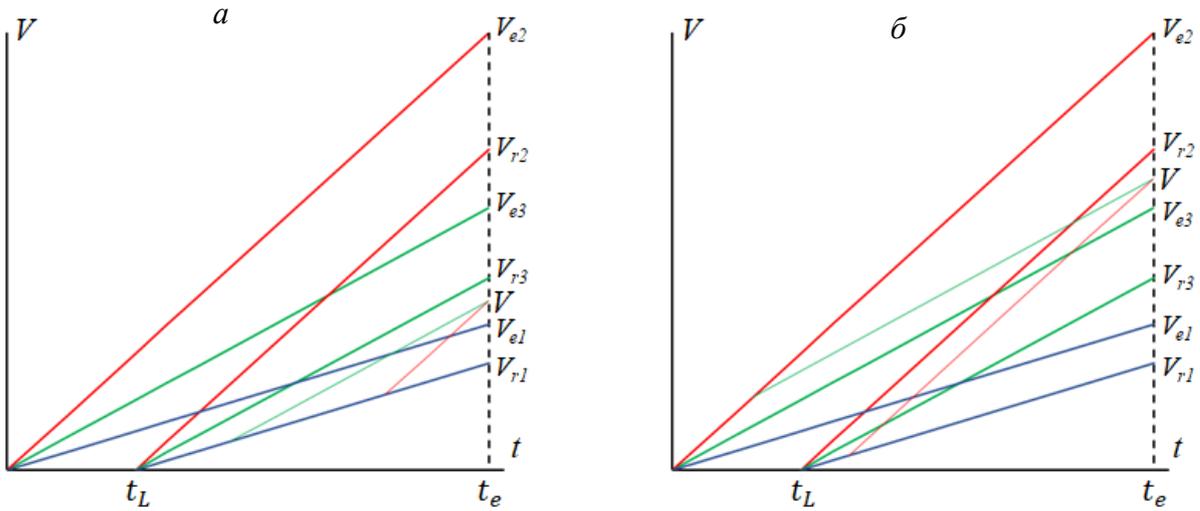

**Рис. 8**. *V-t* диаграмма движения веществ в хроматографической колонке с соотношением скоростей $q_2 \geq q_3 \geq q_1$. Точка $(t_e, V)$ выбрана: *а* – в диапазоне $V_{e1} \div V_{r3}$; *б* – в диапазоне $V_{e3} \div V_{r2}$. Синие, красные и зеленые линии соответствуют движению веществ ①, ② и ③.

Обозначим первую точку $(t_e, V)$ в диапазоне $V_{e1} \div V_{r3}$ (Рис. 8а). Видно, что пути дифференциальных элементов $dV$② и $dV$③ начинаются с тыла $V_{r1}$ и пересекают фронт $V_{e1}$ вещества ①. Из Таблицы 3 выбираем члены $b_k^j$, удовлетворяющие этим условиям (path $dV$② $\cap$ $V_{e1}$, path $dV$② $\cap$ $V_{r1}$, path $dV$③ $\cap$ $V_{e1}$ и path $dV$③ $\cap$ $V_{r1}$), и составляем уравнение для $\frac{dN_3}{dV}$:

$$\frac{dN_3}{dV}(t_e, V) = \lambda_1 \lambda_2 \frac{dN_1^0}{dV} \left( -\frac{e^{-\Delta_1^2}(1-e^{-r_1^2})}{(\lambda_1-\lambda_2)(\lambda_3-\Lambda_1)} - \frac{k_1 e^{-\Delta_1^3}(1-e^{-r_1^3})}{(\lambda_1-\lambda_3)(\Lambda_1-\lambda_3)} \right) \quad (37)$$

Теперь обозначим точку $(t_e, V)$ в диапазоне $V_{e3} \div V_{r2}$ (Рис. 8б). Путь $dV$② по-прежнему начинается с тыла $V_{r1}$ и пересекает фронт $V_{e1}$ вещества ①, а путь $dV$③ начинается с фронта $V_{e2}$ и пересекает тыл $V_{r2}$ вещества ②. Находим в Таблице 3 члены $b_k^j$, удовлетворяющие этим условиям, и составляем уравнение для $\frac{dN_3}{dV}$:

$$\frac{dN_3}{dV}(t_e, V) = \lambda_1 \lambda_2 \frac{dN_1^0}{dV} \left( -\frac{e^{-\Delta_1^2}(1-e^{-r_1^2})}{(\lambda_1-\lambda_2)(\lambda_3-\Lambda_1)} + \frac{(k_0-k_1)e^{-\Delta_2^3}(1-e^{-r_2^3})}{(\Lambda_0-\lambda_3)(\Lambda_1-\lambda_3)} \right) + \lambda_2 \frac{dN_2^0}{dV} \frac{e^{-\Delta_2^3}(1-e^{-r_2^3})}{(\Lambda_0-\lambda_3)} \quad (38)$$

В заключение этого раздела, покажем, как изменятся ур-ия (37, 38) в случае, когда скоростью движения вещества ① в сорбенте, временем $t_L$ загрузки исходного раствора и начальной концентрацией $c_2^0$ можно пренебречь ($q_1 \to 0, t_L \to 0, c_2^0 = 0$). Ранее при выводе ур-ия (16а) было получено, что $\lim_{q_1 \to 0} e^{-\Delta_1^2}(1-e^{-r_1^2}) = (\lambda_2 - \lambda_1)\frac{q_1}{q_2} t_L e^{-\Delta_0^2}$. Поскольку $\Lambda_1 \xrightarrow[q_1 \to 0]{} \Lambda_0$,

$$\lim_{q_1 \to 0} \left( -\frac{e^{-\Delta_1^2}(1-e^{-r_1^2})}{(\lambda_1-\lambda_2)(\lambda_3-\Lambda_1)} \right) = \frac{q_1 t_L e^{-\Delta_0^2}}{q_2(\lambda_3-\Lambda_0)}.$$ Подобным образом, $\lim_{q_1 \to 0} \left( \frac{k_1 e^{-\Delta_1^3}(1-e^{-r_1^3})}{(\lambda_1-\lambda_3)(\Lambda_1-\lambda_3)} \right) = \frac{q_1 t_L e^{-\Delta_0^2-(\lambda_3-\Lambda_0)\frac{V}{q_3}}}{q_2(\lambda_3-\Lambda_0)}.$

Учитывая, что $\frac{dN_1^0}{dV} = \frac{N_1^0}{t_L q_1}$ и $\Delta_0^2 = \lambda_1 t_e + (\lambda_2 - \lambda_1)\frac{V}{q_2}$, ур-ие (37) принимает вид:

$$\frac{dN_3}{dV} = \frac{\lambda_2 \lambda_1}{(\lambda_3 - \Lambda_0)q_2} N_1^0 e^{-\lambda_1 t_e - (\lambda_2-\lambda_1)\frac{V}{q_2}} \left( 1 - e^{-(\lambda_3-\Lambda_0)\frac{V}{q_3}} \right) \tag{37a}$$

Для изменения ур-ия (38) выполним следующие действия:

1) раскладывая в ряд Тейлора и ограничиваясь двумя первыми членами, преобразуем $\lim_{t_L \to 0}(1 - e^{-r_2^3}) = -\left( (\lambda_2 - \lambda_1) + (\lambda_3 - \lambda_2)\frac{q_2}{(q_2-q_3)} \right) t_L$, и $\lim_{q_1 \to 0}(k_0 - k_1) = \frac{q_1}{q_2}\left(1 - \frac{q_3}{q_2}\right)$;

2) выразим $\Delta_2^3$ в виде $\Delta_2^3 = \Delta_0^2 + \frac{(\lambda_3-\Lambda_0)}{(1-k_0)}\left(t_e - \frac{V}{q_2}\right)$.

Подставляя полученные выражения в ур-ие (38), получаем:

$$\frac{dN_3}{dV} = \frac{\lambda_2 \lambda_1}{(\lambda_3 - \Lambda_0)q_2} N_1^0 e^{-\lambda_1 t_e - (\lambda_2-\lambda_1)\frac{V}{q_2}} \left( 1 - e^{-\frac{(\lambda_3-\Lambda_0)}{(1-k_0)}\left(t_e - \frac{V}{q_2}\right)} \right) \tag{38a}$$

Уравнения (37а, 38а) совпадают с ур-иями (5а, 6а), полученными в нашей предыдущей работе [11] для случая, когда материнское вещество удерживается в начальном слое сорбента.

Предложенный подход к кинетике последовательных реакций первого порядка применим и в случае, когда подвижное равновесие не достигается. Кроме того, он может быть расширен на: i) большее число участников реакций; ii) движение в многомерной среде; iii) реакции более высокого порядка, etc. Эти задачи выходят за рамки данной работы и будут изучены позднее.

## 2. Влияние изменения условий хроматографического разделения на профиль концентрации вещества

### 2.1. Изменение скорости пропускания или состава подвижной фазы

Изменение состава подвижной фазы (элюента) – один из основных методов хроматографического разделения. В зависимости от характера взаимодействия с новым элюентом, это приводит к увеличению или уменьшению скоростей движения веществ в сорбенте, когда фронт нового элюента их достигнет. Изменение скорости пропускания подвижной фазы применяется реже в традиционной хроматографии, однако имеет смысл при разделении веществ-участников последовательных реакций, поскольку скорости движения веществ меняются, а константы скорости превращения в цепочке реакций остаются прежними. В терминах подхода, развиваемого в данной работе, дискретное изменение скорости пропускания является частным случаем изменения подвижной фазы, в котором скорости подвижной фазы $Q$ и движущихся веществ $q_i$ меняются в определенный момент $t_1$ в одинаковое число раз.

Ограничившись цепочкой из двух реакций, а также приняв, что $\lambda_1 \ll \lambda_2$, и в растворе до начала движения наступило подвижное равновесие, рассмотрим влияние изменения скорости $Q$ на профиль концентрации вещества для случая $t_1 < t_L$. Используем дополнительный индекс в обозначении скоростей: $Q_1, q_{11}, q_{21}$ – скорости подвижной фазы и веществ ①, ② до $t_1$; и $Q_2, q_{12}, q_{22}$ – после $t_1$.

На Рис. 9 показано, как выглядит *V-t* диаграмма после того, как в момент $t_1$ скорости меняются в $K$ раз: $K = \frac{Q_1}{Q_2} = \frac{q_{11}}{q_{12}} = \frac{q_{21}}{q_{22}}$, при скорости вещества ① больше скорости вещества ②. Концентрация вещества ① в дифференциальном элементе $dV$①, стартующем в момент $t_{s1} < t_1$

(Рис. 9б), равна $\frac{dN_1}{dV} = c_1^0 \frac{Q_1}{q_{11}} e^{-\lambda_1 t_{s1}}$, а в $dV$①, стартующем в момент $t_{s1} > t_1$ (Рис. 9а), равна $\frac{dN_1}{dV} = c_1^0 \frac{Q_2}{q_{12}} e^{-\lambda_1 t_{s1}}$. Поскольку $\frac{Q_1}{q_{11}} = \frac{Q_2}{q_{12}}$, концентрация вещества ① во всей области его существования в сорбенте в момент $t_e$ равна $\frac{dN_1}{dV} = \frac{dN_1^0}{dV} e^{-\lambda_1 t_e}$, т.е. изменение скорости подвижной фазы не влияет на его профиль концентрации.

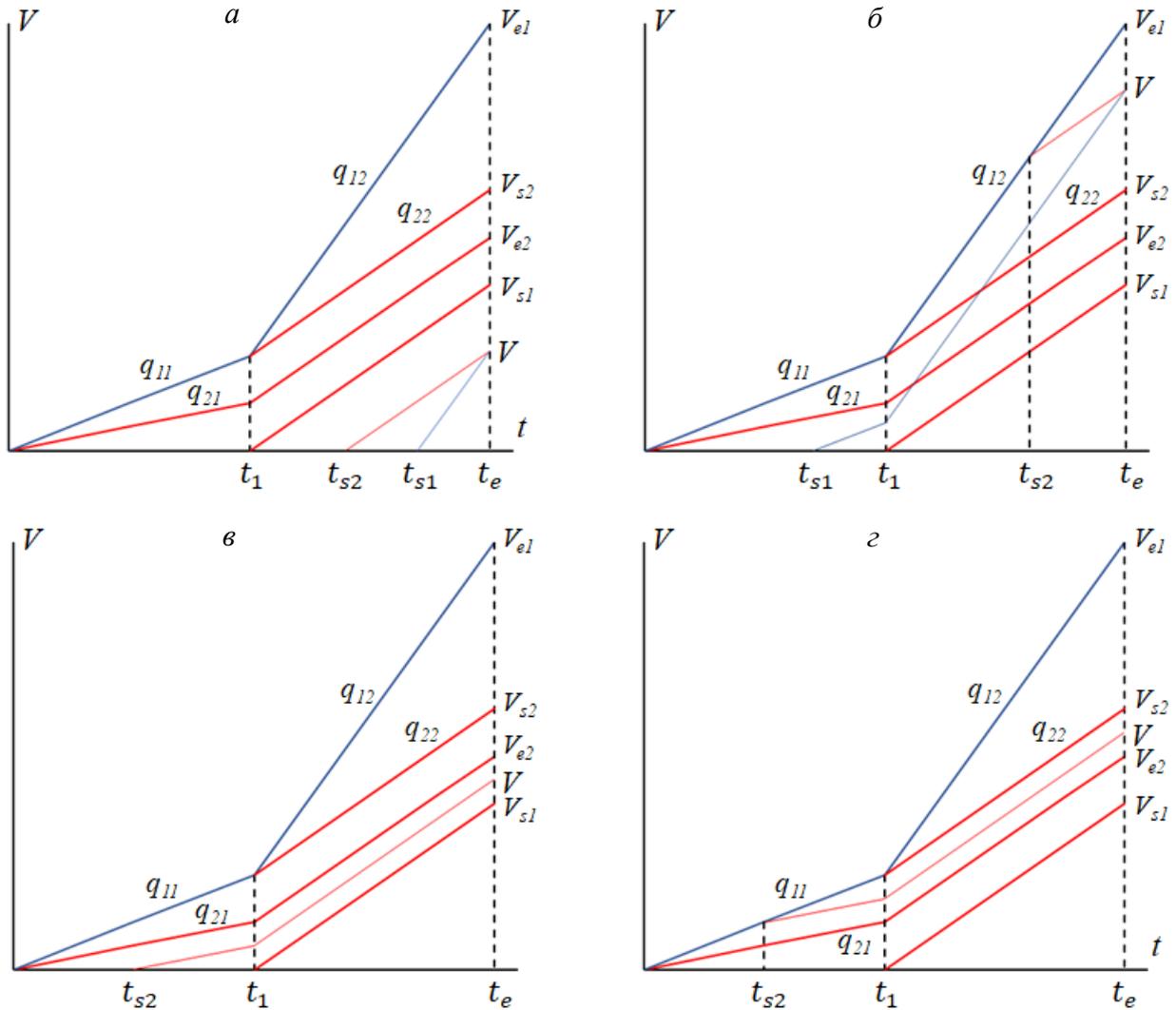

**Рис. 9**. *V-t* диаграммы движения дифференциального элемента $dV$② в хроматографической колонке после дискретного изменения скоростей движения веществ в момент $t_1 < t_L$. Конечная точка пути $dV$② находится в диапазоне: *а* – $0 \div V_{s1}$; *б* – $V_{s2} \div V_{e1}$; *в* – $V_{s1} \div V_{e2}$; *г* – $V_{e2} \div V_{s2}$.
Синие и красные линии соответствуют движению веществ ① и ②.

В области существования вещества ② в момент $t_1$ происходит образование двух вторичных фронтов, движущихся со скоростью $q_{22}$. Фронт $V_{s1}$ начинается в точке $(t_1, 0)$, а фронт $V_{s2}$ – в точке $(t_1, q_{11}t_1)$. Для нахождения уравнений $\frac{dN_2}{dV} = f(t_e, V)$ используем процедуру, описанную в Примере 1 раздела 1.2.

В диапазонах $0 \div V_{s1}$ и $V_{s2} \div V_{e1}$ движение дифференциального элемента $dV$② проходит после $t_1$, схемы движения на Рис. 9а, б аналогичны схемам на Рис. 3 а, б, поэтому и концентрация $\frac{dN_2}{dV}$ определяются похожими уравнениями. Расширим обозначения: $\Delta_0^{2p} = \lambda_1 t_e + (\lambda_2 - \lambda_1) \frac{V}{q_{2p}}$ и

$$\Delta_1^{2p} = \lambda_1 t_e + (\lambda_2 - \lambda_1)\frac{(V_{e1}-V)}{(q_{1p}-q_{2p})}, \text{ где } p=1, \text{ если } dV② \text{ движется до } t_1, \text{ и } p=2, \text{ если } dV②$$

движется после $t_1$. Тогда уравнения для $\frac{dN_2}{dV}$ принимают вид:

$$0 \div V_{s1}: \qquad \frac{dN_2}{dV}(t_e, V) = \frac{dN_1^0}{dV}\frac{\lambda_1}{(\lambda_2-\lambda_1)}\left(e^{-\Delta_0^1} - e^{-\Delta_0^{22}}\right) + \frac{dN_2^0}{dV}e^{-\Delta_0^{22}} \qquad (39)$$

$$V_{s2} \div V_{e1}: \qquad \frac{dN_2}{dV}(t_e, V) = \frac{dN_1^0}{dV}\frac{\lambda_1}{(\lambda_2-\lambda_1)}\left(e^{-\Delta_0^1} - e^{-\Delta_1^{22}}\right) \qquad (40)$$

Путь элемента $dV②$, прибывающего в точку $(t_e, V)$ диапазона $V_{s1} \div V_{e2}$ (Рис. 9в), состоит из двух участков – до и после $t_1$. До $t_1$ концентрация $\frac{dN_2}{dV}$ определяется ур-иями (4-6), а в момент $t_1$ она становится граничным условием для следующего участка пути $dV②$:

$$\left(\frac{dN_2}{dV}\right)_{t_1} = \frac{dN_1^0}{dV}\frac{\lambda_1}{(\lambda_2-\lambda_1)}\left(e^{-\lambda_1 t_1} - e^{-\lambda_1 t_{s2}-\lambda_2(t_1-t_{s2})}\right) + \frac{dN_2^0}{dV}e^{-e^{-\lambda_1 t_{s2}-\lambda_2(t_1-t_{s2})}} \qquad (4\text{гу})$$

Решаем уравнение материального баланса (3) с граничным условием (4гу) и, вводя обозначение $T = (\lambda_2 - \lambda_1)(t_e - t_1)(1 - 1/K)$, приходим к уравнению:

$$V_{s1} \div V_{e2}: \qquad \frac{dN_2}{dV}(t_e, V) = \frac{dN_1^0}{dV}\frac{\lambda_1}{(\lambda_2-\lambda_1)}\left(e^{-\Delta_0^1} - e^{-\Delta_0^{21}-T}\right) + \frac{dN_2^0}{dV}e^{-\Delta_0^{21}-T} \qquad (41)$$

Так же для диапазона $V_{e2} \div V_{s2}$ (Рис. 9г): концентрация предыдущего участка (ур-ие (7)) в момент $t_1$ становится граничным условием для следующего участка пути $dV②$. Решение ур-ия (3) принимает вид:

$$V_{e2} \div V_{s2}: \qquad \frac{dN_2}{dV}(t_e, V) = \frac{dN_1^0}{dV}\frac{\lambda_1}{(\lambda_2-\lambda_1)}\left(e^{-\Delta_0^1} - e^{-\Delta_1^{21}-T}\right) \qquad (42)$$

Общее выражение $\frac{dN_2}{dV}$ для рассмотренного случая изменения скорости пропускания подвижной фазы описывается ур-ием (17), в котором значения члена $b_k^j$ ($i=2$) для $t_e \leq t_1$ приведены в Таблице 1, а в интервале $t_e \geq t_1$ – в Таблице 4. Общее выражение верно и при $q_{1p} \leq q_{2p}$.

**Таблица 4.** Члены ур-ия (17), описывающего профиль концентрации вещества ② после изменения скорости пропускания подвижной фазы ($t_1 \leq t_e < t_L$).

| $j$ $(1 \div i)$ | $\frac{dN_j^0}{dV}\left(\prod_{k=j}^{i-1}\lambda_k\right)$ | $k$ $(j \div i)$ | $\Pi(\lambda_{k,j})$ | $b_k^j$ | наличие в ур-ии (17) occurrence condition |
|---|---|---|---|---|---|
| 1 | $\frac{dN_1^0}{dV}\lambda_1$ | 1 | $\lambda_2 - \lambda_1$ | $e^{-\Delta_0^1}$ | $(t_e, V) \in 0 \div V_{e1}$ |
| | | 2 | $\lambda_1 - \lambda_2$ | $e^{-\Delta_0^{22}}$ | $(t_e, V) \in 0 \div V_{e2}$ path $dV② \not\cap t=t_1$ |
| | | | | $e^{-\Delta_0^{21}-T}$ | $(t_e, V) \in 0 \div V_{e2}$ path $dV② \cap t=t_1$ |
| | | | | $\text{sgn}(q_{12}-q_{22})e^{-\Delta_1^{22}}$ | path $dV② \cap V_{e1}$ path $dV② \not\cap t=t_1$ |
| | | | | $\text{sgn}(q_{11}-q_{21})e^{-\Delta_1^{21}-T}$ | path $dV② \cap V_{e1}$ path $dV② \cap t=t_1$ |
| 2 | $\frac{dN_2^0}{dV}$ | 2 | 1 | $e^{-\Delta_0^{22}}$ | $(t_e, V) \in 0 \div V_{e2}$ path $dV② \not\cap t=t_1$ |
| | | | | $e^{-\Delta_0^{21}-T}$ | $(t_e, V) \in 0 \div V_{e2}$ path $dV② \cap t=t_1$ |

Мы обнаружили, что дискретное изменение подвижной фазы приводит к образованию дополнительных фронтов в профиле концентрации дочернего вещества. Не менее важный вопрос: как на него влияют изменения, связанные с другим компонентом хроматографической среды – с сорбентом.

## 2.2. Изменение неподвижной фазы

Одним из простых примеров изменения неподвижной фазы является выход вымываемых веществ из колонки с сорбентом в элюат. Обычно разделение считается законченным, когда одно вещество оказывается в элюате, а другое остается в колонке. Другой пример – это использование колонок с разными сорбентами с целью вымыть вещество из первой колонки и сконцентрировать его на второй. Для изучения влияния смены сорбента на распределение продуктов последовательных реакций модифицируем систему, состоящую из сосуда и колонки (Рис. 10). Пусть начальная часть колонки заполнена первым сорбентом со свободным объемом $V_c$ (объемом, доступным для протекания подвижной фазы), а остальная колонка – вторым сорбентом. Так же, как в предыдущих случаях, ограничимся цепочкой из двух реакций и условимся, что $\lambda_1 \ll \lambda_2$, и в растворе до начала движения наступило подвижное равновесие.

При движении раствора с объёмной скоростью $Q$ вещества движутся в сорбентах со скоростями $q_{im} \leq Q$, где $i$ – номер вещества, $m$ – номер сорбента; в общем случае соотношение скоростей веществ может быть любым. Фронт $i$-го вещества достигает границы первого сорбента за время $t_{i1} = \frac{V_c}{q_{i1}}$, которое может быть как больше, так и меньше времени $t_L$ загрузки исходного раствора. Рассмотрим случай с соотношением скоростей $q_{1m} \geq q_{2m}$ и времен $t_{21} < t_L$ (Рис. 11).

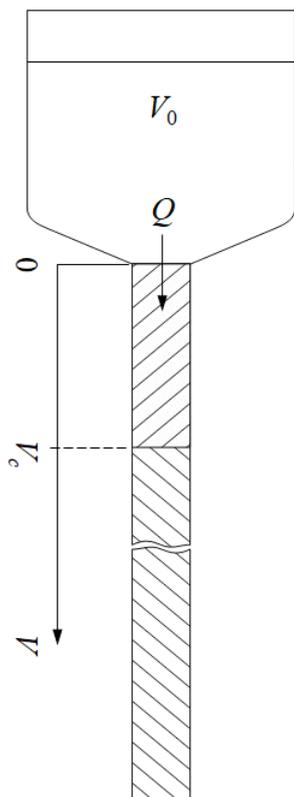

**Рис. 10**. Модель "сосуд и хроматографическая колонка": влияние смены сорбента на движение веществ-участников последовательных реакций 1-го порядка.

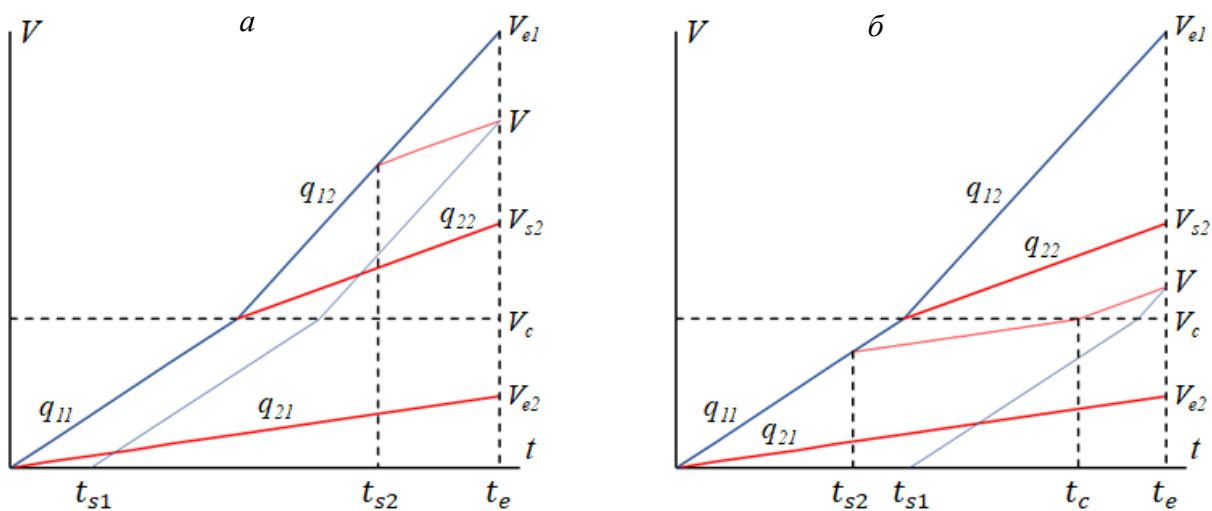

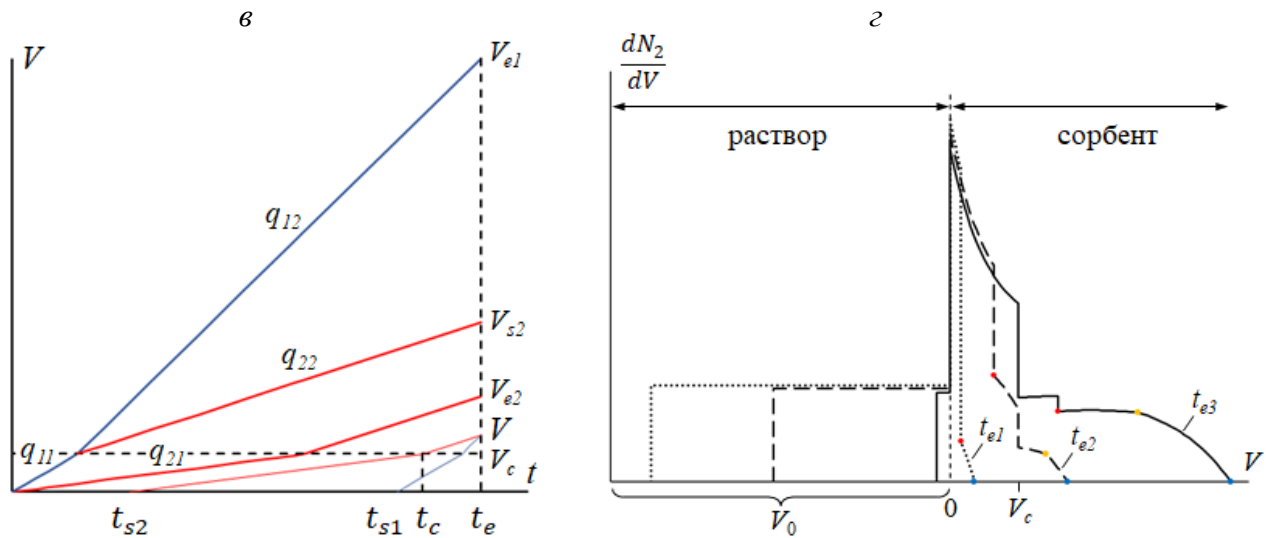

**Рис. 11**. Влияние смены сорбента на распределение вещества ② в хроматографической колонке, заполненной двумя сорбентами.
*а – в*: *V-t* диаграммы движения веществ со скоростями $q_{1m} \geq q_{2m}$. Точка $(t_e, V)$ находится в диапазоне: $а - V_{s2} \div V_{e1}$; $б - V_c \div V_{s2}$; $в - V_c \div V_{e2}$. Синие и красные линии соответствуют движению веществ ① и ②.
*г*: профили концентрации вещества ②, возникающие в разные моменты $t_e$ элюирования: $0 \leq t_{e1} \leq t_{11}$; $t_{11} \leq t_{e2} \leq t_{21}$; $t_{21} \leq t_{e3} \leq t_L$. Цветные точки показывают положения фронтов: синие – $V_{e1}$, красные – $V_{e2}$, желтые – $V_{s2}$. В расчете использованы значения: $V_0 = 25$ мл, $Q = 3$ мл/мин, $q_{11} = 1.8$ мл/мин, $q_{21} = 0.8$ мл/мин, $q_{12} = 3$ мл/мин, $q_{22} = 1.7$ мл/мин, $\lambda_1 = 2 \cdot 10^{-4}$ с$^{-1}$, $\lambda_2 = 5 \cdot 10^{-3}$ с$^{-1}$, $t_{e1} = 1$ мин, $t_{e2} = 4$ мин, $t_{e3} = 8$ мин.

В момент $t_{11} = \frac{V_c}{q_{11}}$, когда фронт $V_{e1}$ вещества ① пересекает $V = V_c$, происходит образование вторичного фронта $V_{s2}$, движущегося со скоростью $q_{22}$. Такой же фронт появляется и в предыдущем случае (Рис. 9) при пересечении $V_{e1}$ с $t = t_1$. Поскольку концентрация является функцией $(t_e, V)$, можно заключить, что любое влияние хроматографической среды, влекущее дискретное изменение скорости материнского вещества, ведет к возникновению вторичного фронта дочернего вещества.

Обозначим $K_1 = \frac{q_{11}}{q_{12}}$ и $K_2 = \frac{q_{21}}{q_{22}}$. Концентрация вещества ① в дифференциальном элементе $dV$①, пересекающем $V = V_c$, изменяется в $K_1$ раз, т.е. его концентрация в момент $t_e > t_{11}$ в диапазоне $0 \div V_c$ равна: $\frac{dN_1}{dV} = \frac{dN_1^0}{dV} e^{-\lambda_1 t_e}$, в диапазоне $V_c \div V_{e1}$: $\frac{dN_1}{dV} = K_1 \frac{dN_1^0}{dV} e^{-\lambda_1 t_e}$.

Схема движения $dV$② в диапазоне $V_{s2} \div V_{e1}$ (Рис. 11а) схожа со схемой на Рис. 9б, поэтому уравнение для $\frac{dN_2}{dV}$ отличается только на коэффициент $K_1$:

$$V_{s2} \div V_{e1}: \quad \frac{dN_2}{dV}(t_e, V) = K_1 \frac{dN_1^0}{dV} \frac{\lambda_1}{(\lambda_2 - \lambda_1)} \left( e^{-\Delta_0^1} - e^{-\Delta_1^{22}} \right) \quad (43)$$

Путь элемента $dV$②, прибывающего в точку $(t_e, V)$ диапазона $V_c \div V_{s2}$ (Рис. 11б), состоит из двух участков – до и после момента $t_c$ пересечения с $V = V_c$. До $t_c$ концентрация $\frac{dN_2}{dV}$ определяется ур-иями (7, 8), а в момент $t_c$ она изменяется в $K_2$ раз и становится граничным условием для следующего участка пути $dV$②:

$$\left( \frac{dN_2}{dV} \right)_{t_c} = K_2 \frac{dN_1^0}{dV} \frac{\lambda_1}{(\lambda_2 - \lambda_1)} \left( e^{-\lambda_1 t_c} - e^{-\lambda_1 t_{s2} - \lambda_2 (t_c - t_{s2})} \right) \quad (7\text{гу})$$

Решением уравнения материального баланса (3) с граничным условием (7гу) является:

$$\frac{dN_2}{dV} = \frac{dN_1^0}{dV} \frac{\lambda_1 e^{-\lambda_1 t_e}}{(\lambda_2 - \lambda_1)} \left( K_1 \left( 1 - e^{-(\lambda_2 - \lambda_1)(t_e - t_c)} \right) + K_2 \left( e^{-(\lambda_2 - \lambda_1)(t_e - t_c)} - e^{-(\lambda_2 - \lambda_1)(t_e - t_{s2})} \right) \right) \quad (44)$$

Из *V-t* диаграммы на Рис. 11б видно, что $t_e - t_c = \frac{V - V_c}{q_{22}}$. Ранее мы ввели величину $\Delta_0^2 = \lambda_1 t_e + (\lambda_2 - \lambda_1) \frac{V}{q_2}$, означающую, что вещество ② начинает движение в первом сорбенте с линии $V = 0$.

По аналогии обозначим $\Delta_c^{22} = \lambda_1 t_e + (\lambda_2 - \lambda_1)\frac{(V-V_c)}{q_{22}}$, подразумевая, что вещество ② начинает движение со скоростью $q_{22}$ во втором сорбенте с линии $V = V_c$. Далее выразим $t_e - t_{s2} = (t_e - t_c) + (t_c - t_{s2}) = \frac{(V-V_c)}{q_{22}} + \frac{q_{11}(V_{s2}-V)}{q_{22}(q_{11}-q_{21})}$ и обозначим $S = (\lambda_2 - \lambda_1)\frac{q_{11}(V_{s2}-V)}{q_{22}(q_{11}-q_{21})}$. Подставляя полученные выражения в ур-ие (44), приходим к уравнению $\frac{dN_2}{dV} = f(t_e, V)$ в этом диапазоне:

$$V_c \div V_{s2}: \qquad \frac{dN_2}{dV}(t_e, V) = \frac{dN_1^0}{dV}\frac{\lambda_1}{(\lambda_2-\lambda_1)}\left(K_1 e^{-\Delta_0^1} - (K_1 - K_2)e^{-\Delta_c^{22}} - K_2 e^{-\Delta_c^{22}-S}\right) \qquad (45)$$

Заметим, что величина $\Delta_c^{22} + S$ появляется при условии, что путь $dV$② имеет два пересечения: сначала с фронтом $V_{e1}$ (стартует с него), затем с линией $V = V_c$. Если $K_1 = K_2$, то ур-ие (45) превращается в ур-ие (43), т.е. вторичный фронт $V_{s2}$ вырождается.

После пересечения фронтом $V_{e2}$ вещества ② границы первого сорбента в момент $t_{21} = \frac{V_c}{q_{21}}$ образуется диапазон $V_c \div V_{e2}$, а ур-ие (45) действует теперь в диапазоне $V_{e2} \div V_{s2}$ (Рис. 11в). Схема движения $dV$②, прибывающего в одну из точек нового диапазона, подобна схеме на Рис. 9в. Решая уравнение материального баланса (3) и учитывая, что скорости веществ ① и ② меняются в $K_1$ и $K_2$ раз при переходе из первого сорбента во второй, приходим к уравнению:

$$V_c \div V_{e2}: \qquad \frac{dN_2}{dV} = \frac{dN_1^0}{dV}\frac{\lambda_1}{(\lambda_2-\lambda_1)}\left(K_1 e^{-\Delta_0^1} - (K_1 - K_2)e^{-\Delta_c^{22}} - K_2 e^{-\Delta_c^{22}-C}\right) + \frac{dN_2^0}{dV}K_2 e^{-\Delta_c^{22}-C} \qquad (46),$$

в котором величина $C = (\lambda_2 - \lambda_1)\frac{V_c}{q_{21}}$ означает, что путь $dV$② имеет пересечения с линиями $V=0$ и $V = V_c$. Сравнивая Рис. 11б и 11в, обнаруживаем, что $\lim_{(t_c - t_{s2}) \to t_{21}} S = C$.

Влияние перехода из одного сорбента в другой на профиль концентрации вещества ② при условиях $q_{1m} \geq q_{2m}$ и $t_{21} < t_L$ показано на Рис. 11г. Использованная процедура нахождения $\frac{dN_2}{dV}$ действенна и при других условиях. Общее выражение $\frac{dN_2}{dV}$ для произвольного соотношения скоростей в обоих сорбентах и $t_e \leq t_L$ описывается ур-ием (17), в котором значения члена $b_k^j$ ($i = 2$) для $V \leq V_c$ приведены в Таблице 1, а для $V \geq V_c$ – в Таблице 5.

**Таблица 5**. Члены ур-ия (17) концентрации вещества ②, описывающего переход из одного сорбента в другой ($V \geq V_c$, $t_e \leq t_L$).

| $j$ $(1 \div i)$ | $\frac{dN_j^0}{dV}\left(\prod_{k=j}^{i-1} \lambda_k\right)$ | $k$ $(j \div i)$ | $\Pi(\lambda_{k,j})$ | $b_k^j$ | наличие в ур-ии (17) occurrence condition |
|---|---|---|---|---|---|
| 1 | $\frac{dN_1^0}{dV}\lambda_1$ | 1 | $\lambda_2 - \lambda_1$ | $K_1 e^{-\Delta_0^1}$ | $(t_e, V) \in V_c \div V_{e1}$ |
| | | 2 | $\lambda_1 - \lambda_2$ | $K_2 e^{-\Delta_c^{22}-C}$ | $(t_e, V) \in V_c \div V_{e2}$ |
| | | | | $(K_1 - K_2)e^{-\Delta_c^{22}}$ | $(t_e, V) \in V_c \div V_{s2}$ |
| | | | | $\mathrm{sgn}(q_{12} - q_{22})K_1 e^{-\Delta_1^{22}}$ | path $dV$② $\cap V_{e1}$ path $dV$② $\cap V = V_c$ |
| | | | | $\mathrm{sgn}(q_{11} - q_{21})K_2 e^{-\Delta_c^{22}-S}$ | path $dV$② $\cap V_{e1}$ path $dV$② $\cap V = V_c$ |
| 2 | $\frac{dN_2^0}{dV}$ | 2 | 1 | $K_2 e^{-\Delta_c^{22}-C}$ | $(t_e, V) \in V_c \div V_{e2}$ |

Таким образом, дискретное изменение компонентов хроматографической среды (как подвижной фазы, так и неподвижной) приводит к образованию вторичных фронтов дочернего вещества. Частный случай распределения веществ ② и ③ в колонке, заполненной двумя сорбентами, когда вещество ① загружено и удерживается в начальном слое первого сорбента, рассмотрен в нашей предыдущей работе [11].

### 2.3. Циркуляция подвижной фазы

Циркуляцию с использованием сорбентов или фильтров применяют для очистки от примесей большого объема раствора. Фильтрат возвращают в емкость с раствором, в результате чего концентрация примесей постепенно уменьшается. В традиционной хроматографии, где объемы растворов обычно невелики, циркуляцию используют редко.

При хроматографическом разделении веществ, участвующих в последовательных реакциях, циркуляция подвижной фазы оказывается весьма полезным приемом, особенно при условии, что константа скорости (постоянная распада) первой реакции много меньше остальных. В этом случае циркуляция приводит к новому, динамическому подвижному равновесию, когда вещества циклично движутся в замкнутой системе. Подбирая параметры, мы можем добиться желаемого распределения веществ. Например, представим себе хроматографическую систему из двух колонок, заполненных разными сорбентами. Материнское вещество удерживается на первой колонке, а вымываемое дочернее вещество адсорбируется на второй. Раствор, вытекающий из второй колонки, направляется на вход первой, образуя замкнутый контур. Спустя определенное время система приходит в состояние, в котором дочернее вещество, находясь в подвижном равновесии с материнским, будет пространственно от него отделено.

Покажем влияние циркуляции подвижной фазы на распределение дочернего вещества в замкнутом контуре на примере цепочки из двух реакций ($\lambda_1 \ll \lambda_2$), приняв, что в растворе до начала движения наступило подвижное равновесие. Скорость движения вещества ① в сорбенте может быть как больше, так и меньше скорости вещества ②, рассмотрим вначале первый случай.

### 2.3.1. Соотношение скоростей движения веществ ① и ② в сорбенте: $q_1 \geq q_2$

Модифицируем систему, состоящую из сосуда и колонки, так, как показано на Рис. 12.

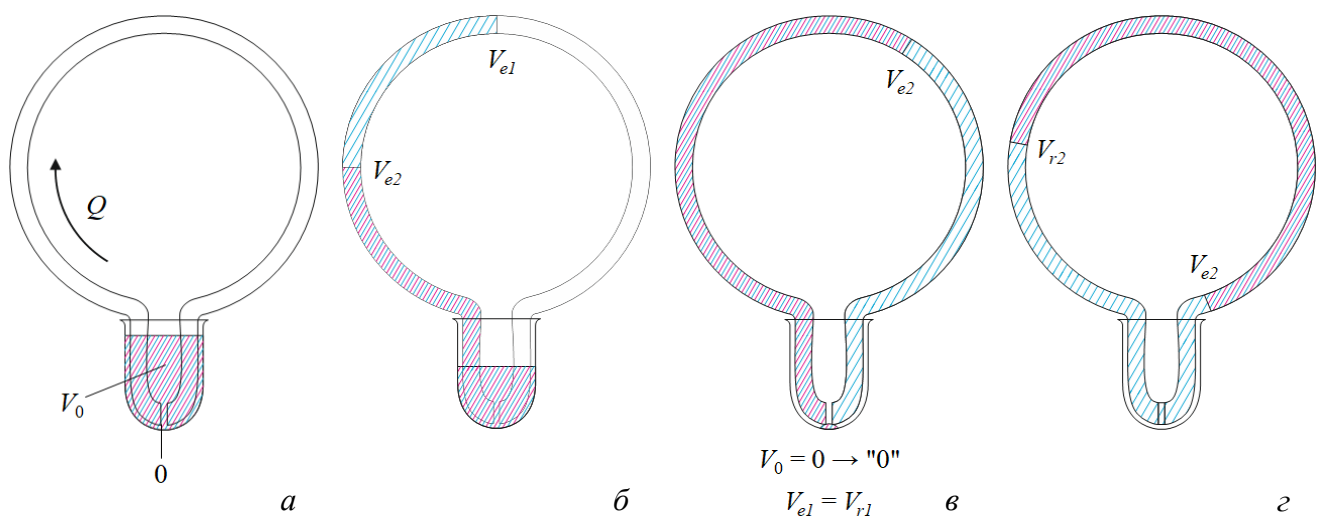

**Рис. 12.** Модель хроматографической колонки в виде замкнутого контура при движении веществ ① и ② со скоростями $q_1 = Q \geq q_2$: *а* – начало движения раствора с объемной скоростью $Q$; *б* – положения фронтов $V_{e1}$ и $V_{e2}$ при частичном заполнении контура раствором; *в* – момент замыкания контура; *г* – одно из положений фронта $V_{e2}$ и тыла $V_{r2}$ вещества ② при движении по замкнутому контуру. Синими и красными линиями показаны вещества ① и ②.

В исходный момент времени раствор начинает поступать в колонку (Рис. 12а), примем вход в колонку за начало оси $V$, направление которой совпадает с направлением движения. Допустим, что вещество ① движется в сорбенте со скоростью раствора $q_1 = Q$. Спустя некоторое время раствор заполнит часть контура, а фронт $V_{e1}$ обгонит фронт $V_{e2}$ (Рис. 12б). Примем, что объем $V_0$ исходного раствора равен свободному объему сорбента в колонке. Тогда в момент $t_L$ окончания загрузки (Рис. 12в) фронт раствора и вещества ① достигнет входа в колонку, контур замкнется, а раствора в сосуде не останется. Фронт $V_{e1}$ соединится со своим тылом $V_{r1}$, и границы вещества ① исчезнут, т.е. после загрузки контур становится равномерно заполнен веществом ①. Тыл $V_{r2}$ вещества ② в момент $t_L$ начинает отдаляться от начала хроматографической колонки. Затем коридор $V_{r2} \div V_{e2}$ вещества ② циркулирует по контуру со скоростью $q_2 = q$ и периодом обращения $t_q = \frac{V_0}{q}$ (Рис. 12г).

$V$-$t$ диаграммы движения веществ представлены на Рис. 13.

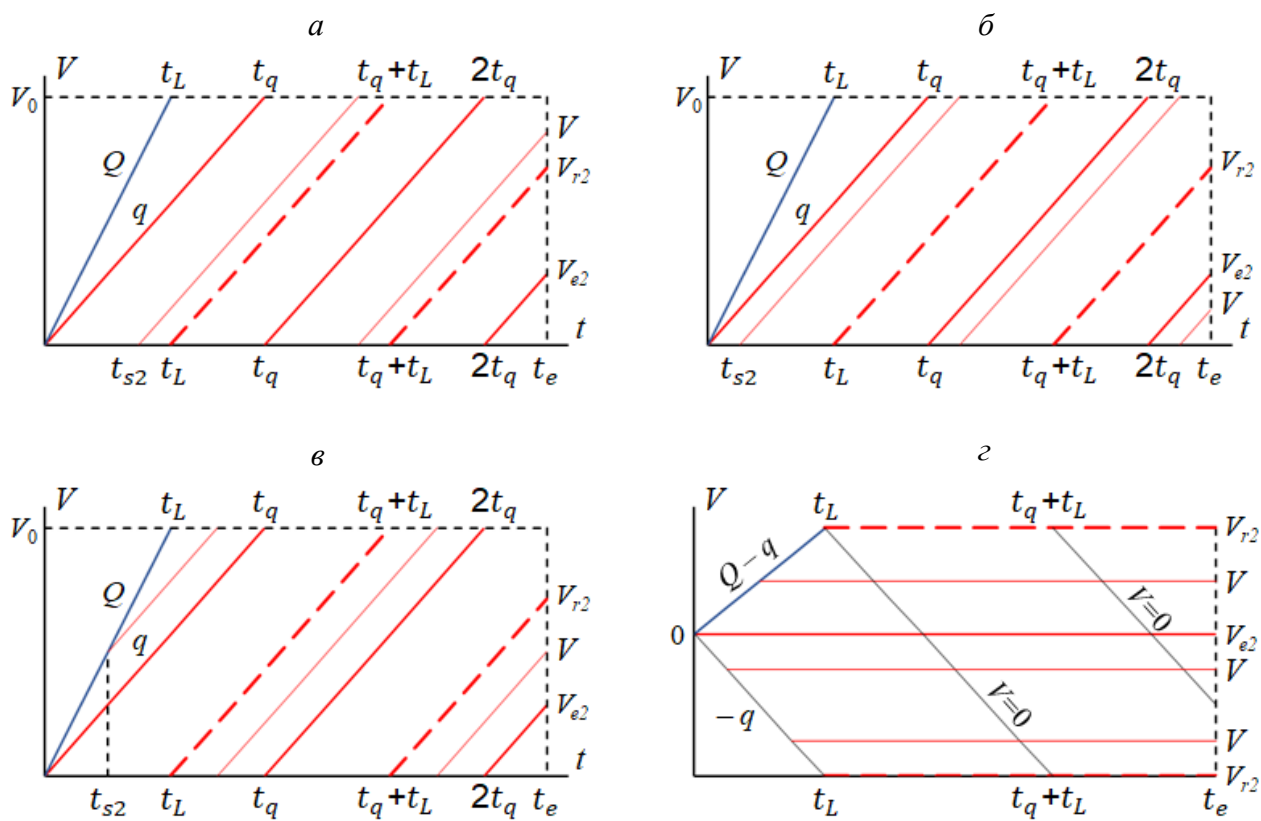

**Рис. 13**. $V$-$t$ диаграммы движения веществ в хроматографической колонке в виде замкнутого контура при скорости $q_1 = Q$ вещества ① больше скорости $q_2 = q$ вещества ②. Точка $(t_e, V)$ находится во временном интервале $2t_q \leq t_e \leq 2t_q + t_L$ в диапазоне: $а - V_{r2} \div V_0$; $б - 0 \div V_{e2}$; $в - V_{r2} \div V_{e2}$; $г - V$-$t$ диаграмма, построенная относительно начала колонки ($V=0$), движущегося со скоростью $-q$. Синие и красные линии соответствуют движению веществ ① и ②.

Положение фронта $V_{e2}$ вещества ② в контуре в момент $t_e$ определяется уравнением:
$$V_{e2} = qt_e - aV_0 \qquad (47),$$
где $a = \left[\frac{t_e}{t_q}\right]$ – число целых циклов, совершенных $V_{e2}$. Подобным образом определяется положение тыла $V_{r2}$:
$$V_{r2} = q(t_e - t_L) - bV_0 \qquad (48),$$

где $b = \left[\frac{t_e - t_L}{t_q}\right]$ – число целых циклов, совершенных $V_{r2}$. Из Рис. 13 видно, что в момент $t_e$ фронт $V_{e2}$ сделал два полных оборота ($a = 2$), а тыл $V_{r2}$ – один полный оборот ($b = 1$), другими словами, при $a = b$: $V_{e2} > V_{r2}$, при $a = b + 1$: $V_{e2} < V_{r2}$.

Контур становится замкнутым в момент $t_L$, это значит, что любой дифференциальный элемент стартует в интервале $0 \div t_L$. Проследим путь элемента $dV②$, начинающего движение со входа в колонку ($V=0$) и прибывающего в точку ($t_e, V$) диапазона $V_{r2} \div V_0$ (Рис. 13а). Его положение в контуре:

$$V = q(t_e - t_{s2}) - \left[\frac{t_e - t_{s2}}{t_q}\right] V_0 \qquad (49),$$

где выражение в квадратных скобках означает число целых циклов, совершенных элементом $dV②$. Устремляя $t_{s2} \to t_L$, мы видим, что $V \to V_{r2}$. Если устремить $t_{s2}$ в противоположную сторону, то в момент $t_{s2} = t_e - (b+1)t_q$ координата $V$ достигает $V_0 = 0$ и переходит на новый виток, изображенный на Рис. 13б (диапазон $0 \div V_{e2}$), из которого следует, что $t_{s2} = t_e - at_q$. То есть, $a = b + 1$, что соответствует взаимному расположению $V_{e2} < V_{r2}$. При устремлении $t_{s2}$ к левой границе интервала: $t_{s2} \to 0$, получаем $V \to V_{e2}$. Концентрация $\frac{dN_2}{dV}$ в рассматриваемом элементе $dV②$ определяется ур-иями (4-6), перепишем ур-ие (4) в виде:

$$\frac{dN_2}{dV} = c_2^0 e^{-\lambda_1 t_e}\left(1 + \left(\frac{Q}{q} - 1\right) e^{-(\lambda_2 - \lambda_1)(t_e - t_{s2})}\right) \qquad (50)$$

Для диапазона $V_{r2} \div V_0$ (Рис. 13а) $t_e - t_{s2} = \frac{V + V_0}{q}$, а для диапазона $0 \div V_{e2}$ (Рис. 13б) $t_e - t_{s2} = \frac{V + 2V_0}{q}$.

Так же проследим путь элемента $dV②$, начинающего движение с фронта $V_{e1}$ вещества ① и прибывающего в точку ($t_e, V$) диапазона $V_{e2} \div V_{r2}$ (Рис. 13в). Его положение в контуре:

$$V = Qt_{s2} + q(t_e - t_{s2}) - \left[\frac{Qt_{s2} + q(t_e - t_{s2})}{V_0}\right] V_0 \qquad (51)$$

Видно, что при $t_{s2} \to 0$: $V \to V_{e2}$, при $t_{s2} \to t_L$: $V \to V_{r2}$. Концентрация $\frac{dN_2}{dV}$ в этом элементе $dV②$ определяется ур-иями (7-8), перепишем ур-ие (7) в виде:

$$\frac{dN_2}{dV} = c_2^0 e^{-\lambda_1 t_e}\left(1 - e^{-(\lambda_2 - \lambda_1)(t_e - t_{s2})}\right) \qquad (52)$$

С помощью $V$-$t$ диаграммы выражаем: $t_e - t_{s2} = \frac{t_e Q - V - 2V_0}{Q - q}$, т.е. на своем пути элемент $dV②$ дважды пересекает отметку $V_0 = 0$.

На Рис. 13г $V$-$t$ диаграмма построена относительно начала колонки (или "0"-отметки $V_0 = 0$), движущегося со скоростью $-q$. В таком представлении, когда фронт и тыл вещества ② неподвижны, легко видеть, что элемент $dV②$, попадающий в диапазон $V_{r2} \div (V_0 = 0)$, пересекает "0"-отметку один раз, а элементы $dV②$, попадающие в диапазоны $(V_0 = 0) \div V_{e2}$ и $V_{e2} \div V_{r2}$ пересекают "0"-отметку по два раза.

Таким образом, для нахождения профиля концентрации вещества ② в произвольный момент $t_e$ достаточно двух уравнений – (50) и (52). Первое уравнение определяет концентрацию $\frac{dN_2}{dV}$ внутри коридора вещества ②, второе – вне его. Чтобы записать эти уравнения в общем виде, выразим $t_e - t_{s2}$ через $t_e$ и $V$ и привяжем ур-ие (50) к числу $a$ целых циклов, совершенных $V_{e2}$, а ур-ие (52) – к числу $b$, при этом $V_{e2}$ и $V_{r2}$ будут служить верхними границами действия уравнений (50) и (52):

$$\frac{dN_2}{dV} = c_2^0 e^{-\lambda_1 t_e}\left(1 + \left(\frac{Q}{q} - 1\right) e^{-(\lambda_2 - \lambda_1)\frac{(V + aV_0)}{q}}\right) \qquad (50а)$$

$$\frac{dN_2}{dV} = c_2^0 e^{-\lambda_1 t_e} \left(1 - e^{-(\lambda_2-\lambda_1)\frac{(t_e Q - V - (b+1)V_0)}{(Q-q)}}\right) \qquad (52a)$$

Заметим, что для числа пересечений пути элемента $dV$②, стартующего с фронта $V_{e1}$, с "0"-отметкой использовано $b+1$, поскольку тыл $V_{r2}$ вещества ② образуется в момент $t_L$.

В общем случае, профиль концентрации вещества ② в произвольный момент $t_e$ состоит из трех диапазонов, поскольку "0"-отметка разделяет на две части либо внутренний коридор вещества ②, либо внешний. В разделенном коридоре для уравнения нижней его части используется счетчик циклов на единицу меньше, чем для верхней. Например (Рис. 13г), в момент $t_e$ "0"-отметка разделяет на две части внутренний коридор $V_{r2} \div V_{e2}$, поэтому для уравнения диапазона $V_{r2} \div (V_0 = 0)$ счетчик $a = 1$, а для уравнения диапазона $(V_0 = 0) \div V_{e2}$ счетчик $a = 2$. Профили концентрации вещества ②, возникающие в разные моменты времени, показаны на Рис. 14.

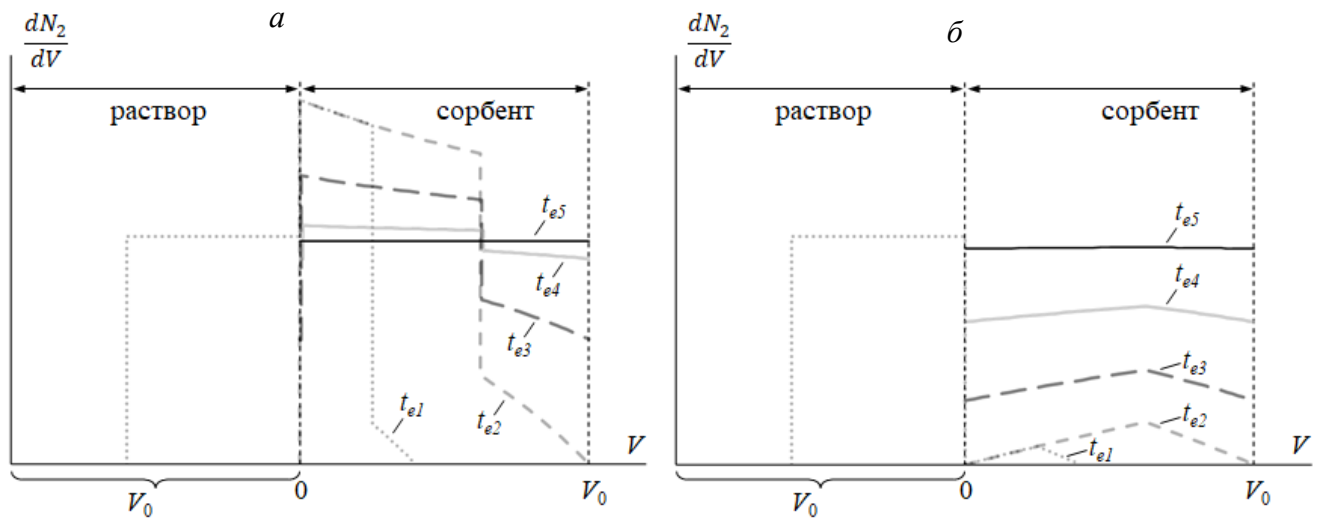

**Рис. 14**. Профили концентрации вещества ②, возникающие в разные моменты $t_e$ движения по контуру, при скорости $Q$ вещества ① больше скорости $q$ вещества ②: $t_{e1} < t_L$; $t_{e2} = t_L$; $t_{e3} = t_q + t_L$; $t_{e4} = 3t_q + t_L$; $t_{e5} = 9t_q + t_L$. Два варианта поступления вещества ② из исходного раствора в контур: *а* – обычный; *б* – с применением воображаемого фильтра, удерживающего вещество ② от попадания в контур. В расчете использованы значения: $V_0 = 10$ мл, $Q = 4$ мл/мин, $q = 2.5$ мл/мин, $\lambda_1 = 7.7 \cdot 10^{-6}$ с$^{-1}$, $\lambda_2 = 3.3 \cdot 10^{-3}$ с$^{-1}$.

Профиль, приведенный на Рис. 14а при $t_{e1} < t_L$, уже встречался на Рисунках 5а и 11г. Следующий профиль соответствует моменту $t_{e2} = t_L$, когда весь исходный раствор загружен, и контур замыкается. Тыл $V_{r2}$ вещества ② находится в положении $0 = V_0$, а его фронт – в положении $qt_L$. Далее показаны профили, возникающие спустя один, три и девять периодов $t_q$ циркуляции. Мы видим, что вещество ② становится равномерно распределенным по контуру, что естественно, поскольку после загрузки контур равномерно заполнен веществом ①. Действительно, ур-ия (50а) и (52а) с ростом $t_e$ стремятся к: $\frac{dN_2}{dV} = c_2^0 e^{-\lambda_1 t_e}$.

Проведем мысленный эксперимент, поместив на входе в контур воображаемый фильтр, препятствующий во время загрузки попаданию вещества ② из исходного раствора в сорбент. Тогда все вещество ② будет образовываться в контуре из вещества ①. Это условие приводит к исчезновению члена $\frac{Q}{q}$ из ур-ия (50а), ур-ие (52а) остается прежним. На Рис. 14б показаны профили концентрации вещества ② в те же моменты времени, что и на Рис. 14а. Мы обнаруживаем, что по достижении подвижного равновесия распределение вещества ② на Рис.

14б становится таким же, как и на Рис. 14а. То есть, равновесное, в данном случае равномерное, распределение вещества ② в замкнутом контуре не зависит от начальных условий.

### 2.3.2. Соотношение скоростей движения веществ ① и ② в сорбенте: $q_2 \geq q_1$

Рассмотрим теперь ту же систему, состоящую из сосуда и хроматографической колонки в виде замкнутого контура, в случае, когда скорость движения вещества ② в сорбенте равна скорости раствора $q_2 = Q$ и больше скорости $q_1 = q$ вещества ① (Рис. 15). Объем $V_0$ исходного раствора по-прежнему равен свободному объему сорбента в колонке.

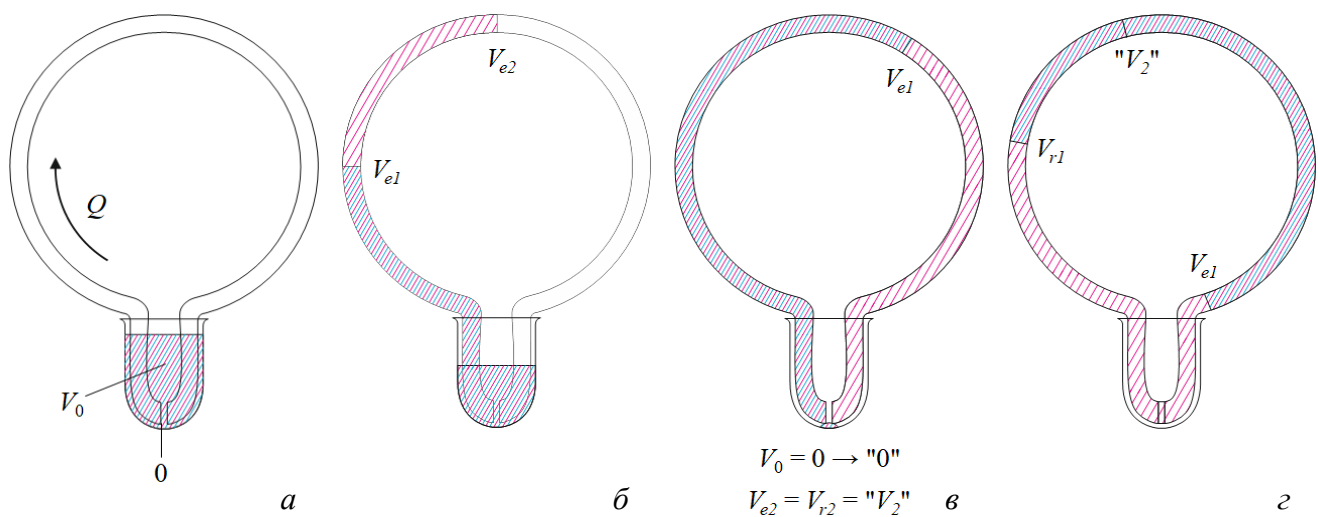

**Рис. 15**. Модель хроматографической колонки в виде замкнутого контура при движении веществ ① и ② со скоростями $q_2 = Q \geq q_1$: *а* – начало движения раствора с объемной скоростью $Q$; *б* – положения фронтов $V_{e1}$ и $V_{e2}$ при частичном заполнении контура раствором; *в* – момент замыкания контура; *г* – одно из положений фронта $V_{e1}$ и тыла $V_{r1}$ вещества ①, а также "$V_2$"-отметки соединения фронта и тыла вещества ② при движении по замкнутому контуру. Синими и красными линиями показаны вещества ① и ②.

На Рис. 15б показано положение фронтов при заполнении раствором части контура. В момент $t_L$ заполнения всего контура (Рис. 15в) фронт раствора и вещества ② достигнет входа в колонку, а раствора в сосуде не останется. Фронт вещества ② соединяется со своим тылом $V_{e2} = V_{r2} =$ "$V_2$", и вещество ② начинает циркулировать по контуру с периодом обращения $t_L$. Тыл $V_{r1}$ вещества ① в момент $t_L$ начинает отдаляться от начала хроматографической колонки, и коридор $V_{r1} \div V_{e1}$ вещества ① циркулирует по контуру с периодом обращения $t_q = \frac{V_0}{q}$. Образуются две основные области (Рис. 15г): «питающий» коридор $V_{r1} \div V_{e1}$ вещества ①, в котором вещество ② накапливается и распадается, и коридор $V_{e1} \div V_{r1}$, где вещество ② только распадается. С учетом отметок "0" и "$V_2$" мы видим, что распределение вещества ② в замкнутом контуре описывается четырьмя диапазонами. *V-t* диаграммы движения веществ представлены на Рис. 16.

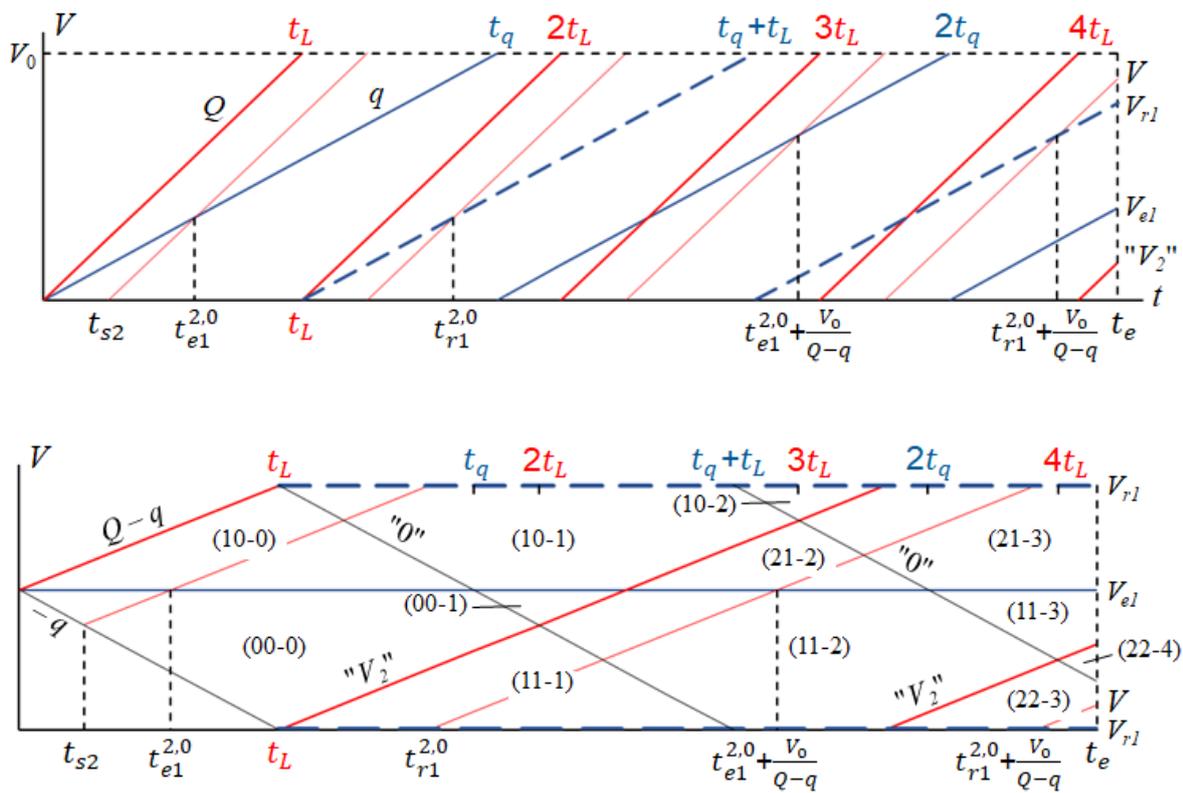

**Рис. 16.** *V-t* диаграммы движения веществ в хроматографической колонке в виде замкнутого контура при скорости $q_2 = Q$ вещества ② больше скорости $q_1 = q$ вещества ①: *а* – относительно неподвижного начала колонки ($V=0$); *б* – относительно начала колонки ($V=0$), движущегося со скоростью $-q$. Синие и красные линии соответствуют движению веществ ① и ②.

Положения фронта $V_{e1}$ и тыла $V_{r1}$ вещества ① в контуре определяются уравнениями:

$$V_{e1} = qt_e - aV_0 \quad (47а)$$
$$V_{r1} = q(t_e - t_L) - bV_0 \quad (48а),$$

где $a = \left[\frac{t_e}{t_q}\right]$ и $b = \left[\frac{t_e - t_L}{t_q}\right]$ – число целых циклов, совершенных $V_{e1}$ и $V_{r1}$, соответственно.

Положение фронта $V_{e2}$, превращающегося после $t_L$ в "$V_2$"-отметку, определяется уравнением:

$$V_2 = Qt_e - lV_0 \quad (53),$$

где $l = \left[\frac{t_e}{t_L}\right]$ – число целых циклов, совершенных "$V_2$"-отметкой. Число пересечений "$V_2$"-отметки с фронтом $V_{e1}$ и тылом $V_{r1}$ в момент $t_e$ равно $n = \left[1 + \frac{t_e(Q-q)}{V_0}\right]$ и $m = \left[1 + \frac{(t_e - t_L)(Q-q)}{V_0}\right]$. В числа $n$ и $m$ входят пересечения "$V_2$" с $V_{e1}$ и $V_{r1}$ в точках $(0,0)$ и $(t_L, 0)$, соответственно.

Из Рис. 16 видно, что вещество ② начинает движение в контуре со входа ($V=0$) в коридоре вещества ①. Для нахождения концентрации $\frac{dN_2}{dV}(t_e, V)$ в момент $t_e$ проследим путь элемента $dV$② , стартующего в момент $t_{s2}$. Его положение в контуре определяется уравнением:

$$V = Q(t_e - t_{s2}) - l_V V_0 \quad (54),$$

где $l_V = \left[\frac{t_e - t_{s2}}{t_L}\right]$ – число целых циклов, совершенных элементом $dV$②. В момент $t_{e1}^{2,0}$ (напомним, верхний индекс означает, что вещество ② начинает движение со входа в колонку ($V=0$), нижний индекс – пересекается с фронтом $V_{e1}$) элемент $dV$② первый раз выходит из коридора вещества ①. Путь $dV$② пересекается с фронтом $V_{e1}$ с периодом $\frac{V_0}{Q-q}$, число его выходов из «питающего» коридора к моменту $t_e$ равно $n_V = \left[1 + \frac{(t_e - t_{e1}^{2,0})(Q-q)}{V_0}\right]$. С той же периодичностью элемент $dV$② входит в коридор вещества ①, в первый раз – в момент $t_{r1}^{2,0} = t_{e1}^{2,0} + t_L$. Число входов к моменту $t_e$ равно $m_V = \left[1 + \frac{(t_e - t_{e1}^{2,0} - t_L)(Q-q)}{V_0}\right]$.

На первом участке пути $dV$② в коридоре вещества ① (интервал $t_{s2} \div t_{e1}^{2,0}$, Рис. 16а) концентрация $\frac{dN_2}{dV}$ определяется ур-иями (4-6), воспользуемся ур-ием (4) в виде:

$$\frac{dN_2}{dV} = \frac{Q}{q} c_2^0 e^{-\lambda_1 t_e} \left(1 + \left(\frac{q}{Q} - 1\right) e^{-(\lambda_2 - \lambda_1)(t_e - t_{s2})}\right) \tag{55}$$

На следующем участке (интервал $t_{e1}^{2,0} \div t_{r1}^{2,0}$) вне коридора вещества ①, когда вещество ② только распадается, концентрация $\frac{dN_2}{dV}$ подчиняется ур-ию (14), выразим его в виде:

$$\frac{dN_2}{dV} = \frac{Q}{q} c_2^0 e^{-\lambda_1 t_e} \left(e^{-(\lambda_2 - \lambda_1)(t_e - t_{e1}^{2,0})} + \left(\frac{q}{Q} - 1\right) e^{-(\lambda_2 - \lambda_1)(t_e - t_{s2})}\right) \tag{56}$$

Экспонента, несущая в степени $t_e - t_{e1}^{2,0}$, говорит о том, что элемент $dV$② вышел первый раз ($n_V = 1$) из коридора вещества ①, обозначим $Ex = (\lambda_2 - \lambda_1)(t_e - t_{e1}^{2,0})$. Затем в момент $t_{r1}^{2,0}$ элемент $dV$② пересекает тыл $V_{r1}$ и снова входит в коридор вещества ①. Ур-ие (56), в котором $t_e = t_{r1}^{2,0}$, становится граничным условием для решения ур-ия (3) материального баланса и нахождения концентрации $\frac{dN_2}{dV}$ на следующем участке пути $dV$② в интервале $t_{r1}^{2,0} \div t_{e1}^{2,0} + \frac{V_0}{Q-q}$:

$$\frac{dN_2}{dV} = \frac{Q}{q} c_2^0 e^{-\lambda_1 t_e} \left(1 + e^{-Ex} - e^{-In} + \left(\frac{q}{Q} - 1\right) e^{-(\lambda_2 - \lambda_1)(t_e - t_{s2})}\right) \tag{57}$$

где $In = (\lambda_2 - \lambda_1)(t_e - t_{r1}^{2,0})$ отражает вход элемента $dV$② в коридор вещества ① ($m_V = 1$). Далее на участке вне «питающего» коридора (интервал $t_{e1}^{2,0} + \frac{V_0}{Q-q} \div t_{r1}^{2,0} + \frac{V_0}{Q-q}$) концентрация $\frac{dN_2}{dV}$ определяется уравнением:

$$\frac{dN_2}{dV} = \frac{Q}{q} c_2^0 e^{-\lambda_1 t_e} \left(e^{-Ex}\left(1 + e^{(\lambda_2 - \lambda_1)\frac{V_0}{Q-q}}\right) - e^{-In} + \left(\frac{q}{Q} - 1\right) e^{-(\lambda_2 - \lambda_1)(t_e - t_{s2})}\right) \tag{58}$$

Член $e^{-Ex}\left(1 + e^{(\lambda_2 - \lambda_1)\frac{V_0}{Q-q}}\right)$ говорит о двух пересечениях пути $dV$② с фронтом $V_{e1}$ ($n_V = 2$). Наконец, на последнем показанном на Рис. 16 участке $dV$② в интервале $t_{r1}^{2,0} + \frac{V_0}{Q-q} \div t_e$ концентрация $\frac{dN_2}{dV}$ выражается уравнением, несущем информацию о двух выходах ($n_V = 2$) и двух входах ($m_V = 2$) в коридор вещества ①:

$$\frac{dN_2}{dV} = \frac{Q}{q} c_2^0 e^{-\lambda_1 t_e} \left(\begin{array}{c} 1 + e^{-Ex}\left(1 + e^{(\lambda_2 - \lambda_1)\frac{V_0}{Q-q}}\right) - e^{-In}\left(1 + e^{(\lambda_2 - \lambda_1)\frac{V_0}{Q-q}}\right) + \\ + \left(\frac{q}{Q} - 1\right) e^{-(\lambda_2 - \lambda_1)(t_e - t_{s2})} \end{array}\right) \tag{59}$$

Таким образом, на участке пути $dV$② внутри коридора вещества ① ($n_V = m_V$, интервал $t_{r1}^{2,0} + (m_V - 1)\frac{V_0}{Q-q} \div t_{e1}^{2,0} + n_V \frac{V_0}{Q-q}$) концентрация $\frac{dN_2}{dV}$ выражается уравнением:

$$\frac{dN_2}{dV} = \frac{Q}{q} c_2^0 e^{-\lambda_1 t_e} \left(\begin{array}{c} 1 + e^{-Ex} \sum_{i=1}^{n_V} e^{(\lambda_2 - \lambda_1)\frac{(i-1)V_0}{Q-q}} - e^{-In} \sum_{i=1}^{m_V} e^{(\lambda_2 - \lambda_1)\frac{(i-1)V_0}{Q-q}} + \\ + \left(\frac{q}{Q} - 1\right) e^{-(\lambda_2 - \lambda_1)(t_e - t_{s2})} \end{array}\right) \tag{60}$$

а на участке вне этого коридора ($n_V = m_V + 1$, интервал $t_{e1}^{2,0} + (n_V - 1)\frac{V_0}{Q-q} \div t_{r1}^{2,0} + m_V \frac{V_0}{Q-q}$) – уравнением:

$$\frac{dN_2}{dV} = \frac{Q}{q} c_2^0 e^{-\lambda_1 t_e} \left(\begin{array}{c} e^{-Ex} \sum_{i=1}^{n_V} e^{(\lambda_2 - \lambda_1)\frac{(i-1)V_0}{Q-q}} - e^{-In} \sum_{i=1}^{m_V} e^{(\lambda_2 - \lambda_1)\frac{(i-1)V_0}{Q-q}} + \\ + \left(\frac{q}{Q} - 1\right) e^{-(\lambda_2 - \lambda_1)(t_e - t_{s2})} \end{array}\right) \tag{61}$$

То есть, в общем виде можно написать:

$$\frac{dN_2}{dV} = \frac{Q}{q} c_2^0 e^{-\lambda_1 t_e} \left( \delta_{(n_V=m_V)} + e^{-Ex} \sum_{i=1}^{n_V} e^{(\lambda_2-\lambda_1)\frac{(i-1)V_0}{Q-q}} - e^{-In} \sum_{i=1}^{m_V} e^{(\lambda_2-\lambda_1)\frac{(i-1)V_0}{Q-q}} + \right. \\ \left. + \left(\frac{q}{Q} - 1\right) e^{-(\lambda_2-\lambda_1)(t_e-t_{s2})} \right) \quad (62)$$

Поскольку любой дифференциальный элемент $dV$②  начинает движение со входа в контур ($V=0$), концентрацию $\frac{dN_2}{dV}$ оказалось возможно выразить одним уравнением. Преобразуем ур-ие (62) к виду $\frac{dN_2}{dV} = f(t_e, V)$, для этого выразим с помощью $V$-$t$ диаграммы: $t_e - t_{s2} = \frac{V+l_V V_0}{Q}$, $t_e - t_{e1}^{2,0} = \frac{t_e q - V - l_V V_0}{q-Q}$. Учитывая, что $t_{r1}^{2,0} = t_{e1}^{2,0} + t_L$, получаем:

$$\frac{dN_2}{dV} = \frac{Q}{q} c_2^0 e^{-\lambda_1 t_e} \left( \delta_{(n_V=m_V)} + e^{-(\lambda_2-\lambda_1)\frac{t_e q-V}{q-Q}} \left( \sum_{i=l_V-n_V+1}^{l_V} e^{-(\lambda_2-\lambda_1)\frac{iV_0}{Q-q}} - e^{(\lambda_2-\lambda_1)t_L} \sum_{i=l_V-m_V+1}^{l_V} e^{-(\lambda_2-\lambda_1)\frac{iV_0}{Q-q}} \right) + \right. \\ \left. + \left(\frac{q}{Q} - 1\right) e^{-(\lambda_2-\lambda_1)(t_e-t_{s2})} \right) \quad (63)$$

В результате циркуляции ур-ие (63) содержит три счетчика. Пронумеруем его в зависимости от счетчиков в виде $(n_V m_V - l_V)$ и проследим (Рис. 16б), как меняются значения счетчиков при движении элемента $dV$②. Началу его пути соответствует набор (00-0), при пересечении фронта $V_{e1}$ увеличивается на единицу значение $n_V$, при пересечении "0"-отметки – значение $l_V$, а при пересечении тыла $V_{r1}$ – значение $m_V$. В произвольный момент $t_e$ положения $V_{e1}$, $V_{r1}$ и "$V_2$" окажутся в произвольном порядке друг относительно друга (всего вариантов 3! = 6). Например, в момент $t_e$, показанный на Рис. 16, вещество ② распределено в следующих четырех диапазонах, каждый из которых характеризуется своим набором значений счетчиков:

| Диапазон | "$V_2$" ÷ $V_{e1}$ | $V_{e1}$ ÷ $V_{r1}$ | $V_{r1}$ ÷ "0" | "0" ÷ "$V_2$" |
|---|---|---|---|---|
| Значения $(n_V m_V - l_V)$ | (11-3) | (21-3) | (22-3) | (22-4) |

Привяжем значения $n_V$, $m_V$ и $l_V$ к значениям $n$, $m$ и $l$, т.е. возьмем "$V_2$"-отметку за отправную точку и будем двигаться вниз от нее по $V$-$t$ диаграмме. "$V_2$" является верхней границей диапазона, в котором, в нашем случае, $n_V = n = 2$, $m_V = m = 2$ и $l_V = l = 4$, а нижняя граница – это "0"-отметка. При ее пересечении $l_V = l - 1 = 3$. Двигаясь дальше вниз, мы пересекаем $V_{r1}$, и $m_V$ становится равным $m - 1 = 1$, затем при пересечении $V_{e1}$: $n_V = n - 1 = 1$, и мы оказываемся в диапазоне, где нижней границей является "$V_2$", цикл замкнулся.

В произвольном случае, у первого диапазона с верхней границей "$V_2$" нижней границей могут быть $V_{e1}$, $V_{r1}$ и "0". На схеме (Рис. 17) показаны все шесть возможных вариантов взаимного расположения диапазонов и соответствующие им значения счетчиков $n_V$, $m_V$, $l_V$.

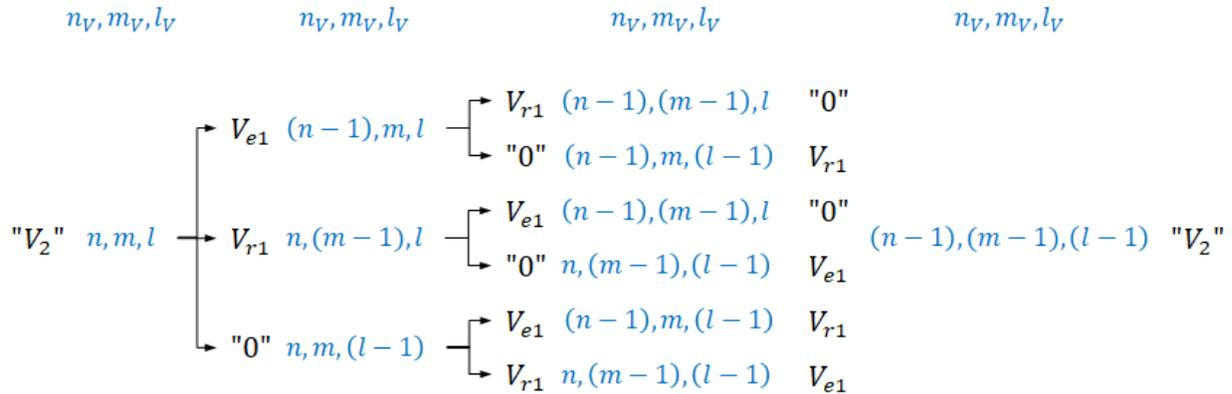

**Рис. 17**. Схема возможных вариантов взаимного расположения диапазонов и соответствующие им значения счетчиков $n_V, m_V, l_V$.

В итоге, построение профиля концентрации вещества ② в контуре в произвольный момент $t_e$ состоит из следующих шагов: i) расчет значений счетчиков $n$, $m$ и $l$; ii) расчет значений $V_{e1}$, $V_{r1}$, "$V_2$" по ур-иям (47а), (48а), (53) и определение взаимного расположения диапазонов; iii) определение значений $n_V$, $m_V$ и $l_V$ по схеме на Рис. 17 и расчет концентрации $\frac{dN_2}{dV}$ по ур-ию (63) для каждого диапазона. На Рис. 18 показаны профили концентрации вещества ②, возникающие в разные моменты времени.

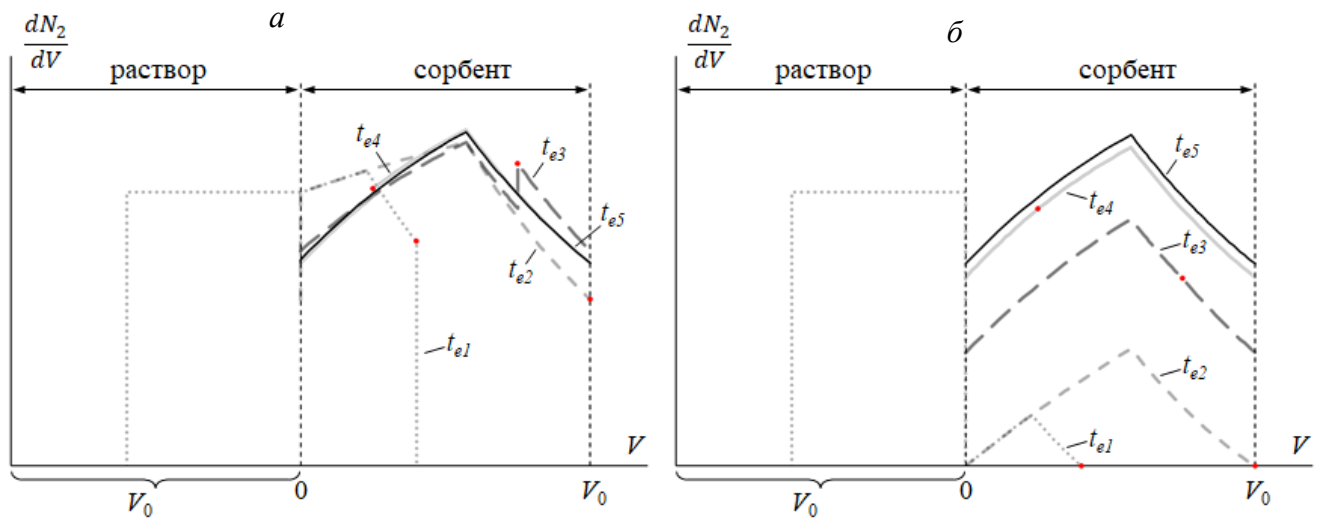

**Рис. 18**. Профили концентрации вещества ②, возникающие в разные моменты $t_e$ движения по контуру, при скорости $Q$ вещества ② больше скорости $q$ вещества ①: $t_{e1} < t_L$; $t_{e2} = t_L$; $t_{e3} = t_q + t_L$; $t_{e4} = 3t_q + t_L$; $t_{e5} = 9t_q + t_L$. Два варианта поступления вещества ② из исходного раствора в контур: *а* – обычный; *б* – с применением воображаемого фильтра, удерживающего вещество ② от попадания в контур. Красные точки показывают положения "$V_2$"-отметки. В расчете использованы значения: $V_0$ = 10 мл, $Q$ = 4 мл/мин, $q$ = 2.3 мл/мин, $\lambda_1 = 7.7 \cdot 10^{-6}$ с$^{-1}$, $\lambda_2 = 3.3 \cdot 10^{-3}$ с$^{-1}$.

Профиль, приведенный на Рис. 18а при $t_{e1} < t_L$, аналогичен построенному ранее (Рис. 5б). В момент $t_{e2} = t_L$ замыкания контура тыл $V_{r1}$ вещества ① находится в положении $0 = V_0$, а его фронт – в положении $qt_L$. Фронт вещества ② соединяется со своим тылом $V_{e2} = V_{r2}$, и превращается в "$V_2$"-отметку, циркулирующую с периодом $t_L$. Далее показаны профили, возникающие спустя один, три и девять периодов $t_q$ циркуляции. Положения "$V_2$"-отметки, заметные по перепаду концентрации $\frac{dN_2}{dV}$ (Рис. 18а), обозначены красными точками. С распадом вещества ② скачок концентрации становится все меньше, а сам профиль становится равновесным, движущимся по контуру с периодом $t_q$. Внутри «питающего» коридора $V_{r1} \div V_{e1}$ происходит нарастание концентрации $\frac{dN_2}{dV}$, достигающее максимального значения в точке $V_{e1}$, вне его – снижение.

Так же, как в предыдущем разделе, проведем мысленный эксперимент и во время загрузки исходного раствора поместим на входе в контур воображаемый фильтр, препятствующий попаданию вещества ② в сорбент. Это условие соответствует отсутствию члена $\frac{q}{Q}$ в ур-ии (63). На Рис. 18б показаны профили концентрации вещества ② в те же моменты времени, что и на Рис. 18а. Снова оказывается, что, несмотря на разные начальные условия, по достижении подвижного равновесия распределение вещества ② одинаково в обоих случаях.

Найдем равновесное распределение вещества ②, преобразуя ур-ия (60, 61). Используя выражение для суммы ряда $\sum_{i=1}^{n} e^{(i-1)x} = \frac{1-e^{nx}}{1-e^x}$ и учитывая, что $t_{r1}^{2,0} = t_{e1}^{2,0} + t_L$, перепишем ур-ие (61) для концентрации $\frac{dN_2}{dV}$ вне «питающего» коридора вещества ① ($n_V = m_V + 1$) в виде:

$$\frac{dN_2}{dV} = \frac{Q}{q} c_2^0 e^{-\lambda_1 t_e} \left( e^{-Ex + (\lambda_2 - \lambda_1)\frac{(n_V-1)V_0}{Q-q}} + e^{-Ex}\left(1 - e^{(\lambda_2-\lambda_1)t_L}\right) \frac{1 - e^{(\lambda_2-\lambda_1)\frac{(n_V-1)V_0}{Q-q}}}{1 - e^{(\lambda_2-\lambda_1)\frac{V_0}{Q-q}}} + \left(\frac{q}{Q} - 1\right)e^{-(\lambda_2-\lambda_1)(t_e - t_{s2})} \right) \quad (61а)$$

Поскольку $l_V - n_V + 1 = a$, воспользуемся ур-ием (47а) для $V_{e1}$ и выразим степень при первой экспоненте в скобках: $-Ex + (\lambda_2 - \lambda_1)\frac{(n_V-1)V_0}{Q-q} = -(\lambda_2 - \lambda_1)\frac{V_{e1}-V}{q-Q}$. Тогда ур-ие (61а) принимает вид:

$$\frac{dN_2}{dV} = \frac{Q}{q} c_2^0 e^{-\lambda_1 t_e} \left( e^{-(\lambda_2-\lambda_1)\frac{V_{e1}-V}{q-Q}}\left(1 - \left(1 - e^{-(\lambda_2-\lambda_1)\frac{(n_V-1)V_0}{Q-q}}\right)\frac{1 - e^{(\lambda_2-\lambda_1)t_L}}{1 - e^{(\lambda_2-\lambda_1)\frac{V_0}{Q-q}}}\right) + \left(\frac{q}{Q} - 1\right)e^{-(\lambda_2-\lambda_1)(t_e - t_{s2})} \right) \quad (61б)$$

С ростом $t_e$ ур-ие (61б) стремится к

$$\frac{dN_2}{dV} = \frac{Q}{q} c_2^0 e^{-\lambda_1 t_e - (\lambda_2-\lambda_1)\frac{V_{e1}-V}{q-Q}} \left(1 - \frac{e^{(\lambda_2-\lambda_1)\frac{V_0}{Q}} - 1}{e^{(\lambda_2-\lambda_1)\frac{V_0}{Q-q}} - 1}\right) = \frac{Q}{q} c_2^0 e^{-\Delta_1^2}\left(1 - \frac{e^{(\lambda_2-\lambda_1)\frac{V_0}{Q}} - 1}{e^{(\lambda_2-\lambda_1)\frac{V_0}{Q-q}} - 1}\right) \quad (61р)$$

Мы получили уравнение для части равновесного профиля концентрации $\frac{dN_2}{dV}$, находящейся вне коридора вещества ① (правая часть профиля при $t_{e5}$ на Рис. 18). Преобразуя таким же образом ур-ие (60), получаем выражение для другой части равновесного профиля, находящейся внутри коридора вещества ① (левая часть профиля при $t_{e5}$ на Рис. 18):

$$\frac{dN_2}{dV} = \frac{Q}{q} c_2^0 \left( e^{-\Delta_0^1} - e^{-\Delta_1^2}\frac{e^{(\lambda_2-\lambda_1)\frac{V_0}{Q}} - 1}{e^{(\lambda_2-\lambda_1)\frac{V_0}{Q-q}} - 1}\right) \quad (60р)$$

Заканчивая анализ циркуляции веществ по замкнутому контуру, рассмотрим равновесное распределение вещества ② в случае, когда скоростью движения вещества ① в сорбенте можно пренебречь ($q \to 0$). Так как $V_{e1} \approx V_{r1} \to 0$, после загрузки исходного раствора в контур вещество ② существует в одном единственном диапазоне $0 \div V_0$ вне коридора вещества ①. Соответствующее этому диапазону ур-ие (61р) упрощается следующим образом:

1) $\lim_{q \to 0} e^{-\Delta_1^2} = e^{-\lambda_1 t_e - (\lambda_2 - \lambda_1)\frac{V}{Q}}$; 2) раскладывая в ряд Тейлора и ограничиваясь двумя первыми членами, преобразуем $\lim_{q \to 0}\left(e^{(\lambda_2 - \lambda_1)\frac{V_0}{Q-q}} - e^{(\lambda_2 - \lambda_1)\frac{V_0}{Q}}\right) = (\lambda_2 - \lambda_1)\frac{V_0 q}{Q^2} e^{(\lambda_2 - \lambda_1)\frac{V_0}{Q}}$. Используя полученные выражения и учитывая, что $c_2^0 = \frac{N_1^0}{V_0} \frac{\lambda_1}{(\lambda_2 - \lambda_1)}$, приходим к уравнению:

$$\frac{dN_2}{dV} = \frac{\lambda_1 N_1^0}{Q} \frac{e^{-\lambda_1 t_e - (\lambda_2 - \lambda_1)\frac{V}{Q}}}{\left(1 - e^{-(\lambda_2 - \lambda_1)\frac{V_0}{Q}}\right)} \tag{64}$$

Коридор вещества ① превращается в точку $V = 0$, в которой концентрация вещества ② равна $\left(\frac{dN_2}{dV}\right)_{V=0} = \frac{\lambda_1 N_1^0}{Q} \frac{e^{-\lambda_1 t_e}}{\left(1 - e^{-(\lambda_2 - \lambda_1)\frac{V_0}{Q}}\right)}$, что следует как из ур-ия (64), так и из ур-ия (60р).

Основным результатом циркуляции подвижной фазы в замкнутой хроматографической системе является достижение подвижного равновесия во времени и пространстве, при котором вещества-участники последовательных реакций движутся циклично. Равновесное распределение веществ зависит от параметров замкнутой системы и не зависит от начальных условий.

Математические модели, рассмотренные в рамках общего подхода к кинетике последовательных реакций первого порядка, протекающих в условиях хроматографического разделения, можно использовать как элементы для конструирования различных хроматографических систем, таких как прямые и обратные генераторные схемы. Например, в замкнутом контуре можно разместить несколько сорбентов, разделив их трубкой для свободного движения подвижной фазы, или скомбинировать модели, описывающие дискретное изменение подвижной и неподвижной фазы, и распространить их на большее число веществ-участников последовательных реакций, движущихся с разными скоростями. Для иллюстрации возможностей математического аппарата, описанного в данной работе, будет построена упрощенная модель радионуклидного $^{225}$Ac/$^{213}$Bi генератора и оценена его эффективность.

## 3. Модель радионуклидного $^{225}$Ac/$^{213}$Bi генератора

На данный момент известны $^{225}$Ac/$^{213}$Bi генераторы на основе ионообменных и экстракционно-хроматографических смол [14-21]. Генератор прямого типа, в котором $^{225}$Ac ($T_{1/2}$ = 9.9 дн.) удерживается макропористой катионообменной смолой AG MP-50, является наиболее распространенным [15]. $^{213}$Bi ($T_{1/2}$ = 46 мин), вымываемый из генератора йодид-содержащими растворами, используется в клинических испытаниях радиофармпрепаратов для терапии лейкемии, лимфомы, меланомы, опухолей мозга и других онкологических заболеваний [22].

Использованию $^{225}$Ac/$^{213}$Bi генераторов в рутинной медицинской практике препятствует дефицит материнского $^{225}$Ac. Сейчас его получают в основном генераторным способом из $^{229}$Th ($T_{1/2}$ = 7340 лет), причем возможности производства недостаточны даже для проведения полноценных клинических испытаний [22]. Перспективный способ получения $^{225}$Ac предложен ИЯИ РАН [23-25], заключающийся в облучении металлического $^{232}$Th протонами средних энергий и позволяющий за 7-10 дней нарабатывать столько же $^{225}$Ac, сколько его производят из $^{229}$Th за год. Этот способ активно развивается научными центрами США (LANL, BNL [26]) и Канады (TRIUMF [27]). Основным недостатком способа считается сопутствующее образование 0.1-0.2% $^{227}$Ac ($T_{1/2}$ = 21.8 лет) в расчете на окончание облучения.

Генератор, в котором в качестве материнского радионуклида используется $^{225}$Ac с примесью $^{227}$Ac, должен гарантировать получение $^{213}$Bi с низким содержанием изотопов актиния, чтобы предотвратить попадание $^{227}$Ac и относительно долгоживущих продуктов его распада – $^{227}$Th (T$_{1/2}$ = 18.7 дн.) и $^{223}$Ra (T$_{1/2}$ = 11.4 дн.), в организм пациента. Поскольку $^{213}$Bi образуется из $^{225}$Ac в результате цепочки радиоактивных превращений $^{225}$Ac → $^{221}$Fr (T$_{1/2}$ = 4.8 мин) → $^{217}$At (T$_{1/2}$ = 32 мс) → $^{213}$Bi →, получение $^{213}$Bi через отделение и распад более короткоживущего промежуточного $^{221}$Fr дает возможность существенно улучшить его радионуклидную чистоту (период полураспада $^{217}$At настолько мал, что он всегда находится в подвижном равновесии с $^{221}$Fr в обычных условиях хроматографического разделения). Ион Fr(I), как ион щелочного металла, не проявляет значительных комплексообразующих или ионообменных свойств и слабо удерживается многими сорбентами даже из разбавленных растворов минеральных кислот. Поэтому имеет смысл рассмотреть схему генератора, в которой $^{221}$Fr служит транспортным агентом, состоящую, например, из двух колонок, связанных объемом, не заполненным сорбентом. Из первой колонки, содержащей $^{225}$Ac, непрерывно вымывают $^{221}$Fr, раствор поступает в промежуточный сосуд для распада $^{221}$Fr в $^{213}$Bi и далее - на вторую колонку, где $^{213}$Bi концентрируется. Предварительные эксперименты с колонками, заполненными экстракционно-хроматографической смолой Actinide Resin (Triskem Int.) показали перспективность такой схемы [12]. Предлагаемая теоретическая модель позволяет оценить и выбрать оптимальные параметры получения $^{213}$Bi без проведения трудоемких экспериментов.

### 3.1. Прямоточный двух-колоночный $^{225}$Ac/$^{213}$Bi генератор

На Рис. 19а показана система из двух колонок, содержащих смолу Actinide Resin со свободными объемами $V_{c1}$ и $V_{c2}$, разделенных трубкой объемом $V_p$. Доля свободного объема ряда экстракционно-хроматографических смол, включая Actinide Resin, находится в пределах 0.65-0.69 [28-30]. Значения коэффициента удерживания $k'$ (capacity factor) смолой ионов Ac(III) и Bi(III) из разбавленных растворов минеральных кислот превышают $10^5$ и $10^4$ [17, 31], поэтому

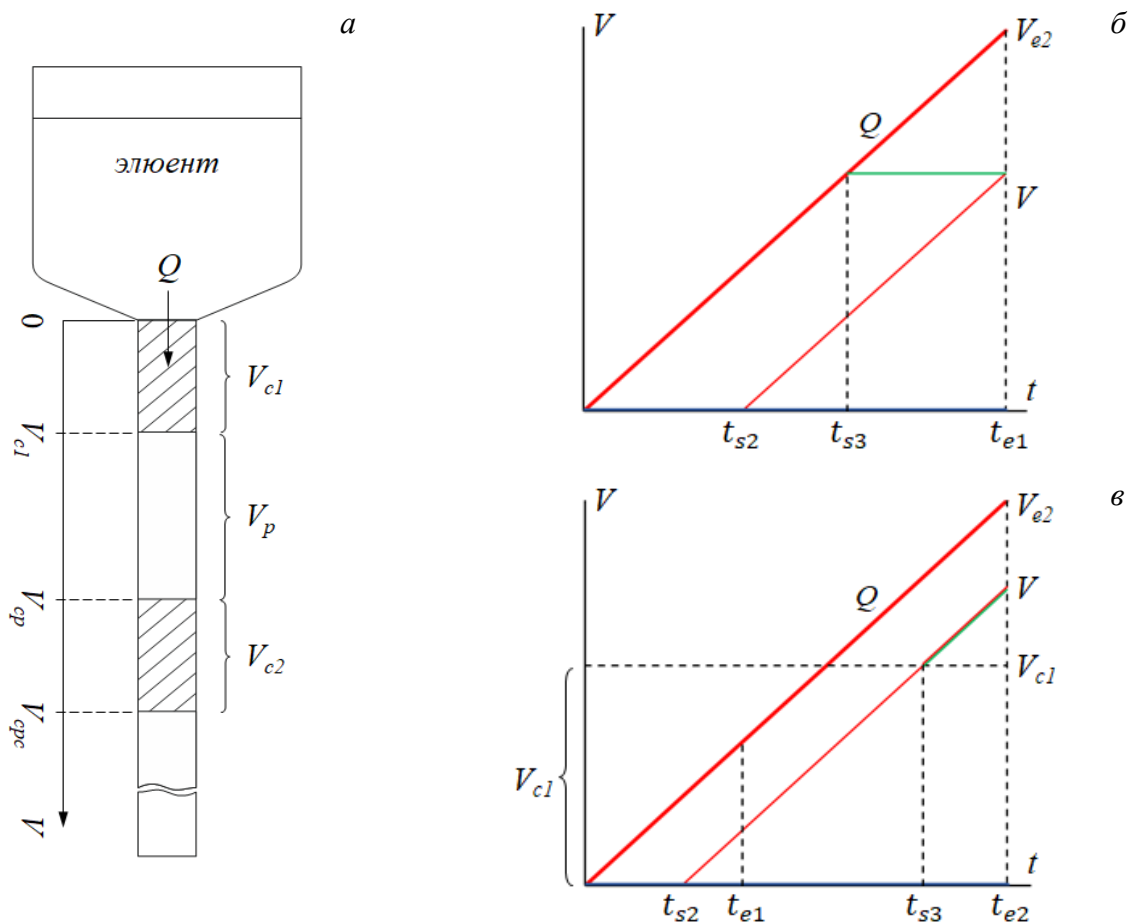

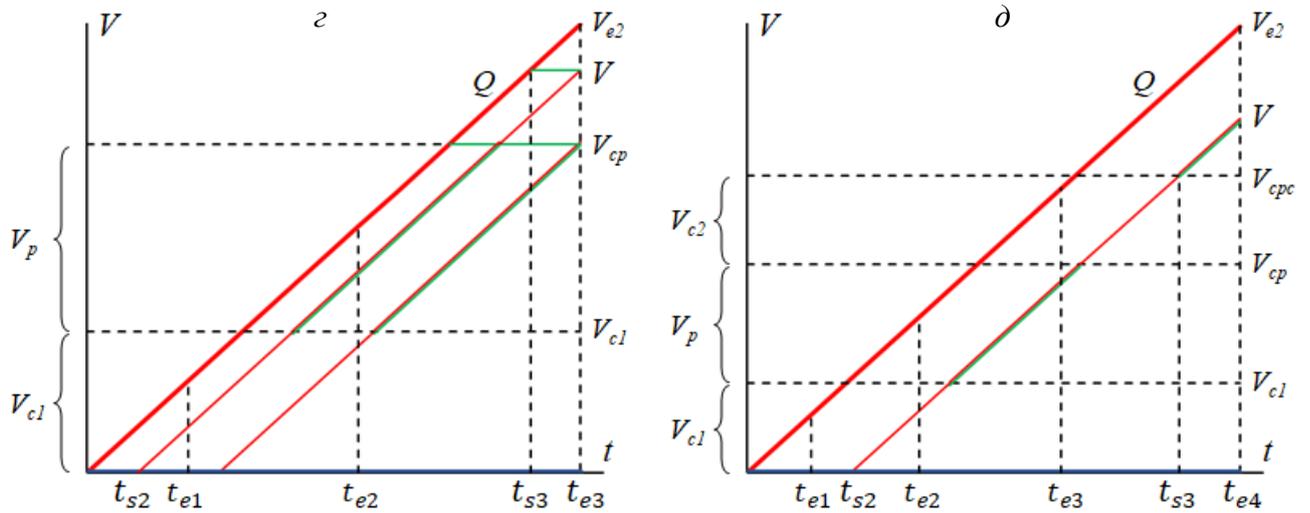

**Рис. 19**. Схема (*а*) и *V-t* диаграммы (*б – д*) прямоточного двух-колоночного $^{225}$Ac/$^{213}$Bi генератора для разных моментов $t_e$ элюирования: *б* – $0 \leq t_{e1} \leq \frac{V_{c1}}{Q}$; *в* – $\frac{V_{c1}}{Q} \leq t_{e2} \leq \frac{V_{cp}}{Q}$; *г* – $\frac{V_{cp}}{Q} \leq t_{e3} \leq \frac{V_{cpc}}{Q}$; *д* – $t_{e4} \geq \frac{V_{cpc}}{Q}$. Синие, красные и зеленые линии соответствуют движению $^{225}$Ac, $^{221}$Fr и $^{213}$Bi (вещества ①, ② и ③).

скорости их движения в сорбенте можно считать равными нулю: $q_1 = q_3 = 0$. В тех же условиях *k'* Fr(I) < 1 [11], т.е. в первом приближении Fr(I) движется в сорбенте со скоростью подвижной фазы: $q_2 = Q$.

После загрузки исходного раствора материнский радионуклид $^{225}$Ac (вещество ①) адсорбирован в начальном слое смолы первой колонки ($V = 0$). Находясь в покое определенное время, система приходит в подвижное равновесие, в котором $^{221}$Fr (вещество ②) и $^{213}$Bi (вещество ③) сконцентрированы там же, где и $^{225}$Ac. Затем в некоторый момент начинается движение элюента с объёмной скоростью $Q$. Обозначим количество $^{225}$Ac в этот момент $N_1^0$, а равновесные (исходные) количества $^{221}$Fr и $^{213}$Bi, определяемые ур-ием (2): $N_2^0 = \frac{\lambda_1}{\lambda_2 - \lambda_1} N_1^0$ и $N_3^0 = \frac{\lambda_1 \lambda_2}{(\lambda_2 - \lambda_1)(\lambda_3 - \lambda_1)} N_1^0$, соответственно. Поскольку скорость $^{221}$Fr в сорбенте $q_2 = Q$, его исходное количество движется вместе с фронтом элюента $V_{e2} = V_e$, тогда как исходное количество $^{213}$Bi остается вместе с $^{225}$Ac на старте первой колонки.

$\mathbf{0 \leq t_{e1} \leq \frac{V_{c1}}{Q}}$. *V-t* диаграмма для первого временного интервала, когда фронт элюента еще не вышел из первой колонки, показана на Рис. 19б. Концентрация $^{221}$Fr в дифференциальном элементе $dV$②, начинающем движение в момент $t_{s2}$, равна: $\left(\frac{dN_2}{dV}\right)_{t_{s2}} = \frac{\lambda_1}{Q} N_1(t_{s2}) = \frac{\lambda_1}{Q} N_1^0 e^{-\lambda_1 t_{s2}}$. В момент $t_{e1}$, когда элемент $dV$② оказывается в точке *V*, концентрация $^{221}$Fr в нем составит: $\frac{dN_2}{dV} = \frac{\lambda_1}{Q} N_1^0 e^{-\lambda_1 t_{s2} - \lambda_2 (t_e - t_{s2})}$. Подставляя $t_e - t_{s2} = \frac{V}{Q}$, приходим к полученному ранее выражению для концентрации вещества ②:

$$\frac{dN_2}{dV} = \frac{\lambda_1}{Q} N_1^0 e^{-\lambda_1 t_e - (\lambda_2 - \lambda_1)\frac{V}{Q}} \tag{16а}$$

В момент $t_{s3} = t_e - t_{s2} = \frac{V}{Q}$, когда фронт $V_{e2}$ достигает точки *V*, начинается накопление $^{213}$Bi (Рис. 19б), оно продолжается в течение $t_3 = t_e - t_{s3} = t_{s2}$. К моменту $t_{s3}$ исходного количества $^{221}$Fr, движущегося с фронтом, остается $(N_2)_{t_{s3}} = N_2^0 e^{-\lambda_2 \frac{V}{Q}}$. Концентрация $^{213}$Bi, образующегося в результате его распада: $\left(\frac{dN_3}{dV}\right)_{t_{s3}} = -\left(\frac{dN_2}{dV}\right)_{t_{s3}} = \frac{\lambda_2 N_2^0}{Q} e^{-\lambda_2 \frac{V}{Q}}$, служит граничным условием для уравнения материального баланса, записываемого в виде:

$$\frac{d^2 N_3}{dV dt} = \lambda_2 \frac{dN_2}{dV} - \lambda_3 \frac{dN_3}{dV} = \frac{\lambda_2 \lambda_1}{Q} N_1^0 e^{-\lambda_1 t_3 - \lambda_2 t_{s3}} - \lambda_3 \frac{dN_3}{dV} \tag{21}$$

Решая это уравнение относительно $t_3$, получаем уравнение концентрации $^{213}$Bi в первой колонке (диапазон $0 \leq V_{e2} \leq V_{c1}$):

$$\frac{dN_3}{dV} = \frac{\lambda_2 \lambda_1 N_1^0}{Q(\lambda_3 - \lambda_1)} e^{-\lambda_2 t_{s3}} \left( e^{-\lambda_1 t_3} - e^{-\lambda_3 t_3} \right) + \left( \frac{dN_3}{dV} \right)_{t_{s3}} e^{-\lambda_3 t_3} =$$

$$= \frac{\lambda_2 \lambda_1 N_1^0}{Q(\lambda_3 - \lambda_1)} e^{-\lambda_2 t_{s3}} \left( e^{-\lambda_1 t_3} - \frac{(\lambda_2 - \lambda_3)}{(\lambda_2 - \lambda_1)} e^{-\lambda_3 t_3} \right) = \frac{\lambda_2 \lambda_1 N_1^0}{Q(\lambda_3 - \lambda_1)} \left( e^{-\lambda_1 t_e - (\lambda_2 - \lambda_1)\frac{V}{Q}} - \frac{(\lambda_2 - \lambda_3)}{(\lambda_2 - \lambda_1)} e^{-\lambda_3 t_e - (\lambda_2 - \lambda_3)\frac{V}{Q}} \right) \tag{65}$$

Общее количество $^{213}$Bi в первой колонке в произвольный момент $t_e \leq \frac{V_{c1}}{Q}$ элюирования складывается из интеграла ур-ия (65) и исходного количества $^{213}$Bi, распавшегося к моменту $t_e$ ($N_3^0 e^{-\lambda_3 t_e}$).

$\frac{V_{c1}}{Q} \leq t_{e2} \leq \frac{V_{cp}}{Q}$. В момент $\frac{V_{c1}}{Q}$ фронт подвижной фазы выходит из первой колонки в трубку, в которой $^{221}$Fr и $^{213}$Bi движутся со скоростью $q_2 = q_3 = Q$. Количество $^{213}$Bi, образующегося из исходного $^{221}$Fr и тоже движущегося вместе с фронтом, определяется ур-ием Бейтмана:

$$N_3(V_{e2}) = \frac{\lambda_2}{(\lambda_3 - \lambda_2)} N_2^0 e^{-\lambda_2 \frac{V_{c1}}{Q}} \left( e^{-\lambda_2 \left( t_e - \frac{V_{c1}}{Q} \right)} - e^{-\lambda_3 \left( t_e - \frac{V_{c1}}{Q} \right)} \right) \tag{1а}$$

$V$-$t$ диаграмма движения дифференциального элемента $dV\text{③}$ во втором временном интервале показана на Рис. 19в. Накопление $^{213}$Bi в элементе $dV\text{③}$, отделяющемся от $V_{c1}$, начинается в момент $t_{s3} = t_{s2} + \frac{V_{c1}}{Q}$ и продолжается в течение $t_3 = t_e - t_{s3} = t_e - t_{s2} - \frac{V_{c1}}{Q}$ до прибытия в точку $V$. В уравнении (16а) для $\frac{dN_2}{dV}$ выразим $t_{s2}$ и $t_e - t_{s2}$ через $t_{s3}$ и $t_3$ и, решая ур-ие (21) относительно $t_3$ с граничным условием $t_3 = 0$: $\left( \frac{dN_3}{dV} \right)_{t_{s3}} = 0$, получим уравнение концентрации $^{213}$Bi в трубке объемом $V_p$, соединяющей две колонки (диапазон $V_{c1} \leq V_{e2} \leq V_{cp}$):

$$\frac{dN_3}{dV} = \frac{\lambda_2 \lambda_1 N_1^0}{Q(\lambda_3 - \lambda_2)} e^{-\lambda_1 t_{s3} - (\lambda_2 - \lambda_1)\frac{V_{c1}}{Q}} \left( e^{-\lambda_2 t_3} - e^{-\lambda_3 t_3} \right) =$$

$$= \frac{\lambda_2 \lambda_1 N_1^0}{Q(\lambda_3 - \lambda_2)} e^{-\lambda_1 t_e - (\lambda_2 - \lambda_1)\frac{V_{c1}}{Q}} \left( e^{-(\lambda_2 - \lambda_1)\frac{V - V_{c1}}{Q}} - e^{-(\lambda_3 - \lambda_1)\frac{V - V_{c1}}{Q}} \right) \tag{66}$$

Количество $^{213}$Bi в трубке складывается из интеграла ур-ия (66) и $^{213}$Bi, образующегося из исходного $^{221}$Fr (ур-ие (1а)). Проводя параллель с рассмотренной выше моделью движения в бесконечно длинной хроматографической колонке (Раздел 1.2), отметим, что ур-ие (66) при $V_{c1} = 0$ является частным случаем ур-ия (37а) при $q_3 = Q$.

Во втором временном интервале $^{213}$Bi содержится в двух диапазонах (Рис. 19в и 20а): в первой колонке, причем ур-ие (65) интегрируется теперь в пределах $0 \div V_{c1}$, и в трубке.

$\frac{V_{cp}}{Q} \leq t_{e3} \leq \frac{V_{cpc}}{Q}$. В момент $\frac{V_{cp}}{Q}$ подвижная фаза входит во вторую колонку. Количество $^{213}$Bi, достигающего входа во вторую колонку вместе с фронтом, равно ($V_p = V_{cp} - V_{c1}$):

$$N_3(V_{e2} = V_{cp}) = \frac{\lambda_2 \lambda_1}{(\lambda_3 - \lambda_2)(\lambda_2 - \lambda_1)} N_1^0 e^{-\lambda_2 \frac{V_{c1}}{Q}} \left( e^{-\lambda_2 \frac{V_p}{Q}} - e^{-\lambda_3 \frac{V_p}{Q}} \right) \tag{1гу}$$

Поскольку скорость $^{213}$Bi в сорбенте $q_3 = 0$, происходит скачок количества $^{213}$Bi в точке $V_{cp}$ в начале третьего временного интервала (Рис. 20а). С этого момента количество $^{213}$Bi в трубке (интеграл ур-ия (66) в пределах $V_{c1} \div V_{cp}$) зависит от времени как $e^{-\lambda_1 t_e}$, т.е. становится практически постоянным. Из $V$-$t$ диаграммы (Рис. 19г) видно, что время накопления $^{213}$Bi в точке $V_{cp}$ равно $t_3 = t_e - \frac{V_{cp}}{Q}$, а концентрация $^{213}$Bi на входе во вторую колонку:

$$\left( \frac{dN_3}{dV} \right)_{V_{cp}} = \frac{\lambda_2 \lambda_1 N_1^0}{Q(\lambda_3 - \lambda_2)} e^{-\lambda_1 t_3 - \lambda_2 \frac{V_{c1}}{Q}} \left( e^{-\lambda_2 \frac{V_p}{Q}} - e^{-\lambda_3 \frac{V_p}{Q}} \right) \tag{66а}$$

Количество $^{213}$Bi в точке $V_{cp}$ определяется уравнением материального баланса, записанным в виде: $\frac{dN_3}{dt} = Q\left(\frac{dN_3}{dV}\right)_{V_{cp}} - \lambda_3 N_3$. Его решение относительно $t_3$ с использованием ур-ия (1гу) в качестве граничного условия приводит к выражению:

$$N_3(V_{cp}) = \frac{\lambda_2 \lambda_1 N_1^0}{(\lambda_3 - \lambda_1)} e^{-\lambda_2 \frac{V_{c1}}{Q}} \left(e^{-\lambda_2 \frac{V_p}{Q}} - e^{-\lambda_3 \frac{V_p}{Q}}\right) \left(\frac{e^{-\lambda_1\left(t_e - \frac{V_{cp}}{Q}\right)}}{(\lambda_3 - \lambda_2)} - \frac{e^{-\lambda_3\left(t_e - \frac{V_{cp}}{Q}\right)}}{(\lambda_1 - \lambda_2)}\right) \qquad (67)$$

Накопление $^{213}$Bi в произвольной точке $V$ сорбента второй колонки аналогично накоплению в первой колонке (Рис. 19б и 19г), т.е. ур-ие (65) действительно для концентрации $^{213}$Bi во второй колонке. Количество $^{213}$Bi во второй колонке складывается из интеграла ур-ия (65) в пределах диапазона $V_{cp} \div V_{cpc}$ и количества $^{213}$Bi на входе во вторую колонку (точка $V_{cp}$), определенного по ур-ию (67).

В третьем временном интервале $^{213}$Bi содержится в трех диапазонах (Рис. 20а): в первой колонке, в трубке (ур-ие (66) интегрируется в пределах $V_{c1} \div V_{cp}$) и во второй колонке.

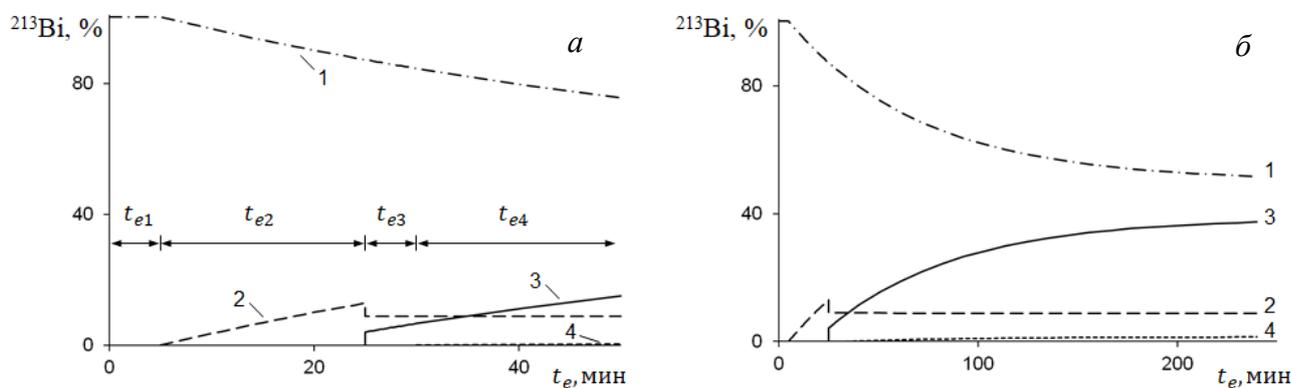

**Рис. 20**. Распределение $^{213}$Bi в элементах хроматографической системы в разных временных интервалах: $а$ – начальная стадия накопления $^{213}$Bi; $б$ – достижение равновесного распределения (~ 4 часа). Обозначения: 1 – первая колонка (диапазон $0 \div V_{c1}$); 2 – трубка между колонками (диапазон $V_{c1} \div V_{cp}$); 3 – вторая колонка (диапазон $V_{cp} \div V_{cpc}$); 4 – элюат (диапазон $V_{cpc} \div V_{e2}$). В расчете использованы значения: $Q = 1$ мл/мин; $V_{c1} = V_{c2} = 5$ мл; $V_p = 20$ мл.

$\boldsymbol{t_{e4} \geq \frac{V_{cpc}}{Q}}$. Последний интервал, начинающийся в момент выхода элюата из второй колонки, представлен $V$-$t$ диаграммой, изображенной на Рис. 19д. $^{213}$Bi движется одинаково ($q_3 = Q$) в диапазонах $V_{c1} \div V_{cp}$ и $V_{cpc} \div V_{e2}$ (вне сорбента, Рис. 19в и 19д), поэтому уравнение для концентрации $^{213}$Bi в диапазоне $V_{cpc} \div V_{e2}$ отличается от ур-ия (66) только границей, с которой начинается накопление:

$$\frac{dN_3}{dV} = \frac{\lambda_2 \lambda_1 N_1^0}{Q(\lambda_3 - \lambda_2)} e^{-\lambda_1 t_e - (\lambda_2 - \lambda_1)\frac{V_{cpc}}{Q}} \left(e^{-(\lambda_2 - \lambda_1)\frac{V - V_{cpc}}{Q}} - e^{-(\lambda_3 - \lambda_1)\frac{V - V_{cpc}}{Q}}\right) \qquad (68)$$

Количество $^{213}$Bi в элюате, вышедшем из второй колонки, складывается из интеграла ур-ия (68) и $^{213}$Bi, образующегося из исходного $^{221}$Fr (ур-ие (1а)).

Диапазоны, в которых распределены $^{221}$Fr и $^{213}$Bi с момента начала четвертого временного интервала (Рис. 20а):
$0 \div V_{c1}$ – первая колонка объемом $V_{c1}$;
$V_{c1} \div V_{cp}$ – трубка объемом $V_p$, соединяющая обе колонки;
$V_{cp} \div V_{cpc}$ – вторая колонка объемом $V_{c2}$;
$V_{cpc} \div V_{e2}$ – элюат.
Модель прямоточного двух-колоночного $^{225}$Ac/$^{213}$Bi генератора построена. Общее количество $^{221}$Fr и $^{213}$Bi в хроматографической системе в любой момент времени равно $N_2^0 e^{-\lambda_1 t_e}$ и $N_3^0 e^{-\lambda_1 t_e}$, соответственно. Другими словами, система остается в состоянии "интегрального" подвижного равновесия, тогда как в ее элементах происходит перераспределение веществ.

Спустя примерно 5-6 $T_{1/2}$ ($^{213}$Bi) или 4-5 часов после начала движения элюента количества $^{213}$Bi в элементах системы становятся практически постоянными (Рис. 20б), можно сказать, что система приходит к новому "дифференциальному" подвижному равновесию по $^{213}$Bi. В этом состоянии уравнение концентрации $^{213}$Bi в первой колонке (ур-ие (65)) упрощается и приходит к виду:

$$\frac{dN_3}{dV} = \frac{\lambda_2 \lambda_1 N_1^0}{Q(\lambda_3 - \lambda_1)} e^{-\lambda_1 t_e - (\lambda_2 - \lambda_1)\frac{V}{Q}} = \frac{\lambda_2}{(\lambda_3 - \lambda_1)} \frac{dN_2}{dV} \qquad (65р)$$

Интегрируя ур-ие (65р), находим равновесное количество $^{213}$Bi в первой колонке: $N_3(0 \div V_{c1}) = N_3^0 e^{-\lambda_1 t_e} \left(1 - e^{-(\lambda_2 - \lambda_1)\frac{V_{c1}}{Q}}\right)$. Количество $^{213}$Bi в трубке становится практически постоянным раньше, в момент $\frac{V_{cp}}{Q}$ входа подвижной фазы во вторую колонку. Уравнение количества $^{213}$Bi на входе во вторую колонку (ур-ие (67)) в новом состоянии принимает вид:

$$N_3(V_{cp}) \approx \frac{\lambda_2 \lambda_1 N_1^0}{(\lambda_1 - \lambda_3)(\lambda_2 - \lambda_3)} e^{-\lambda_1 t_e - \lambda_2 \frac{V_{c1}}{Q}} \left(e^{-\lambda_2 \frac{V_p}{Q}} - e^{-\lambda_3 \frac{V_p}{Q}}\right) \qquad (67р),$$

а равновесное количество $^{213}$Bi во второй колонке складывается из интеграла ур-ия (65р) в пределах $V_{cp} \div V_{cpc}$ и ур-ия (67р). В элюате, объем которого непрерывно увеличивается, количество $^{213}$Bi стремится к равновесному значению, определяемому из ур-ия (68): $N_3(V_{cpc} \div V_{e2}) \approx N_3^0 e^{-\lambda_1 t_e - (\lambda_2 - \lambda_1)\frac{V_{cpc}}{Q}}$.

Для взятых в пробный расчет параметров (Рис. 20: $Q = 1$ мл/мин; $V_{c1} = V_{c2} = 5$ мл; $V_p = 20$ мл) оказывается, что в состоянии "дифференциального" подвижного равновесия более 50% $^{213}$Bi остается на первой колонке, менее 10% - в промежуточной трубке и около 40% накапливается на второй колонке, а объем потраченного элюента составляет порядка 250 мл. Накопленный $^{213}$Bi извлекают из второй колонки, например, небольшим количеством (1-2 колоночных объема) 1М раствора HCl [17]. Очевидно, чтобы повысить содержание $^{213}$Bi на второй колонке, надо уменьшить объем первой. С другой стороны, смолы в первой колонке должно быть достаточно, чтобы удерживать $^{225}$Ac в течение всего срока действия генератора, примерно 2-3 $T_{1/2}$ ($^{225}$Ac) или 20-30 дней. Предполагая с запасом, что генератор используется дважды в день, и на один цикл получения $^{213}$Bi расходуется 0.5 л элюента, получаем, что общий объем пропущенного элюента составит 20-30 л. Исходя из значения коэффициента удерживания $k'$ смолой Actinide Resin ионов Ac(III), объем смолы можно уменьшить до 0.5 мл, при этом свободный объем будет равен: $V_{c1} = 0.34$ мл. Небольшая часть $^{225}$Ac, вымываемого из первой колонки, будет удерживаться на второй, которая будет предотвращать попадание $^{225}$Ac в извлекаемый из генератора $^{213}$Bi, обеспечивая высокую радионуклидную чистоту продукта.

Размер второй колонки тоже имеет смысл уменьшить по двум соображениям: i) $^{213}$Bi концентрируется преимущественно в начале колонки (точка $V_{cp}$); ii) для дальнейшего использования желательно извлечение $^{213}$Bi из генератора в возможно меньшем объеме раствора. Уменьшим объем смолы до 0.3 мл (свободный объем $V_{c2} = 0.2$ мл).

Как оценить оптимальный объем $V_p$ трубки, соединяющей обе колонки? Для этого продифференцируем количество $^{213}$Bi, накапливаемое в точке $V_{cp}$ (ур-ие (67)), по времени $\frac{V_p}{Q}$ нахождения (residence time) $^{213}$Bi в трубке. Приравнивая производную нулю: $\frac{dN_3(V_{cp})}{d(V_p/Q)} = 0$, находим значение $\frac{V_p}{Q}$, соответствующее максимуму $N_3(V_{cp})$: $\left(\frac{V_p}{Q}\right)_{max} = \frac{\ln(\lambda_2/\lambda_3)}{(\lambda_2 - \lambda_3)}$. Величина максимума $N_3(V_{cp})$, как и количества $^{213}$Bi во всей второй колонке, оказывается немногим более 70%, даже при условии $V_{c1} = 0$. Рис. 21 объясняет физический смысл максимума накопления $^{213}$Bi на второй колонке.

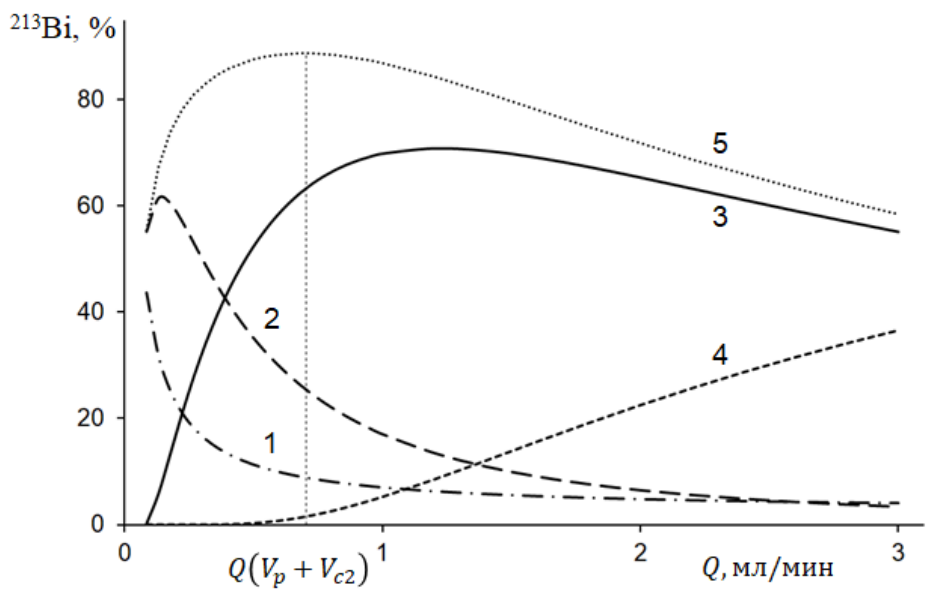

**Рис. 21**. Распределение $^{213}$Bi в элементах хроматографической системы в зависимости от скорости пропускания элюента (время накопления $^{213}$Bi 4 часа). Обозначения доли $^{213}$Bi: 1 – в первой колонке ($V_{c1} = 0.34$ мл); 2 – в трубке между колонками ($V_p = 20$ мл); 3 – во второй колонке ($V_{c2} = 0.2$ мл); 4 – в элюате; 3 – суммарно в трубке между колонками и во второй колонке ($V_p + V_{c2} = 20.2$ мл).

При заданном объеме $V_p$ трубки, пропускание элюента с низкой скоростью ведет к высокому содержанию $^{213}$Bi в первой колонке и в трубке, а при высокой скорости – растет доля $^{213}$Bi в элюате, поскольку время пребывания $^{221}$Fr в трубке (и, следовательно, степень его распада в $^{213}$Bi) уменьшается. Максимальное количество $^{213}$Bi в трубке определяется и объясняется подобным образом.

Анализируя район максимума накопления $^{213}$Bi на второй колонке (Рис. 21), замечаем, что значительное количество $^{213}$Bi остается в растворе в трубке. Например, при $Q = 1$ мл/мин в трубке находится ~17% $^{213}$Bi. Суммарное содержание $^{213}$Bi в трубке и во второй колонке достигает 90% при $Q(V_p + V_{c2}) \sim 0.7$ мл/мин. Для повышения эффективности извлечения $^{213}$Bi из генератора можно предусмотреть слив раствора из трубки, чтобы соединить его с $^{213}$Bi, смытым со второй колонки. Но это ведет к нежелательному росту объема конечного раствора $^{213}$Bi. Можно пойти другим путем, использовав описанный в Разделе 2.1 прием дискретного изменения скорости пропускания элюента.

### 3.2. Использование дискретного повышения скорости пропускания элюента в прямоточном двух-колоночном $^{225}$Ac/$^{213}$Bi генераторе

Повысим скорость пропускания элюента в $K$ раз ($K = \frac{Q_1}{Q_2}$) в выбранный момент времени $t_1 > \frac{V_{cpc}}{Q_1}$. Дискретное изменение хроматографической среды (подвижной или неподвижной фазы), рассмотренное в Разделах 2.1 и 2.2, приводит к образованию вторичных фронтов, разделяющих область существования дочерних веществ на дополнительные диапазоны. Как показано на Рис. 9 для вещества ②, изменяющего скорость движения в момент $t_1$, происходит образование двух вторичных фронтов. В данном случае, когда материнское вещество не движется и сконцентрировано в начале первой колонки, в области существования $^{221}$Fr образуется один вторичный фронт (Рис. 22а).

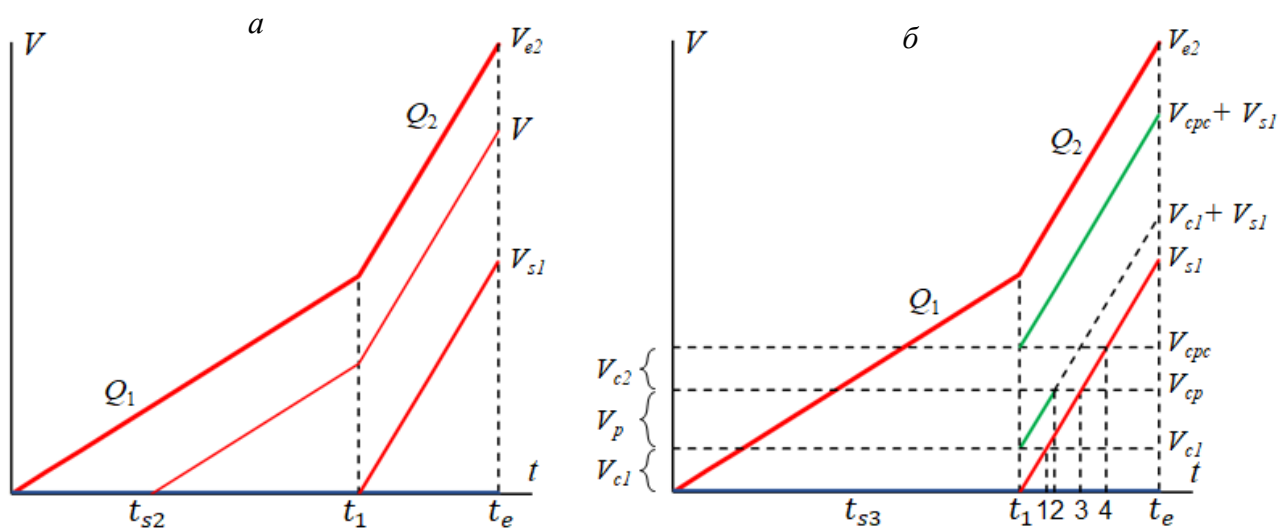

**Рис. 22**. *V-t* диаграммы образования вторичных фронтов при дискретном изменении скорости подвижной фазы в прямоточном двух-колоночном $^{225}$Ac/$^{213}$Bi генераторе: *а* – $^{221}$Fr (один вторичный фронт $V_{s1}$); *б* – $^{213}$Bi (три вторичных фронта: $V_{s1}$, $V_{c1} + V_{s1}$ и $V_{cpc} + V_{s1}$), цифрами обозначены границы временных интервалов: 1 – $t_1 + \frac{V_{c1}}{Q_2}$; 2 – $t_1 + \frac{V_p}{Q_2}$; 3 – $t_1 + \frac{V_{cp}}{Q_2}$; 4 – $t_1 + \frac{V_{cpc}}{Q_2}$. Синие, красные и зеленые линии соответствуют движению $^{225}$Ac, $^{221}$Fr и $^{213}$Bi (вещества ①, ② и ③).

Концентрация $^{221}$Fr в дифференциальном элементе $dV$② , начинающем движение в момент $t_{s2} > t_1$ и оказывающемся в момент $t_e$ в некоторой точке $V$ диапазона $0 \div V_{s1}$, определяется ур-ием (16а): $\frac{dN_2}{dV} = \frac{\lambda_1}{Q_2} N_1^0 e^{-\lambda_1 t_{s2} - \lambda_2(t_e - t_{s2})} = \frac{\lambda_1}{Q_2} N_1^0 e^{-\lambda_1 t_e - (\lambda_2 - \lambda_1)\frac{V}{Q_2}}$.

С помощью *V-t* диаграммы на Рис. 22а выразим концентрацию $^{221}$Fr в диапазоне $V_{s1} \div V_{e2}$ уравнением:

$$\frac{dN_2}{dV} = \frac{\lambda_1}{Q_1} N_1^0 e^{-\lambda_1 t_{s2} - \lambda_2(t_e - t_{s2})} = \frac{\lambda_1}{Q_1} N_1^0 e^{-\lambda_1 t_1 - \Lambda_2(t_e - t_1) - (\lambda_2 - \lambda_1)\frac{V}{Q_1}} \quad (69),$$

где $\Lambda_2 = \lambda_2 + \frac{\lambda_1 - \lambda_2}{K}$.

Скорость движения $^{213}$Bi меняется трижды при переходе подвижной фазы из сорбента в раствор и наоборот, в результате в момент $t_1$ появляются три вторичных фронта (Рис. 22б). Их пересечение с границами сорбентов образует временные интервалы, в которых мы наблюдаем реакцию распределения $^{213}$Bi в хроматографической системе на дискретное повышение скорости подвижной фазы. Рассмотрим, как изменяется концентрация $^{213}$Bi в каждом элементе системы.

**Первая колонка ($0 \div V_{c1}$)**. При условии $t_1 > \frac{V_{cpc}}{Q_1}$ (т.е. фронт $V_{e2}$ заведомо вышел из первой колонки), во временном интервале $t_1 \leq t_e \leq t_1 + \frac{V_{c1}}{Q_2}$ вторичный фронт $V_{s1}$ разделяет объем колонки на два диапазона (Рис. 22).

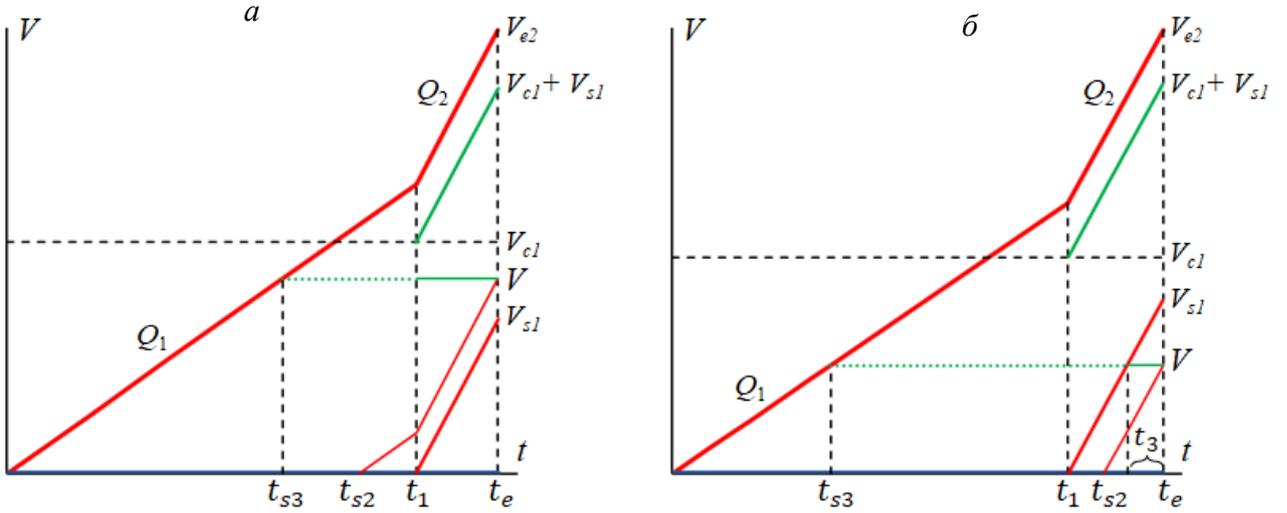

**Рис. 23**. $V$-$t$ диаграммы накопления $^{213}$Bi в первой колонке прямоточного двух-колоночного $^{225}$Ac/$^{213}$Bi генератора после дискретного повышения скорости элюента во временном интервале $t_1 \leq t_e \leq t_1 + \frac{V_{c1}}{Q_2}$: $а$ – диапазон $V_{s1} \div V_{c1}$; $б$ – диапазон $0 \div V_{s1}$. Синие, красные и зеленые линии соответствуют движению $^{225}$Ac, $^{221}$Fr и $^{213}$Bi (вещества ①, ② и ③).

Концентрация $^{213}$Bi в точке $V$ диапазона $V_{s1} \div V_{c1}$ (Рис. 23а) до момента $t_1$ подчиняется ур-ию (65), а при $t_e = t_1$ концентрация $\left(\frac{dN_3}{dV}\right)_{t_1}$ служит граничным условием для следующего участка. Концентрация $^{221}$Fr в точке $V$ определяется ур-ием (69). Решая уравнение материального баланса (21) относительно текущего времени накопления $t_3 = t_e - t_1$, приходим к выражению для концентрации $^{213}$Bi:

$$V_{s1} \div V_{c1}: \quad \frac{dN_3}{dV} = \frac{\lambda_2 \lambda_1 N_1^0}{Q_1(\lambda_3 - \Lambda_2)} e^{-\lambda_1 t_1 - (\lambda_2 - \lambda_1)\frac{V}{Q_1}} \left(e^{-\Lambda_2(t_e - t_1)} - e^{-\lambda_3(t_e - t_1)}\right) + \left(\frac{dN_3}{dV}\right)_{t_1} e^{-\lambda_3(t_e - t_1)} \quad (70)$$

В момент $t_e = t_1 + \frac{V}{Q_2}$, когда фронт $V_{s1}$ достигает точки $V$ (Рис. 23б), концентрация $\left(\frac{dN_3}{dV}\right)_{\left(t_1 + \frac{V}{Q_2}\right)}$, определенная по ур-ию (70) становится граничным условием для следующего участка. В диапазоне $0 \div V_{s1}$ концентрация $^{221}$Fr дается ур-ием (16а, $Q = Q_2$), а решением ур-ия (21) относительно $t_3 = t_e - t_1 - \frac{V}{Q_2}$ является уравнение для концентрации $^{213}$Bi:

$$0 \div V_{s1}: \quad \frac{dN_3}{dV} = \frac{\lambda_2 \lambda_1 N_1^0}{Q_2(\lambda_3 - \lambda_1)} e^{-\lambda_1 t_1} \left(e^{-\lambda_1(t_e - t_1) - (\lambda_2 - \lambda_1)\frac{V}{Q_2}} - e^{-\lambda_3(t_e - t_1) - (\lambda_2 - \lambda_3)\frac{V}{Q_2}}\right) +$$
$$+ \left(\frac{dN_3}{dV}\right)_{\left(t_1 + \frac{V}{Q_2}\right)} e^{-\lambda_3\left(t_e - t_1 - \frac{V}{Q_2}\right)} \quad (71)$$

В момент $t_1 + \frac{V_{c1}}{Q_2}$ исчезает диапазон $V_{s1} \div V_{c1}$, и далее для любого $t_e \geq t_1 + \frac{V_{c1}}{Q_2}$ во всей первой колонке концентрация $^{213}$Bi описывается ур-ием (71).

**Трубка между колонками ($V_{c1} \div V_{cp}$).** При соблюдении условий: i) $V_p > V_{c1}$; ii) $t_1 > \frac{V_{cpc}}{Q_1}$; в трубке во временном интервале $t_1 \leq t_e \leq t_1 + \frac{V_{c1}}{Q_2}$ образуются два диапазона: $V_{c1} \div V_{c1} + V_{s1}$ и $V_{c1} + V_{s1} \div V_{cp}$ (Рис. 24а, в). В момент $t_1 + \frac{V_{c1}}{Q_2}$ вторичный фронт $V_{s1}$ выходит из первой колонки в трубку, и во временном интервале $t_1 + \frac{V_{c1}}{Q_2} \leq t_e \leq t_1 + \frac{V_p}{Q_2}$ объем трубки делится на три диапазона: $V_{c1} \div V_{s1}$, $V_{s1} \div V_{c1} + V_{s1}$ и $V_{c1} + V_{s1} \div V_{cp}$ (Рис. 24д).

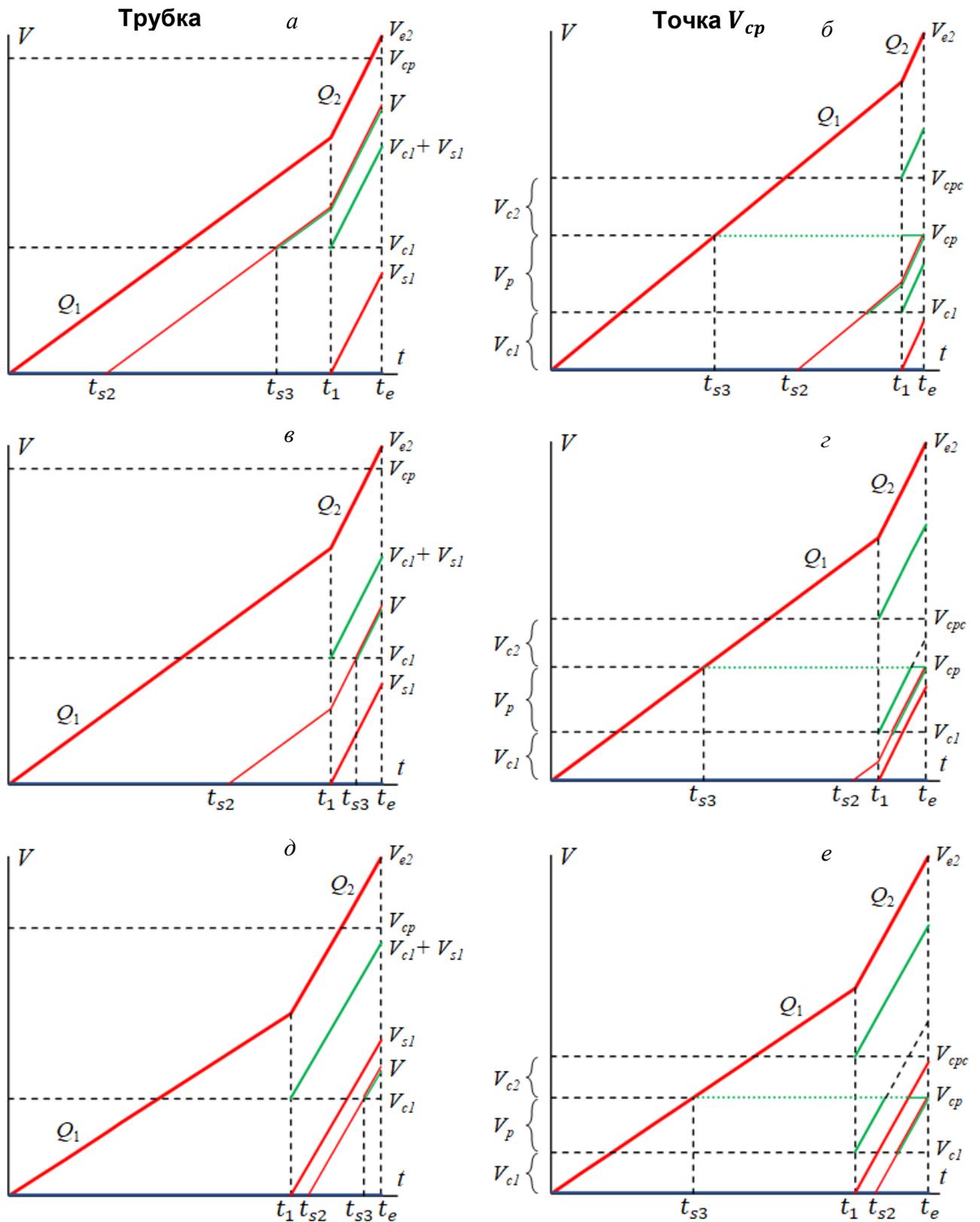

**Рис. 24**. *V-t* диаграммы движения и накопления $^{213}$Bi в трубке (*а, в, д*) и на входе во вторую колонку в точке $V_{cp}$ (*б, г, е*) прямоточного двух-колоночного $^{225}$Ac/$^{213}$Bi генератора после дискретного повышения скорости элюента. Трубка: временной интервал $t_1 \leq t_e \leq t_1 + \frac{V_{c1}}{Q_2}$, диапазоны: *а* – $V_{c1} + V_{s1} \div V_{cp}$; *в* – $V_{c1} \div V_{c1} + V_{s1}$; *д* – временной интервал $t_1 + \frac{V_{c1}}{Q_2} \leq t_e \leq t_1 + \frac{V_p}{Q_2}$, диапазон $V_{c1} \div V_{s1}$.

Точка $V_{cp}$: временные интервалы: *б* – $t_1 \leq t_e \leq t_1 + \frac{V_p}{Q_2}$; *г* – $t_1 + \frac{V_p}{Q_2} \leq t_e \leq t_1 + \frac{V_{cp}}{Q_2}$; *е* – $t_e \geq t_1 + \frac{V_{cp}}{Q_2}$.

Синие, красные и зеленые линии соответствуют движению $^{225}$Ac, $^{221}$Fr и $^{213}$Bi (вещества ①, ② и ③).

Накопление $^{213}$Bi в дифференциальном элементе $dV$③, прибывающем в точку $V$ любого из этих диапазонов (Рис. 24а, в, д), начинается в момент $t_{s3}$ выхода из первой колонки и продолжается в течение $t_3 = t_e - t_{s3}$. Нахождение концентрации $^{213}$Bi сводится к решению ур-ия (21) относительно $t_3$ с граничным условием $t_3 = 0$: $\left(\frac{dN_3}{dV}\right)_{t_{s3}} = 0$. Для диапазонов $V_{c1} \div V_{c1} + V_{s1}$ и $V_{c1} + V_{s1} \div V_{cp}$, подставляя в ур-ие (21) выражение $\frac{dN_2}{dV}$ из ур-ия (69), получаем:

$$\frac{dN_3}{dV} = \frac{\lambda_2 \lambda_1 N_1^0}{Q_1(\lambda_3 - \lambda_2)} e^{-\lambda_1 t_{s2} - \lambda_2 (t_{s3} - t_{s2})} \left(e^{-\lambda_2 t_3} - e^{-\lambda_3 t_3}\right) \tag{72-1}$$

С помощью $V$-$t$ диаграммы (Рис. 24а, в) выражаем концентрацию в виде функции $\frac{dN_3}{dV} = f(t_e, V)$:

$V_{c1} \div V_{c1} + V_{s1}$: $\quad \frac{dN_3}{dV} = \frac{\lambda_2 \lambda_1 N_1^0}{Q_1(\lambda_3 - \lambda_2)} e^{-\lambda_1 t_1 - \Lambda_2(t_e - t_1) - (\lambda_2 - \lambda_1)\frac{V}{Q_1}} \left(1 - e^{-(\lambda_3 - \lambda_2)\frac{V - V_{c1}}{Q_2}}\right) \tag{73}$

$V_{c1} + V_{s1} \div V_{cp}$: $\quad \frac{dN_3}{dV} = \frac{\lambda_2 \lambda_1 N_1^0}{Q_1(\lambda_3 - \lambda_2)} e^{-\lambda_1 t_1 - (\lambda_2 - \lambda_1)\frac{V}{Q_1}} \left(e^{-\Lambda_2(t_e - t_1)} - e^{-\Lambda_3(t_e - t_1) - (\lambda_3 - \lambda_2)\frac{V - V_{c1}}{Q_1}}\right) \tag{74}$,

где $\Lambda_3 = \lambda_3 + \frac{\lambda_1 - \lambda_3}{K}$.

Во временном интервале $t_1 + \frac{V_{c1}}{Q_2} \le t_e \le t_1 + \frac{V_p}{Q_2}$ появляется еще один диапазон $V_{c1} \div V_{s1}$ (Рис. 24д), при нахождении концентрации $^{213}$Bi в нем действуем так же, как для двух предыдущих диапазонов, но в ур-ие (21) подставляем выражение $\frac{dN_2}{dV}$ из ур-ия (16а, $Q = Q_2$):

$$\frac{dN_3}{dV} = \frac{\lambda_2 \lambda_1 N_1^0}{Q_2(\lambda_3 - \lambda_2)} e^{-\lambda_1 t_{s2} - \lambda_2 (t_{s3} - t_{s2})} \left(e^{-\lambda_2 t_3} - e^{-\lambda_3 t_3}\right) \tag{72-2}$$

Ур-ия (72-1) и (72-2) отличаются только нижним индексом при $Q$. С помощью $V$-$t$ диаграммы (Рис. 24д) выражаем концентрацию в виде функции $\frac{dN_3}{dV} = f(t_e, V)$:

$V_{c1} \div V_{s1}$: $\quad \frac{dN_3}{dV} = \frac{\lambda_2 \lambda_1 N_1^0}{Q_2(\lambda_3 - \lambda_2)} e^{-\lambda_1 t_e - (\lambda_2 - \lambda_1)\frac{V_{c1}}{Q_2}} \left(e^{-(\lambda_2 - \lambda_1)\frac{V - V_{c1}}{Q_2}} - e^{-(\lambda_3 - \lambda_1)\frac{V - V_{c1}}{Q_2}}\right) \tag{66-2}$

В момент $t_1 + \frac{V_p}{Q_2}$ исчезает диапазон $V_{c1} + V_{s1} \div V_{cp}$, и в интервале $t_1 + \frac{V_p}{Q_2} \le t_e \le t_1 + \frac{V_{cp}}{Q_2}$ объем трубки делится на два диапазона: $V_{c1} \div V_{s1}$ (ур-ие (66-2)) и $V_{s1} \div V_{cp}$ (ур-ие (73)). В момент $t_1 + \frac{V_{cp}}{Q_2}$ исчезает диапазон $V_{s1} \div V_{cp}$, и далее для любого $t_e \ge t_1 + \frac{V_{cp}}{Q_2}$ во всем объеме трубки $V_{c1} \div V_{cp}$ концентрация $^{213}$Bi описывается ур-ием (66-2).

**Вход во вторую колонку (точка $V_{cp}$).** До повышения скорости элюента количество $^{213}$Bi в начальном слое второй колонки $N_3(V_{cp})$ описывается ур-ием (67). В момент $t_e = t_1$ количество $\left(N_3(V_{cp})\right)_{t_1}$ становится граничным условием для следующего временного интервала $t_1 \le t_e \le t_1 + \frac{V_p}{Q_2}$. Схема накопления $^{213}$Bi в точке $V_{cp}$ (Рис. 26б) умышленно расположена напротив схемы движения $^{221}$Fr и $^{213}$Bi в диапазоне $V_{c1} + V_{s1} \div V_{cp}$ (Рис.24а), чтобы проще было увидеть, что концентрация $^{213}$Bi на входе во вторую колонку определяется ур-ием (74, $V = V_{cp}$). Решая уравнение $\frac{dN_3}{dt} = Q\left(\frac{dN_3}{dV}\right)_{V_{cp}} - \lambda_3 N_3$ относительно $t_3 = t_e - t_1$ с граничным условием $\left(N_3(V_{cp})\right)_{t_1}$ приходим к выражению для количества $^{213}$Bi в точке $V_{cp}$ в этом временном интервале:

$N_3(V_{cp}) = \frac{\lambda_2 \lambda_1 N_1^0}{(\lambda_3 - \lambda_2)} e^{-\lambda_1 t_1 - (\lambda_2 - \lambda_1)\frac{V_{cp}}{Q_1}} \left(\frac{e^{-\Lambda_2(t_e - t_1)} - e^{-\lambda_3(t_e - t_1)}}{K(\lambda_3 - \Lambda_2)} - \frac{e^{-\Lambda_3(t_e - t_1)} - e^{-\lambda_3(t_e - t_1)}}{(\lambda_3 - \lambda_1)} e^{-(\lambda_3 - \lambda_2)\frac{V_p}{Q_1}}\right) +$

$+ \left(N_3(V_{cp})\right)_{t_1} e^{-\lambda_3(t_e - t_1)}$ \hfill (75)

Далее, в момент $t_e = t_1 + \frac{V_p}{Q_2}$ количество $\left(N_3(V_{cp})\right)_{\left(t_1+\frac{V_p}{Q_2}\right)}$ из ур-ия (75) становится граничным условием для следующего временного интервала $t_1 + \frac{V_p}{Q_2} \leq t_e \leq t_1 + \frac{V_{cp}}{Q_2}$ (Рис. 24г). Как видно из Рис. 24в, концентрация $^{213}$Bi на входе во вторую колонку теперь определяется ур-ием (73, $V = V_{cp}$). Решение уравнения материального баланса в точке $V_{cp}$ с граничным условием $\left(N_3(V_{cp})\right)_{\left(t_1+\frac{V_p}{Q_2}\right)}$, полученное относительно $t_3 = t_e - t_1 - \frac{V_p}{Q_2}$, принимает вид:

$$N_3(V_{cp}) = \frac{\lambda_2 \lambda_1 N_1^0}{K(\lambda_3 - \Lambda_2)(\lambda_3 - \lambda_2)} e^{-\lambda_1 t_1 - (\lambda_2 - \lambda_1)\frac{V_{c1}}{Q_1}} \left(e^{-\lambda_2 \frac{V_p}{Q_2}} - e^{-\lambda_3 \frac{V_p}{Q_2}}\right)\left(e^{-\Lambda_2\left(t_e - t_1 - \frac{V_p}{Q_2}\right)} - e^{-\lambda_3\left(t_e - t_1 - \frac{V_p}{Q_2}\right)}\right) +$$
$$+ \left(N_3(V_{cp})\right)_{\left(t_1+\frac{V_p}{Q_2}\right)} e^{-\lambda_3\left(t_e - t_1 - \frac{V_p}{Q_2}\right)} \tag{76}$$

Наконец, для любого $t_e \geq t_1 + \frac{V_{cp}}{Q_2}$ (Рис. 24е) концентрация $^{213}$Bi на входе во вторую колонку определяется ур-ием (66-2, $V = V_{cp}$), а уравнение количества $^{213}$Bi, накопленного в точке $V_{cp}$ в течение $t_3 = t_e - t_1 - \frac{V_{cp}}{Q_2}$, полученное в результате решения уравнения материального баланса с граничным условием $\left(N_3(V_{cp})\right)_{\left(t_1+\frac{V_{cp}}{Q_2}\right)}$ из ур-ия (76), имеет вид:

$$N_3(V_{cp}) = \frac{\lambda_2 \lambda_1 N_1^0}{(\lambda_3 - \lambda_2)(\lambda_3 - \lambda_1)} e^{-\lambda_1 t_1 - \lambda_2 \frac{V_{c1}}{Q_2}} \left(e^{-\lambda_2 \frac{V_p}{Q_2}} - e^{-\lambda_3 \frac{V_p}{Q_2}}\right)\left(e^{-\lambda_1\left(t_e - t_1 - \frac{V_{cp}}{Q_2}\right)} - e^{-\lambda_3\left(t_e - t_1 - \frac{V_{cp}}{Q_2}\right)}\right) +$$
$$+ \left(N_3(V_{cp})\right)_{\left(t_1+\frac{V_{cp}}{Q_2}\right)} e^{-\lambda_3\left(t_e - t_1 - \frac{V_{cp}}{Q_2}\right)} \tag{77}$$

**Вторая колонка ($V_{cp} \div V_{cpc}$).** При условии $t_1 > \frac{V_{cpc}}{Q_1}$ во временном интервале $t_1 \leq t_e \leq t_1 + \frac{V_{cp}}{Q_2}$ для концентрации $^{213}$Bi во второй колонке действует ур-ие (70). В момент $t_1 + \frac{V_{cp}}{Q_2}$ вторичный фронт $V_{s1}$ достигает входа во вторую колонку и во временном интервале $t_1 + \frac{V_{cp}}{Q_2} \leq t_e \leq t_1 + \frac{V_{cpc}}{Q_2}$ разделяет объем колонки на два диапазона: $V_{cp} \div V_{s1}$ и $V_{s1} \div V_{cpc}$ (Рис. 22б). В первом диапазоне концентрация $^{213}$Bi определяется ур-ием (71), а во втором – ур-ием (70). Затем в момент $t_1 + \frac{V_{cpc}}{Q_2}$ исчезает диапазон $V_{s1} \div V_{cpc}$, и для любого $t_e \geq t_1 + \frac{V_{cpc}}{Q_2}$ во всей второй колонке концентрация $^{213}$Bi описывается ур-ием (71). Общее количество $^{213}$Bi во второй колонке в произвольный момент $t_e \geq t_1$ складывается из интегралов ур-ий (70) и (71) в соответствующих пределах и количества $^{213}$Bi на входе во вторую колонку (точка $V_{cp}$).

**Элюат ($V_{cpc} \div V_{e2}$).** В момент $t_1$ образуется вторичный фронт $V_{cpc} + V_{s1}$ (Рис. 22б), разделяющий объем элюата на два диапазона $V_{cpc} \div V_{cpc} + V_{s1}$ и $V_{cpc} + V_{s1} \div V_{e2}$ во временном интервале $t_1 \leq t_e \leq t_1 + \frac{V_{cp}}{Q_2}$. Концентрация $^{213}$Bi в этих диапазонах описывается ур-иями (73) и (74), в которых $V_{c1}$ заменен на $V_{cpc}$. В момент $t_1 + \frac{V_{cp}}{Q_2}$ другой вторичный фронт $V_{s1}$ выходит из второй колонки (Рис. 22б), и далее для $t_e \geq t_1 + \frac{V_{cp}}{Q_2}$ элюат состоит из трех диапазонов: $V_{cpc} \div V_{s1}$, $V_{s1} \div V_{cpc} + V_{s1}$ и $V_{cpc} + V_{s1} \div V_{e2}$. Концентрация $^{213}$Bi в новом диапазоне описывается ур-ием (66-2), в котором также, как в уравнениях для двух предыдущих диапазонов, $V_{c1}$ заменен на $V_{cpc}$.

Таким образом, мы получили уравнения, необходимые для описания дискретного повышения скорости пропускания элюента в прямоточном двух-колоночном $^{225}$Ac/$^{213}$Bi генераторе, и теперь с их помощью можем проанализировать, как изменяется распределение $^{213}$Bi в элементах генератора до и после повышения скорости элюента. Возьмем в расчет те же размеры обеих колонок и соединяющей трубки, что были использованы для Рис. 21, и будем пропускать элюент со скоростью $Q_1 = 0.7$ мл/мин. Спустя 4 часа хроматографическая система приходит к "дифференциальному" подвижному равновесию (Рис. 25а), при котором на второй колонке накоплено 63% $^{213}$Bi, а суммарное содержание $^{213}$Bi в трубке и во второй колонке максимально и достигает 90% (Рис. 21).

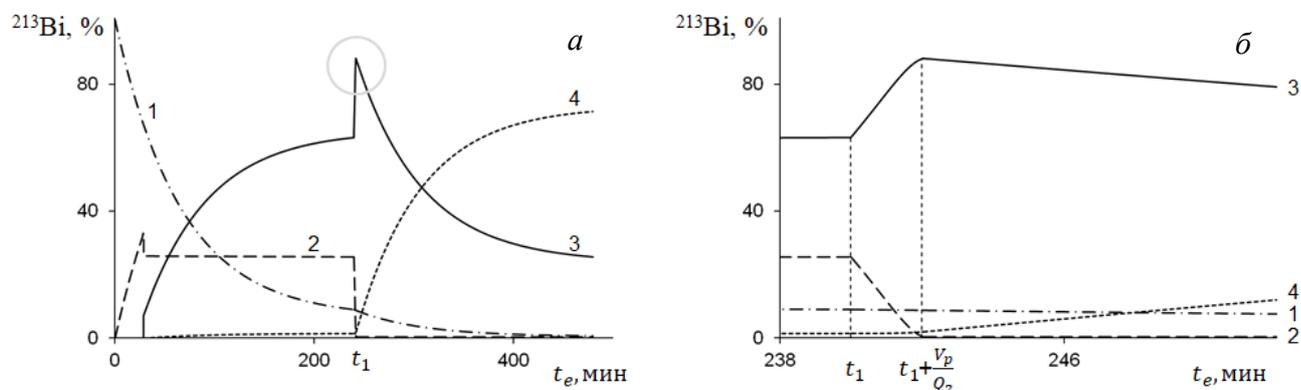

**Рис. 25**. Динамика распределения $^{213}$Bi в элементах прямоточного двух-колоночного $^{225}$Ac/$^{213}$Bi генератора до и после дискретного повышения скорости пропускания элюента: *а* – общий график, описывающий достижение первого равновесного распределения для скорости $Q_1$, изменение скорости $Q_1 \to Q_2$ в момент $t_1$ и достижение второго равновесного распределения для скорости $Q_2$; *б* – район момента $t_1$, обозначенный на графике (*а*) серой окружностью. Обозначения: 1 – первая колонка; 2 – трубка между колонками; 3 – вторая колонка; 4 – элюат. В расчете использованы значения: $Q_1 = 0.7$ мл/мин; $Q_2 = 10$ мл/мин; $t_1 = 240$ мин; $V_{c1} = 0.34$ мл; $V_{c2} = 0.2$ мл; $V_p = 20$ мл; общее время элюирования $t_e = 480$ мин.

В этот момент $t_1$ увеличим скорость элюента до $Q_2 = 10$ мл/мин. Вообще говоря, чем $Q_2$ больше, тем лучше. Выбранное значение достаточно велико, и в то же время не вызывает значительного роста гидравлического сопротивления при пропускании раствора через колонки с сорбентом и может быть реализовано с помощью перистальтического насоса и стандартных соединений.

Сразу после повышения скорости элюента в момент $t_1$ наблюдается резкий рост количества $^{213}$Bi до ~90% на второй колонке (Рис. 25а) и практически зеркальное падение содержания $^{213}$Bi в трубке. Как следует из подробного графика на Рис. 25б, такие изменения длятся до момента $t_1 + \frac{V_p}{Q_2}$, т.е. в течение двух минут. Фактически, во временном интервале $t_1 \leq t_e \leq t_1 + \frac{V_p}{Q_2}$ в трубке происходит вытеснение более концентрированного раствора $^{213}$Bi менее концентрированным. В момент $t_1 + \frac{V_p}{Q_2}$ концентрированный раствор заканчивается, и резкий рост количества $^{213}$Bi на второй колонке прекращается. Другими словами, мы быстро, со скоростью $Q_2$, пропускаем раствор объемом $V_p = 20$ мл из трубки через вторую колонку и адсорбируем находящийся в нем $^{213}$Bi.

Далее, после того, как вторичные фронты вышли в элюат, хроматографическая система стремится ко второму "дифференциальному" подвижному равновесию, характерному для скорости $Q_2$, и спустя еще 4 часа достигает его. При этом выражение для концентрации $^{213}$Bi в колонках (ур-ие (71)) упрощается и принимает вид ур-ия (65р), выражение для количества $^{213}$Bi, накопленного на входе во вторую колонку (ур-ие (77)) превращается в ур-ие (67р), а количество $^{213}$Bi в элюате стремится к полученной в предыдущем разделе величине $N_3(V_{cpc} \div V_{e2}) \approx N_3^0 e^{-\lambda_1 t_e - (\lambda_2 - \lambda_1)\frac{V_{cpc}}{Q_2}}$. То есть, равновесное распределение $^{213}$Bi определяется конфигурацией хроматографической системы и скоростью движения веществ в ее элементах и не зависит от исходных условий.

Для получения максимального (~90%) количества $^{213}$Bi на второй колонке надо закончить пропускание элюента в момент $t_1 + \frac{V_p}{Q_2}$. Затем основная часть $^{213}$Bi может быть десорбирована небольшим (0.5-1 мл) объемом 1М раствора HCl [12]. Описанная процедура накопления $^{213}$Bi, состоящая из достижения "дифференциального" подвижного равновесия при небольшой скорости элюента, дискретного повышения скорости и быстрой адсорбции равновесного $^{213}$Bi из раствора в трубке, позволяет достичь эффективности предложенного $^{225}$Ac/$^{213}$Bi генератора, сравнимой с эффективностью наиболее распространенного [15], используемого в клинических испытаниях. Однако рассмотренная генераторная схема менее компактна и требует пропускания 200-250 мл элюента на стадии накопления $^{213}$Bi. Для уменьшения размеров и повышения технологичности $^{225}$Ac/$^{213}$Bi генератора применим прием циркуляции подвижной фазы, т.е. направим раствор, вытекающий из второй колонки, на вход первой, образовав замкнутый контур.

### 3.3. Частные случаи распределения веществ ② и ③ в замкнутом контуре

В Разделе 2.3 на примере цепочки из двух реакций был разработан подход к описанию распределения дочернего вещества в контуре с циркулирующей подвижной фазой. Расширим его на цепочку из трех реакций ① → ② → ③ → ($^{225}$Ac → $^{221}$Fr → $^{213}$Bi →) и рассмотрим частный случай замкнутой хроматографической системы, в которой вещество ① удерживается в тонком слое сорбента: $q_1 = 0$. Совместим вещество ① с координатой $V = 0$ (Рис. 26).

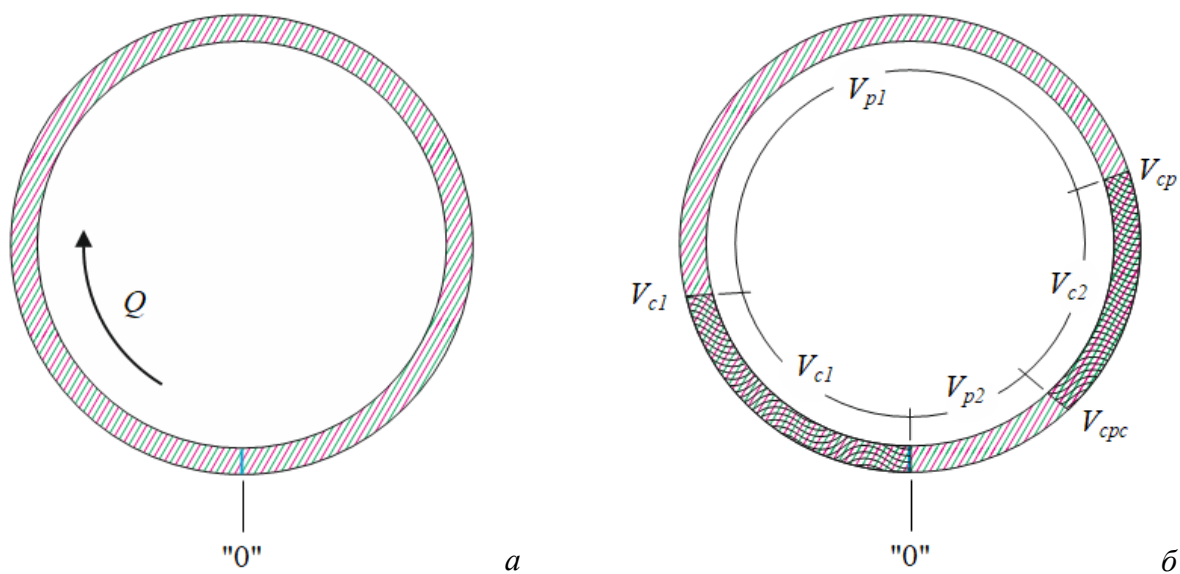

**Рис. 26.** Схема движения веществ ② и ③ в замкнутом хроматографическом контуре объемом $V_0$, в котором вещество ① удерживается в точке $V = 0$: *а* – контур, заполненный однородной средой, за исключением точки $V = 0$; *б* – контур $^{225}$Ac/$^{213}$Bi генератора, состоящий из двух колонок, заполненных сорбентом со свободным объемом $V_{c1}$ и $V_{c2}$ и соединенных трубками объемом $V_{p1}$ и $V_{p2}$ ($V_0 = V_{c1} + V_{p1} + V_{c2} + V_{p2}$). Синей, красными и зелеными линиями показаны вещества ①, ② и ③ ($^{225}$Ac, $^{221}$Fr и $^{213}$Bi).

Рассмотрим сначала контур, заполненный однородной хроматографической средой со свободным объемом $V_0$ (Рис. 26а), и найдем распределения дочерних веществ ② и ③ для следующих комбинаций скоростей: i) $q_2 = q_3 = Q$; ii) $q_2 = Q$; $q_3 = 0$ в точке "0", $q_3 = Q$ вне точки "0"; iii) $q_2 = Q$, $q_3 = 0$. Затем используем полученные закономерности для построения $^{225}$Ac/$^{213}$Bi генератора, изображенного на Рис. 26б.

### 3.3.1. Соотношение скоростей веществ ② и ③: $q_2 = q_3 = Q$

Положим, что в контуре (Рис. 26а), находящемся в покое, достигнуто подвижное равновесие, и равновесные (исходные) количества веществ ② и ③ сконцентрированы там же, где и вещество ①. Затем в некоторый момент начинается движение раствора, веществ ② и ③ с объёмной скоростью $Q$. $V$-$t$ диаграмма движения веществ показана на Рис. 27.

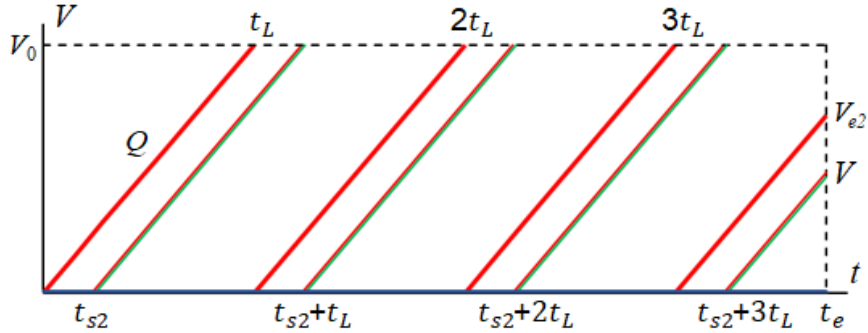

**Рис. 27**. $V$-$t$ диаграмма движения веществ в замкнутом контуре, заполненном однородной хроматографической средой, в которой $q_1 = 0$, $q_2 = q_3 = Q$. Синие, красные и зеленые линии соответствуют движению веществ ①, ② и ③.

Положение фронта $V_{e2}$ определяется ур-ием (53):
$$V_{e2} = Qt_e - lV_0 \tag{53}$$
где $l = \left[\frac{t_e}{t_L}\right]$ – число целых циклов с периодом $t_L = \frac{V_0}{Q}$, совершенных $V_{e2}$. Число пересечений $V_{e2}$ с веществом ①, сосредоточенным в координате $V = 0$ в момент $t_e$ равно $n = \left[1 + \frac{t_e}{t_L}\right] = 1 + l$, включая точку $(0, 0)$. Вместе с фронтом $V_{e2}$ движутся исходные вещества ② и ③, количество которых в момент $t_e$ равно: $N_2(V_{e2}) = N_2^0 e^{-\lambda_2 t_e}$ и $N_3(V_{e2}) = \frac{\lambda_2}{\lambda_3 - \lambda_2} N_2^0 \left(e^{-\lambda_2 t_e} - e^{-\lambda_3 t_e}\right) + N_3^0 e^{-\lambda_3 t_e}$.

Положение дифференциального элемента $dV$②, стартующего в момент $t_{s2}$, определяется ур-ием (54):
$$V = Q(t_e - t_{s2}) - l_V V_0 \tag{54}$$
где $l_V = \left[\frac{t_e - t_{s2}}{t_L}\right]$ – число целых циклов, совершенных элементом $dV$②. Число пересечений пути $dV$② с веществом ① в момент $t_e$ равно $n_V = \left[1 + \frac{t_e - t_{s2}}{t_L}\right] = 1 + l_V$, включая точку $(t_{s2}, 0)$.

Рассмотрим формирование распределения вещества ② в контуре (Рис. 27). Во временном интервале $t_{s2} \div t_{s2} + t_L$ концентрация вещества ② описывается ур-ием (16а):
$$\frac{dN_2}{dV} = \frac{\lambda_1}{Q} N_1^0 e^{-\lambda_1 t_{s2} - \lambda_2(t_e - t_{s2})} = \frac{\lambda_1}{Q} N_1^0 e^{-\lambda_1 t_e - (\lambda_2 - \lambda_1)\frac{V}{Q}} \tag{16а}$$
В момент $t_e = t_{s2} + t_L$ (счетчик $l_V$ увеличивается на единицу $l_V = 1$) элемент $dV$② пересекает "0"-отметку, и концентрация вещества ② в нем равна: $\left(\frac{dN_2}{dV}\right)_0 = \frac{\lambda_1}{Q} N_1^0 e^{-\lambda_1(t_{s2} + t_L) - (\lambda_2 - \lambda_1)t_L}$. Из вещества ①, находящегося в этой точке, в него добавляется: $\left(\frac{dN_2}{dV}\right)_1 = \frac{\lambda_1}{Q} N_1^0 e^{-\lambda_1(t_{s2} + t_L)}$, и в интервале $t_{s2} + t_L \div t_{s2} + 2t_L$ концентрация вещества ② равна:

$$\frac{dN_2}{dV} = \left(\left(\frac{dN_2}{dV}\right)_0 + \left(\frac{dN_2}{dV}\right)_1\right)e^{-\lambda_2(t_e - t_{s2} - t_L)} = \frac{\lambda_1}{Q} N_1^0 e^{-\lambda_1 t_e - (\lambda_2 - \lambda_1)\frac{V}{Q}} \left(1 + e^{-(\lambda_2 - \lambda_1)t_L}\right).$$ Рассуждая так же, получаем, что в произвольном временном интервале $t_{s2} + l_V t_L \div t_{s2} + (l_V + 1)t_L$ концентрация вещества ②  равна:

$$\frac{dN_2}{dV} = \frac{\lambda_1}{Q} N_1^0 e^{-\lambda_1(t_{s2} + l_V t_L) - \lambda_2(t_e - t_{s2} - l_V t_L)} \sum_{i=0}^{l_V} e^{-(\lambda_2 - \lambda_1) i t_L} = \frac{\lambda_1}{Q} N_1^0 e^{-\lambda_1 t_e} \sum_{i=0}^{l_V} e^{-(\lambda_2 - \lambda_1)\left(i t_L + \frac{V}{Q}\right)} \tag{78}$$

Введем обозначение: $S_{ab}^c = \sum_{i=0}^c e^{-(\lambda_a - \lambda_b)t_L}$, где $a$, $b$ и $c$ – целые числа. Сумму $S_{21}^{l_V}$ ряда можно выразить в виде: $S_{21}^{l_V} = \frac{1 - e^{-(\lambda_2 - \lambda_1)(l_V + 1)t_L}}{1 - e^{-(\lambda_2 - \lambda_1)t_L}}$. Заметим, что ур-ие (78) отличается от исходного ур-ия (16а) только наличием суммы $S_{21}^{l_V}$ ряда. С ростом $t_e$ ур-ие (78) превращается в ур-ие (64):

$$\frac{dN_2}{dV} = \frac{\lambda_1 N_1^0}{Q} \frac{e^{-\lambda_1 t_e - (\lambda_2 - \lambda_1)\frac{V}{Q}}}{\left(1 - e^{-(\lambda_2 - \lambda_1)t_L}\right)} \tag{64}$$

подтверждая еще раз, что равновесное распределение в системе определяется ее параметрами и не зависит от стартовых условий.

Получим теперь выражение для концентрации вещества ③. Как видно из $V$-$t$ диаграммы (Рис. 27), дифференциальные элементы $dV$② и $dV$③ движутся одинаково ($t_{s2} = t_{s3}$). Для временного интервала $t_{s3} \div t_{s3} + t_L$ подставляем в ур-ие (21) концентрацию $\frac{dN_2}{dV}$ из ур-ия (78, $l_V = 0$), заменяя $t_{s2}$ на $t_{s3}$, решаем его относительно $t_3 = t_e - t_{s3}$ с граничным условием $t_3 = 0$: $\left(\frac{dN_3}{dV}\right)_{t_{s3}} = 0$ и находим $\frac{dN_3}{dV}$:

$$\frac{dN_3}{dV} = \frac{\lambda_2 \lambda_1 N_1^0}{Q(\lambda_3 - \lambda_2)} e^{-\lambda_1 t_{s3}} \left(e^{-\lambda_2 t_3} - e^{-\lambda_3 t_3}\right) = \frac{\lambda_2 \lambda_1 N_1^0}{Q(\lambda_3 - \lambda_2)} e^{-\lambda_1 t_e} \left(e^{-(\lambda_2 - \lambda_1)\frac{V}{Q}} - e^{-(\lambda_3 - \lambda_1)\frac{V}{Q}}\right) \tag{79}$$

Ур-ие (79) сводится к ур-ию (66) при $V_{c1} = 0$. В момент $t_e = t_{s3} + t_L$ ($l_V = 1$) концентрация $\frac{dN_3}{dV}$ равна: $\left(\frac{dN_3}{dV}\right)_1 = \frac{\lambda_2 \lambda_1 N_1^0}{Q(\lambda_3 - \lambda_2)} e^{-\lambda_1(t_{s3} + t_L)} \left(e^{-(\lambda_2 - \lambda_1)t_L} - e^{-(\lambda_3 - \lambda_1)t_L}\right)$. Для следующего временного интервала $t_{s3} + t_L \div t_{s3} + 2t_L$ используем в ур-ие (21) концентрацию $\frac{dN_2}{dV}$ из ур-ия (78, $l_V = 1$) и, решая его относительно $t_3 = t_e - t_{s3} - t_L$ с граничным условием $t_3 = 0$: $\left(\frac{dN_3}{dV}\right)_{(t_{s3} + t_L)} = \left(\frac{dN_3}{dV}\right)_1$, находим $\frac{dN_3}{dV}$: $\frac{dN_3}{dV} = \frac{\lambda_2 \lambda_1 N_1^0}{Q(\lambda_3 - \lambda_2)} e^{-\lambda_1(t_{s3} + t_L)} \left(S_{21}^1 e^{-\lambda_2 t_3} - S_{31}^1 e^{-\lambda_3 t_3}\right)$. Распространяя рассуждения на произвольный временной интервал $t_{s3} + l_V t_L \div t_{s3} + (l_V + 1)t_L$, находим концентрацию вещества ③:

$$\frac{dN_3}{dV} = \frac{\lambda_2 \lambda_1 N_1^0}{Q(\lambda_3 - \lambda_2)} e^{-\lambda_1(t_{s3} + l_V t_L)} \left(S_{21}^{l_V} e^{-\lambda_2 t_3} - S_{31}^{l_V} e^{-\lambda_3 t_3}\right) =$$
$$= \frac{\lambda_2 \lambda_1 N_1^0}{Q(\lambda_3 - \lambda_2)} e^{-\lambda_1 t_e} \left(S_{21}^{l_V} e^{-(\lambda_2 - \lambda_1)\frac{V}{Q}} - S_{31}^{l_V} e^{-(\lambda_3 - \lambda_1)\frac{V}{Q}}\right) \tag{79-$l_V$}$$

Сравнивая ур-ия (79) и (79-$l_V$), замечаем, что движение по замкнутому контуру приводит к появлению двух сумм рядов при соответствующих экспонентах, а сама структура уравнения остается прежней. С ростом $t_e$ устанавливается равновесное распределение вещества ③:

$$\frac{dN_3}{dV} = \frac{\lambda_2 \lambda_1 N_1^0}{Q(\lambda_3 - \lambda_2)} e^{-\lambda_1 t_e} \left(\frac{e^{-(\lambda_2 - \lambda_1)\frac{V}{Q}}}{\left(1 - e^{-(\lambda_2 - \lambda_1)t_L}\right)} - \frac{e^{-(\lambda_3 - \lambda_1)\frac{V}{Q}}}{\left(1 - e^{-(\lambda_3 - \lambda_1)t_L}\right)}\right) \tag{79p}$$

В произвольный момент $t_e$ весь объем $V_0$, в котором распределены вещества ② и ③, делится "0"-отметкой на два диапазона: "0" $\div V_{e2}$ и $V_{e2} \div$ "0" (в моменты $t_e = i t_L$, когда $V_{e2}$ совпадает с "0"-отметкой, остается один диапазон "0" $\div V_0 = $ "0"). Привяжем значение $l_V$ к значению $l$ так же, как это сделано в разделе 2.3.2, т.е. возьмем положение фронта $V_{e2}$ за отправную точку и будем двигаться вниз от него по *V-t* диаграмме (Рис. 27). $V_{e2}$ является верхней границей диапазона "0" ÷ $V_{e2}$, в котором $l_V = l$. Двигаясь дальше вниз, мы пересекаем "0"-отметку и попадаем в диапазон $V_{e2}$ ÷ "0", в котором $l_V = l - 1$. Графики концентраций $\frac{dN_2}{dV}$ и $\frac{dN_3}{dV}$ в единицах объемной активности, построенные по ур-иям (78) и (79-$l_V$) для разных моментов циклического движения, кратных периоду $t_L$, показаны на Рис. 28.

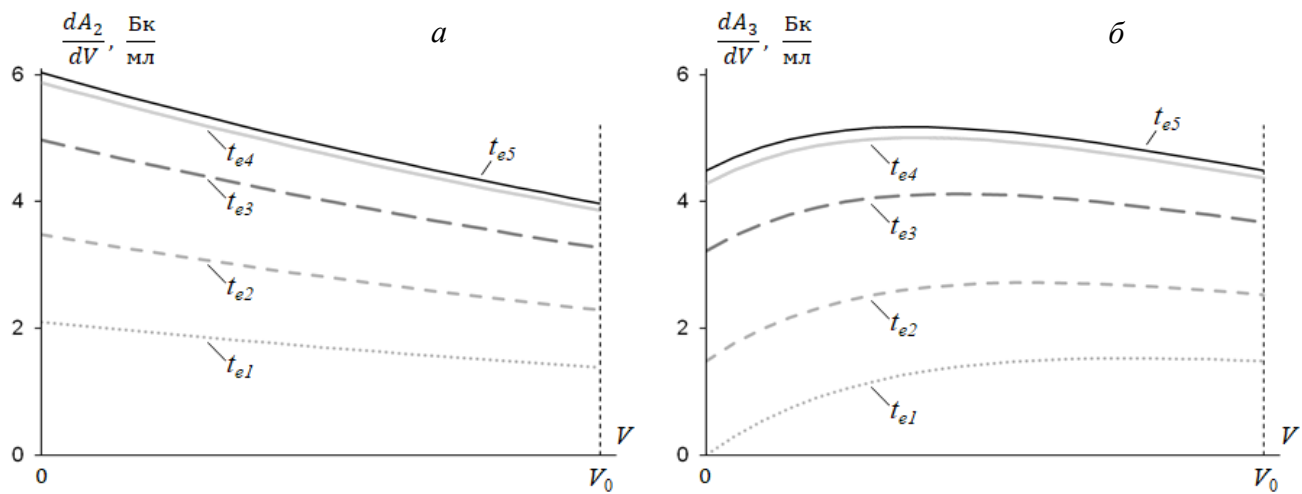

**Рис. 28**. Распределения по контуру веществ ② (*а*) и ③ (*б*), выраженные в единицах объемной активности (Бк/мл), возникающие в разные моменты $t_e$ движения веществ со скоростями $q_1 = 0$, $q_2 = q_3 = Q$: $t_{e1} = t_L$; $t_{e2} = 2t_L$; $t_{e3} = 4t_L$; $t_{e4} = 8t_L$; $t_{e5} = 16t_L$. В расчете использованы значения: $A_1^0 = 100$ Бк, $V_0$ = 20 мл, $Q$ = 1 мл/мин, $\lambda_1 = 7.7 \cdot 10^{-7}$ с$^{-1}$, $\lambda_2 = 3.5 \cdot 10^{-4}$ с$^{-1}$, $\lambda_3 = 2.5 \cdot 10^{-3}$ с$^{-1}$.

Для иллюстрации зависимости профиля концентрации вещества ③ от времени (Рис. 28б) выбрано соотношение констант реакции $\lambda_2 < \lambda_3$. При обратном соотношении констант зависимость от времени останется качественно такой же.

### 3.3.2. Соотношение скоростей веществ ② и ③: $q_2 = Q$; $q_3 = 0$ в точке "0", $q_3 = Q$ вне точки "0"

Модифицируем предыдущую задачу и наложим условие, что вещество ③ движется в контуре со скоростью $q_3 = Q$ везде, кроме точки "0". В этом случае равновесное (исходное, накопленное к началу движения) количество вещества ② движется вместе с фронтом $V_{e2}$, а исходное количество вещества ③ остается в точке "0". *V-t* диаграмма остается такой же (Рис. 27), а область существования вещества ③ состоит теперь из двух диапазонов "0" ÷ $V_{e2}$ и $V_{e2}$ ÷ "0" и точки "0".

Как было показано в предыдущем разделе, концентрация вещества ③ в дифференциальном элементе $dV$③, достигающем точки "0" в момент $t_e = t_{s3} + t_L$ равна:
$$\left(\frac{dN_3}{dV}\right)_{"0"} = \left(\frac{dN_3}{dV}\right)_1 = \frac{\lambda_2 \lambda_1 N_1^0}{Q(\lambda_3 - \lambda_2)} e^{-\lambda_1(t_e - t_L)} \left(e^{-\lambda_2 t_L} - e^{-\lambda_3 t_L}\right) \quad (80)$$
Поступление вещества ③ в точку "0" начинается в момент $t_L$ (Рис. 27), а время накопления: $t_3 = t_e - t_L$. Число полных циклов накопления вещества ③ в точке "0": $l_0 = \left[\frac{t_e - t_L}{t_L}\right] = l - 1$. Для нахождения количества вещества ③ в точке "0" в интервале $t_L \div 2t_L$ решаем уравнение $\frac{dN_3}{dt} = Q\left(\frac{dN_3}{dV}\right)_{"0"} - \lambda_3 N_3$ относительно $t_3$. Граничным условием будет количество вещества ③ в момент $t_3 = 0$. Оно складывается из исходного вещества ③, оставшегося в точке "0", и накопленного из исходного вещества ② за время $t_L$ его движения с фронтом $V_{e2}$:
$(N_3("0"))_{t_L} = N_3^0 e^{-\lambda_3 t_L} + \frac{\lambda_2}{\lambda_3-\lambda_2} N_2^0 (e^{-\lambda_2 t_L} - e^{-\lambda_3 t_L})$. Запишем полученное решение в виде:

$$N_3("0") = \frac{\lambda_2 \lambda_1 N_1^0}{(\lambda_3-\lambda_1)}(e^{-\lambda_2 t_L} - e^{-\lambda_3 t_L})\left(\frac{e^{-\lambda_1(t_e-t_L)}}{(\lambda_3-\lambda_2)} - \frac{e^{-\lambda_3(t_e-t_L)}}{(\lambda_1-\lambda_2)}\right) + N_3^0 e^{-\lambda_3 t_e} \quad (81)$$

Аналогично, в следующем интервале $2t_L \div 3t_L$ граничным условием для $t_3 = t_e - 2t_L = 0$ будет вещество ③ в точке "0" (ур-ие 81) и вещество ③, накопленное из исходного вещества ② за время $t_L$ следующего цикла с фронтом $V_{e2}$:

$$(N_3("0"))_{2t_L} = \frac{\lambda_2 \lambda_1 N_1^0}{(\lambda_3-\lambda_1)}(e^{-\lambda_2 t_L} - e^{-\lambda_3 t_L})\left(\frac{e^{-\lambda_1 t_L}}{(\lambda_3-\lambda_2)} - \frac{e^{-\lambda_3 t_L}}{(\lambda_1-\lambda_2)}\right) + N_3^0 e^{-\lambda_3 2t_L}$$
$$+ \frac{\lambda_2 N_2^0 e^{-\lambda_2 t_L}}{\lambda_3 - \lambda_2}(e^{-\lambda_2 t_L} - e^{-\lambda_3 t_L})$$

Из *V-t* диаграммы на Рис. 27 видно, что поступление вещества ③ в точку "0" в этом интервале обеспечивают дифференциальные элементы $dV$③, стартующие в интервале $t_L \leq t_{s3} \leq 2t_L$ ($t_{s3} = t_{s2} + t_L$; $t_3 = t_e - t_{s3}$). Концентрацию вещества ③ в них находим так же, как в предыдущем разделе, беря в ур-ие (21) концентрацию $\frac{dN_2}{dV}$ из ур-ия (78, $l_V = 1$), но теперь граничным условием будет $t_3 = 0$: $\left(\frac{dN_3}{dV}\right)_{t_{s3}} = 0$, поскольку стартуют они без вещества ③, которое остается в точке "0":

$$\frac{dN_3}{dV} = \frac{\lambda_2 \lambda_1 N_1^0}{Q(\lambda_3-\lambda_2)} e^{-\lambda_1(t_e-t_3)} S_{21}^1 (e^{-\lambda_2 t_3} - e^{-\lambda_3 t_3})$$

Спустя период $t_L$, когда элемент $dV$③ снова достигает точки "0", концентрация вещества ③ в нем равна:

$$\left(\frac{dN_3}{dV}\right)_{"0"} = \left(\frac{dN_3}{dV}\right)_2 = \frac{\lambda_2 \lambda_1 N_1^0}{Q(\lambda_3-\lambda_2)} e^{-\lambda_1(t_e-t_L)} S_{21}^1 (e^{-\lambda_2 t_L} - e^{-\lambda_3 t_L}) \quad (80\text{-}1)$$

Сравнивая ур-ия (80) и (80-1), мы видим, что во втором уравнении сумма $S_{21}^{l_V}$ ряда увеличилась на один член ($S_{21}^0 = 1$). Зная скорость $Q\left(\frac{dN_3}{dV}\right)_{"0"}$ накопления вещества ③ в точке "0", находим его количество в интервале $2t_L \div 3t_L$:

$$N_3("0") = \frac{\lambda_2 \lambda_1 N_1^0}{(\lambda_3-\lambda_1)}(e^{-\lambda_2 t_L} - e^{-\lambda_3 t_L})\left(S_{21}^1 \frac{e^{-\lambda_1(t_e-t_L)}}{(\lambda_3-\lambda_2)} - S_{23}^1 \frac{e^{-\lambda_3(t_e-t_L)}}{(\lambda_1-\lambda_2)}\right) + N_3^0 e^{-\lambda_3 t_e} \quad (81\text{-}1)$$

Так же сравнивая ур-ия (81) и (81-1), мы видим, что во втором уравнении суммы $S_{21}^{l_V}$ и $S_{23}^{l_V}$ рядов увеличились каждая на один член.

Переходя к произвольному интервалу $t_{s3} \div t_{s3} + t_L$ ( $t_{s3} = t_{s2} + l_V t_L$; $t_3 = t_e - t_{s3}$ ) движения элемента $dV$③, для концентрации $\frac{dN_3}{dV}$ в точке $V$ контура получаем выражение:

$$\frac{dN_3}{dV} = \frac{\lambda_2 \lambda_1 N_1^0}{Q(\lambda_3-\lambda_2)} e^{-\lambda_1 t_e} S_{21}^{l_V} \left(e^{-(\lambda_2-\lambda_1)\frac{V}{Q}} - e^{-(\lambda_3-\lambda_1)\frac{V}{Q}}\right) \quad (80\text{-}l_V)$$

Связь $l_V$ и $l$ описана в предыдущем разделе. Также в контуре вместе с фронтом $V_{e2}$ движется вещество ③, накапливаемое из исходного вещества ②:

$$N_3(V_{e2}) = \frac{\lambda_2}{(\lambda_3-\lambda_2)} N_2^0 e^{-\lambda_2 l t_L} (e^{-\lambda_2(t_e-l t_L)} - e^{-\lambda_3(t_e-l t_L)})$$

В соответствующем интервале $l t_L \div (l+1) t_L$ ($t_3 = t_e - l t_L$) накопление вещества ③ в точке "0" описывается уравнением:

$$N_3("0") = \frac{\lambda_2 \lambda_1 N_1^0}{(\lambda_3-\lambda_1)}(e^{-\lambda_2 t_L} - e^{-\lambda_3 t_L})\left(S_{21}^{(l-1)} \frac{e^{-\lambda_1(t_e-t_L)}}{(\lambda_3-\lambda_2)} - S_{23}^{(l-1)} \frac{e^{-\lambda_3(t_e-t_L)}}{(\lambda_1-\lambda_2)}\right) + N_3^0 e^{-\lambda_3 t_e} \quad (81\text{-}l_0)$$

С ростом $t_e$ устанавливается равновесное распределение вещества ③, концентрация в контуре становится равной:

$$\frac{dN_3}{dV} = \frac{\lambda_2 \lambda_1 N_1^0}{Q(\lambda_3 - \lambda_2)} e^{-\lambda_1 t_e} \frac{\left(e^{-(\lambda_2 - \lambda_1)\frac{V}{Q}} - e^{-(\lambda_3 - \lambda_1)\frac{V}{Q}}\right)}{(1 - e^{-(\lambda_2 - \lambda_1)t_L})} \qquad (80p)$$

Равновесное количество вещества ③ в точке "0" стремится к:

$$N_3("0") = \frac{\lambda_2 \lambda_1 N_1^0 e^{-\lambda_1 t_e}}{(\lambda_3 - \lambda_1)(\lambda_3 - \lambda_2)(1 - e^{-(\lambda_2 - \lambda_1)t_L})} \left(e^{-(\lambda_2 - \lambda_1)t_L} - e^{-(\lambda_3 - \lambda_1)t_L}\right) \qquad (81p)$$

Графики концентрации $\frac{dN_3}{dV}$ (в единицах объемной активности) и количества вещества ③ в точке "0" показаны на Рис. 29.

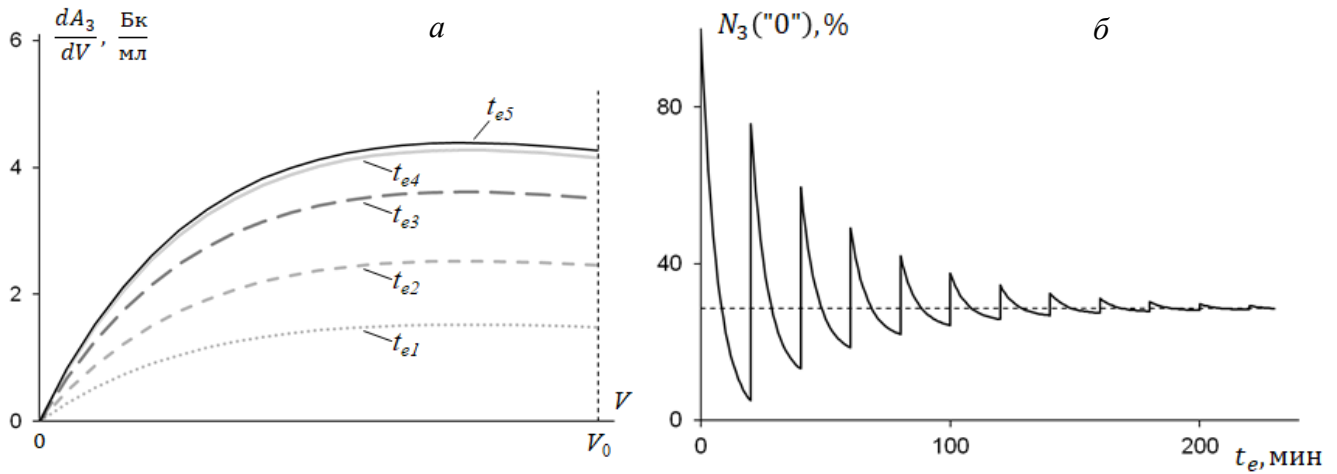

**Рис. 29**. Распределение по контуру вещества ③, выраженное в единицах объемной активности (Бк/мл) (*а*), и зависимость количества вещества ③ в точке "0" от времени (*б*) при движении веществ со скоростями $q_1 = 0$, $q_2 = Q$; $q_3 = 0$ в точке "0", $q_3 = Q$ вне точки "0": $t_{e1} = t_L$; $t_{e2} = 2t_L$; $t_{e3} = 4t_L$; $t_{e4} = 8t_L$; $t_{e5} = 16t_L$. В расчете использованы значения: $A_1^0 = 100$ Бк, $V_0 = 20$ мл, $Q = 1$ мл/мин, $\lambda_1 = 7.7 \cdot 10^{-7}$ с$^{-1}$, $\lambda_2 = 3.5 \cdot 10^{-4}$ с$^{-1}$, $\lambda_3 = 2.5 \cdot 10^{-3}$ с$^{-1}$.

Сравнивая профили концентрации $\frac{dN_3}{dV}$ на Рис. 28б и 29а, построенные по ур-иям (79-$l_V$) и (80-$l_V$), видим, что они качественно похожи с той разницей, что профили на Рис. 29а выходят из нуля. Это связано с наложением условия неподвижности вещества ③ в точке "0", которое привело к замене суммы $S_{31}^{lv}$ в ур-ии (79-$l_V$) на сумму $S_{21}^{lv}$ в ур-ии (80-$l_V$).

Всплески на графике Рис. 29б связаны с поглощением в точке "0" вещества ③, движущегося вместе с фронтом $V_{e2}$ и накапливаемого из исходного вещества ② в течение одного периода $t_L$. Всплески затухают с распадом вещества ②, а количество вещества ③ в точке "0" стремится к равновесному. Для графиков на Рис. 29 выбрано соотношение констант реакции $\lambda_2 < \lambda_3$. При обратном соотношении констант зависимость от времени останется качественно такой же.

### 3.3.3. Соотношение скоростей веществ ② и ③: $q_2 = Q$, $q_3 = 0$

С помощью *V-t* диаграммы на Рис. 30а рассмотрим влияние циркуляции подвижной фазы на распределение вещества ③ в замкнутом контуре в случае, когда вещество ② движется со скоростью потока ($q_2 = Q$), а вещество ③ не движется ($q_3 = 0$), и исходное количество вещества ③ остается в точке "0".

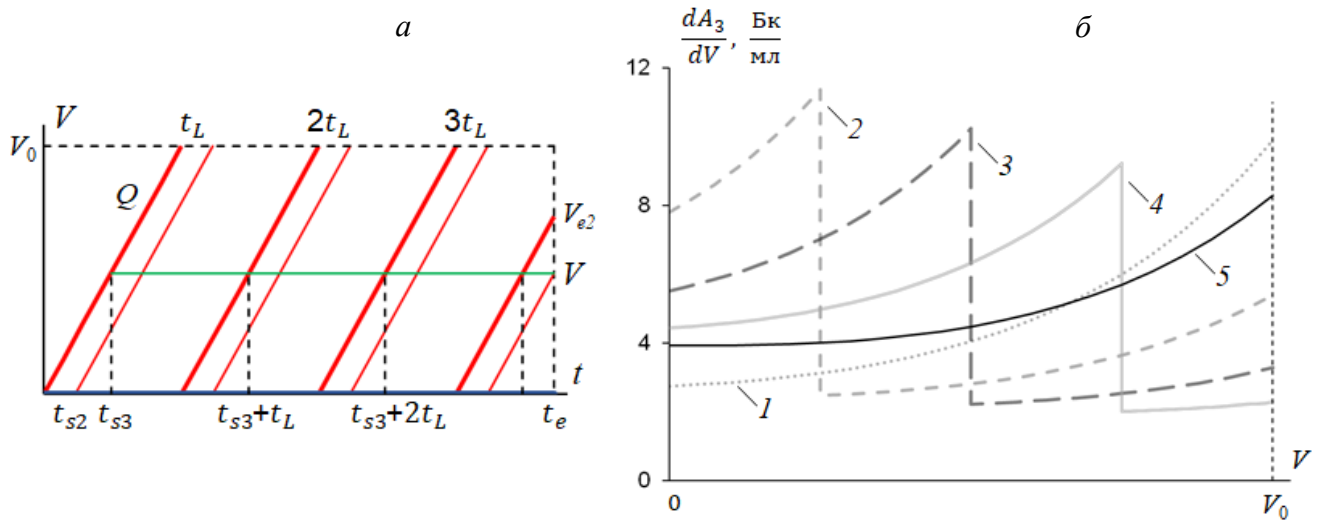

**Рис. 30.** *V-t* диаграмма (*а*) движения веществ в замкнутом контуре, заполненном однородной хроматографической средой и распределение по контуру вещества ③ во временном интервале $t_L \div 2t_L$ (*б*), выраженное в единицах объемной активности (Бк/мл), при движении веществ со скоростями $q_1 = 0$, $q_2 = Q$, $q_3 = 0$: *1* – 20 мин ($t_L$); *2* – 25 мин; *3* – 30 мин; *4* – 35 мин; *5* – 40 мин ($2t_L$). В расчете использованы значения: $A_1^0 = 100$ Бк, $V_0 = 20$ мл, $Q = 1$ мл/мин, $\lambda_1 = 7.7 \cdot 10^{-7}$ с$^{-1}$, $\lambda_2 = 3.5 \cdot 10^{-4}$ с$^{-1}$, $\lambda_3 = 2.5 \cdot 10^{-3}$ с$^{-1}$. Синие, красные и зеленые линии соответствуют движению веществ ①, ② и ③.

Накопление вещества ③ в точке *V* начинается в момент $t_{s3} = \frac{V}{Q}$ и во временном интервале $t_{s3} \div t_{s3} + t_L$ описывается ур-ием (65). В момент $t_e = t_{s3} + t_L$ концентрация вещества ③ в точке *V* равна:

$$\left(\frac{dN_3}{dV}\right)_1 = \frac{\lambda_2 \lambda_1 N_1^0}{Q(\lambda_3 - \lambda_1)} e^{-\lambda_2 t_{s3}} \left( e^{-\lambda_1 t_L} - \frac{(\lambda_2 - \lambda_3)}{(\lambda_2 - \lambda_1)} e^{-\lambda_3 t_L} \right)$$

В этот же момент точки *V* достигает фронт $V_{e2}$, вместе с которым движется исходное вещество ② (Рис. 30а). Концентрация образующегося из него вещества ③ (см. раздел 3.1): $\left(\frac{dN_3}{dV}\right)_{V_{e2}} = \frac{\lambda_2 N_2^0}{Q} e^{-\lambda_2(t_{s3} + t_L)}$. Граничное условие для нахождения концентрация вещества ③ в следующем интервале $t_{s3} + t_L \div t_{s3} + 2t_L$ складывается из этих составляющих: $\left(\frac{dN_3}{dV}\right)_{(t_{s3}+t_L)} = \left(\frac{dN_3}{dV}\right)_1 + \left(\frac{dN_3}{dV}\right)_{V_{e2}}$. Решение ур-ия (21) относительно $t_3 = t_e - t_{s3} - t_L$ приводит к выражению:

$$\frac{dN_3}{dV} = \frac{\lambda_2 \lambda_1 N_1^0}{Q(\lambda_3 - \lambda_1)} e^{-\lambda_2 t_{s3}} \left( S_{21}^1 e^{-\lambda_1(t_3 + t_L)} - \frac{(\lambda_2 - \lambda_3)}{(\lambda_2 - \lambda_1)} S_{23}^1 e^{-\lambda_3(t_3 + t_L)} \right) \qquad (65\text{-}1)$$

Сравнивая ур-ия (65) и (65-1), мы видим, что во втором уравнении суммы $S_{21}^{l_V}$ и $S_{23}^{l_V}$ рядов увеличились каждая на один член. Графики распределения вещества ③ в интервале $t_L \div 2t_L$ показаны на Рис. 30б. Скачок концентрации происходит в точке фронта $V_{e2}$, значение $l_V = 0$ соответствует части кривой справа от $V_{e2}$, а $l_V = 1$ – слева от $V_{e2}$.

Переходя к произвольному интервалу $t_{s3} + l_V t_L \div t_{s3} + (l_V + 1)t_L$ ( $t_{s3} = \frac{V}{Q}$; $t_3 = t_e - t_{s3} - l_V t_L = t_{s2}$), для концентрации $\frac{dN_3}{dV}$ в точке *V* контура получаем выражение:

$$\frac{dN_3}{dV} = \frac{\lambda_2 \lambda_1 N_1^0}{Q(\lambda_3 - \lambda_1)} e^{-\lambda_2 t_{s3}} \left( S_{21}^{l_V} e^{-\lambda_1(t_3 + l_V t_L)} - \frac{(\lambda_2 - \lambda_3)}{(\lambda_2 - \lambda_1)} S_{23}^{l_V} e^{-\lambda_3(t_3 + l_V t_L)} \right) =$$

$$= \frac{\lambda_2 \lambda_1 N_1^0}{Q(\lambda_3 - \lambda_1)} \left( S_{21}^{l_V} e^{-\lambda_1 t_e - (\lambda_2 - \lambda_1)\frac{V}{Q}} - S_{23}^{l_V} e^{-\lambda_3 t_e - (\lambda_2 - \lambda_3)\frac{V}{Q}} \right) \qquad (65\text{-}l_V)$$

Связь $l_V$ и $l$ описана в разделе 3.3.1.

С ростом $t_e$ устанавливается равновесное распределение вещества ③, концентрация в контуре становится равной:

$$\frac{dN_3}{dV} = \frac{\lambda_2 \lambda_1 N_1^0}{Q(\lambda_3 - \lambda_1)} \frac{e^{-\lambda_1 t_e - (\lambda_2 - \lambda_1)\frac{V}{Q}}}{(1 - e^{-(\lambda_2 - \lambda_1)t_L})} = \frac{\lambda_2}{(\lambda_3 - \lambda_1)} \frac{dN_2}{dV} \quad (65\text{рк})$$

Интересно отметить, что равновесные распределения веществ ② (ур-ие 64) и ③, выраженные в виде объемной активности, практически равны: $\frac{dA_3}{dV} \approx \frac{dA_2}{dV} = \frac{\lambda_2 A_1^0}{Q} \frac{e^{-\lambda_1 t_e - (\lambda_2 - \lambda_1)\frac{V}{Q}}}{(1 - e^{-(\lambda_2 - \lambda_1)t_L})}$.

Графики концентрации $\frac{dN_3}{dV}$ в единицах объемной активности, построенные для разных моментов циклического движения, кратных периоду $t_L$, показаны на Рис. 31.

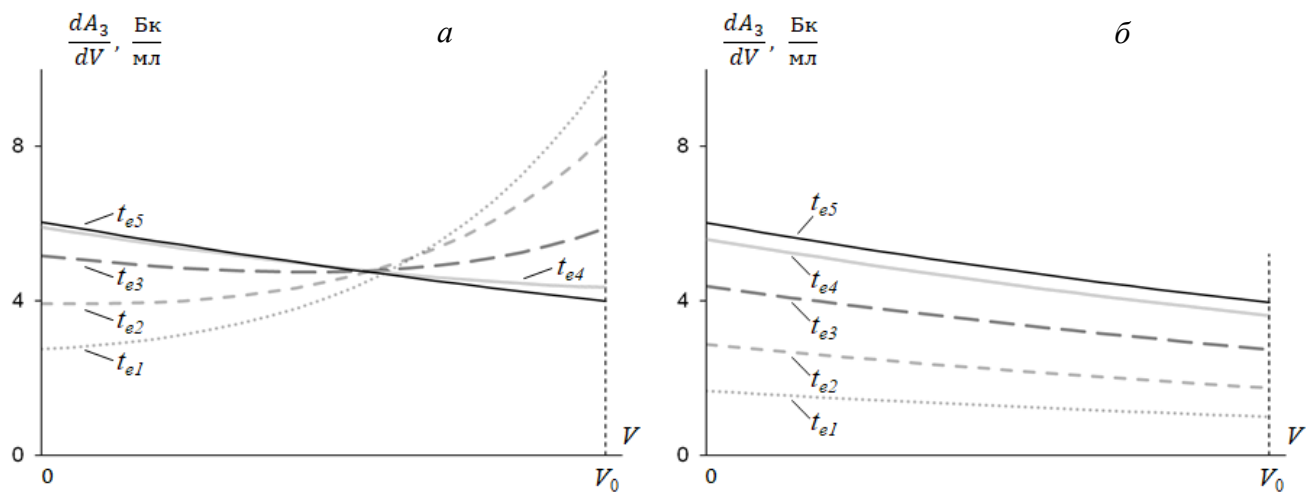

**Рис. 31**. Распределения по контуру вещества ③, выраженные в единицах объемной активности (Бк/мл), возникающие в разные моменты $t_e$ движения веществ со скоростями $q_1 = 0$, $q_2 = Q$, $q_3 = 0$: $t_{e1} = t_L$; $t_{e2} = 2t_L$; $t_{e3} = 4t_L$; $t_{e4} = 8t_L$; $t_{e5} = 16t_L$. В расчете использованы значения: $A_1^0 = 100$ Бк, $V_0 = 20$ мл, $Q = 1$ мл/мин, $\lambda_1 = 7.7 \cdot 10^{-7}$ с$^{-1}$, $\lambda_2 = 3.5 \cdot 10^{-4}$ с$^{-1}$; $а - \lambda_3 = 2.5 \cdot 10^{-3}$ с$^{-1}$; $б - \lambda_3 = 2.5 \cdot 10^{-4}$ с$^{-1}$.

На Рис. 31а использовано соотношение констант реакции $\lambda_2 < \lambda_3$, а на Рис. 31б – обратное: $\lambda_2 > \lambda_3$. Независимо от соотношения, равновесные распределения вещества ③ стремятся к одной и той же кривой, практически равной равновесному распределению вещества ② (Рис. 28а). То есть, в условии равновесия профиль концентрации вещества ③ является "слепком" вещества ②, что весьма полезно для экспериментального изучения распределений веществ ② и ③.

Изучив закономерности распределения дочерних веществ ② и ③ в замкнутом контуре, заполненном однородной хроматографической средой, переходим к построению циркулирующего $^{225}$Ac/$^{213}$Bi генератора, изображенного на Рис. 26б.

### 3.3.4. Циркулирующий двух-колоночный $^{225}$Ac/$^{213}$Bi генератор

Рассмотрим замкнутый контур $^{225}$Ac/$^{213}$Bi генератора, состоящий из двух колонок, заполненных сорбентом Actinide Resin со свободным объемом $V_{c1}$ и $V_{c2}$ и соединенных трубками объемом $V_{p1}$ и $V_{p2}$ (Рис. 26б). Материнский радионуклид $^{225}$Ac (вещество ①) адсорбирован в начальном слое ($V = 0$) сорбента первой колонки $V_{c1}$. Так же, как и в прямоточном варианте, в трубке $V_{p1}$ происходит распад большей части $^{221}$Fr, а в колонке $V_{c2}$ – накопление $^{213}$Bi. Трубка $V_{p2}$, соединяющая выход второй колонки со входом первой, играет вспомогательную роль. Как будет показано ниже, объем $V_{p2}$ должен быть минимальным для достижения максимального выхода $^{213}$Bi из генератора. Общий объем системы: $V_0 = V_{c1} + V_{p1} + V_{c2} + V_{p2}$. Из $V$-$t$ диаграммы, показанной на Рис. 32, следует, что $^{213}$Bi содержится в следующих элементах контура: точка "0", диапазон "0" ÷ $V_{c1}$, диапазон $V_{c1}$ ÷ $V_{cp}$, точка $V_{cp}$, диапазон $V_{cp}$ ÷ $V_{cpc}$ и диапазон $V_{cpc}$ ÷ "0".

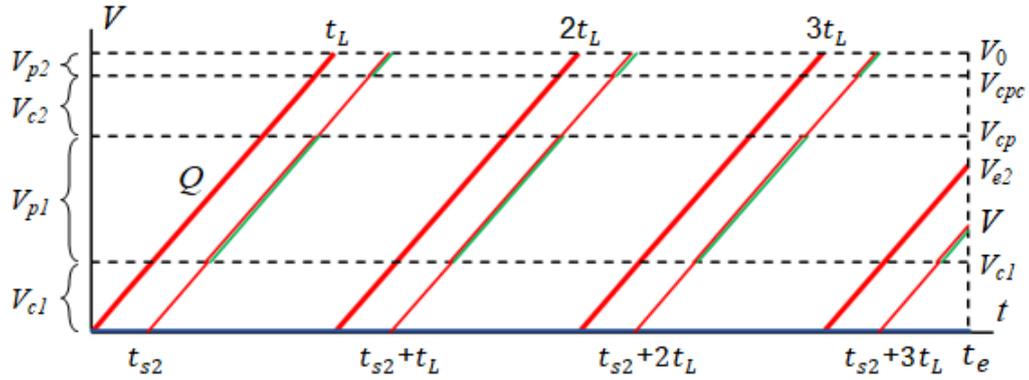

**Рис. 32**. *V-t* диаграмма движения веществ в циркулирующем двух-колоночном $^{225}$Ac/$^{213}$Bi генераторе. Синие, красные и зеленые линии соответствуют движению $^{225}$Ac, $^{221}$Fr и $^{213}$Bi (вещества ①, ② и ③).

До момента $t_L = \frac{V_0}{Q}$ формирование распределения дочерних веществ $^{221}$Fr и $^{213}$Bi происходит так же, как и в прямоточном генераторе, описанном в разделе 3.1. В момент $t_L$, когда фронт $V_{e2}$ достигает "0"-отметки ($V_0 = 0$), начинается накопление $^{213}$Bi в точке "0". Число полных циклов накопления $^{213}$Bi в точке "0": $l_0 = \left[\frac{t_e - t_L}{t_L}\right] = l - 1$. Из *V-t* диаграммы на Рис. 32 видно, что поступление $^{213}$Bi в точку "0" в интервале $t_L \div 2t_L$ обеспечивают дифференциальные элементы $dV$③, в которых концентрация $^{213}$Bi определяется ур-ием (68) при $V = V_0$. В произвольном интервале $lt_L \div (l+1)t_L$ ($t_3 = t_e - lt_L$) накопление $^{213}$Bi в точке "0" описывается уравнением, сходным по структуре с ур-ием (81-$l_0$):

$$N_3("0") = \frac{\lambda_2 \lambda_1 N_1^0}{(\lambda_3 - \lambda_1)} e^{-\lambda_2 \frac{V_{cpc}}{Q}} \left( e^{-\lambda_2 \frac{V_{p2}}{Q}} - e^{-\lambda_3 \frac{V_{p2}}{Q}} \right) \left( S_{21}^{(l-1)} \frac{e^{-\lambda_1(t_e - t_L)}}{(\lambda_3 - \lambda_2)} - S_{23}^{(l-1)} \frac{e^{-\lambda_3(t_e - t_L)}}{(\lambda_1 - \lambda_2)} \right) + N_3^0 e^{-\lambda_3 t_e} \quad (82)$$

В обеих колонках $V_{c1}$ и $V_{c2}$ (диапазоны "0" $\div V_{c1}$ и $V_{cp} \div V_{cpc}$) во временном интервале $t_{s3} + l_V t_L \div t_{s3} + (l_V + 1)t_L$ ( $t_{s3} = \frac{V}{Q}$; $t_3 = t_e - t_{s3} - l_V t_L = t_{s2}$ ) для концентрации $^{213}$Bi действует ур-ие (65-$l_V$).

В первой трубке $V_{p1}$ (диапазон $V_{c1} \div V_{cp}$) в интервале $t_{s3} \div t_{s3} + t_L$ ($t_{s3} = t_{s2} + l_V t_L$; $t_3 = t_e - t_{s3} = \frac{V - V_{c1}}{Q}$) уравнение концентрации $^{213}$Bi является синтезом ур-ий (66) и (80-$l_V$):

$$\frac{dN_3}{dV} = \frac{\lambda_2 \lambda_1 N_1^0}{Q(\lambda_3 - \lambda_2)} e^{-\lambda_1 t_e - (\lambda_2 - \lambda_1)\frac{V_{c1}}{Q}} S_{21}^{l_V} \left( e^{-(\lambda_2 - \lambda_1)\frac{V - V_{c1}}{Q}} - e^{-(\lambda_3 - \lambda_1)\frac{V - V_{c1}}{Q}} \right) \quad (83)$$

Если, как показано на Рис. 32, в произвольный момент $t_e$ фронт $V_{e2}$ находится в этом диапазоне, то к общему количеству $^{213}$Bi в первой трубке добавляется $^{213}$Bi, образующийся из исходного $^{221}$Fr:

$$N_3(V_{e2}) = \frac{\lambda_2}{(\lambda_3 - \lambda_2)} N_2^0 e^{-\lambda_2 \left(lt_L + \frac{V_{c1}}{Q}\right)} \left( e^{-\lambda_2 \left(t_e - lt_L - \frac{V_{c1}}{Q}\right)} - e^{-\lambda_3 \left(t_e - lt_L - \frac{V_{c1}}{Q}\right)} \right) \quad (1a\text{-}l)$$

Аналогично, во второй трубке $V_{p2}$ (диапазон $V_{cpc} \div$ "0") в интервале $t_{s3} \div t_{s3} + t_L$ ($t_{s3} = t_{s2} + \frac{V_{cpc}}{Q} + l_V t_L$; $t_3 = t_e - t_{s3} = \frac{V - V_{cpc}}{Q}$) уравнение концентрации $^{213}$Bi является синтезом ур-ий (68) и (80-$l_V$):

$$\frac{dN_3}{dV} = \frac{\lambda_2 \lambda_1 N_1^0}{Q(\lambda_3 - \lambda_2)} e^{-\lambda_1 t_e - (\lambda_2 - \lambda_1)\frac{V_{cpc}}{Q}} S_{21}^{l_V} \left( e^{-(\lambda_2 - \lambda_1)\frac{V - V_{cpc}}{Q}} - e^{-(\lambda_3 - \lambda_1)\frac{V - V_{cpc}}{Q}} \right) \quad (84)$$

Так же, если в произвольный момент $t_e$ фронт $V_{e2}$ находится в этом диапазоне, то добавляется $^{213}$Bi, образующийся из исходного $^{221}$Fr (ур-ие (1a-$l$), в котором $V_{c1}$ заменяется на $V_{cpc}$.

Наконец, накопление $^{213}$Bi в точке $V_{cp}$ начинается в момент $\frac{V_{cp}}{Q}$, а число полных циклов накопления $^{213}$Bi в этой точке: $l_{cp} = \left[\frac{t_e - \frac{V_{cp}}{Q}}{t_L}\right]$. По аналогии с точкой "0", в произвольном интервале $\frac{V_{cp}}{Q} + l_{cp} t_L \div \frac{V_{cp}}{Q} + (l_{cp} + 1)t_L$ ($t_3 = t_e - \frac{V_{cp}}{Q} - l_{cp} t_L$) накопление $^{213}$Bi в точке $V_{cp}$ описывается уравнением, сходным по структуре с ур-иями (81-$l_0$) и (82):

$$N_3(V_{cp}) = \frac{\lambda_2 \lambda_1 N_1^0}{(\lambda_3 - \lambda_1)} e^{-\lambda_2 \frac{V_{c1}}{Q}} \left(e^{-\lambda_2 \frac{V_{p1}}{Q}} - e^{-\lambda_3 \frac{V_{p1}}{Q}}\right) \left(S_{21}^{l_{cp}} \frac{e^{-\lambda_1(t_e - t_L)}}{(\lambda_3 - \lambda_2)} - S_{23}^{l_{cp}} \frac{e^{-\lambda_3(t_e - t_L)}}{(\lambda_1 - \lambda_2)}\right) \qquad (85)$$

Таким образом, мы получили все уравнения для модели циркулирующего двух-колоночного $^{225}$Ac/$^{213}$Bi генератора. На Рис. 33а показана начальная стадия накопления $^{213}$Bi. Отдельно отмечены временные интервалы первого периода $t_L$ циркуляции. Проследим поведение кривых 2-4. Кривая 2 (количество $^{213}$Bi в первой трубке $V_{p1}$) начинает расти в момент $\frac{V_{c1}}{Q}$, когда фронт $V_{e2}$ выходит из первой колонки. Количество $^{213}$Bi складывается из интеграла ур-ия (83, $l_V = 0$) и $^{213}$Bi, образующегося из исходного $^{221}$Fr и движущегося вместе с фронтом. Интервал $t_{e2}$ заканчивается в момент $\frac{V_{cp}}{Q}$. Поскольку скорость $^{213}$Bi в сорбенте $q_3 = 0$, количество $^{213}$Bi, достигающего входа во вторую колонку вместе с фронтом, поглощается. Кривая 2 совершает скачок вниз, а кривая 3 начинает рост с зеркального скачка вверх количества $^{213}$Bi в точке $V_{cp}$. После прохождения второй колонки за время $t_{e3}$, фронт $V_{e2}$ достигает второй трубки, и начинается рост кривой 4 (интервал $t_{e4}$). Ее форма повторяет форму кривой 2, а скачок вниз происходит, когда кривая 4 достигает точки "0", и начинается второй цикл движения по контуру. Будет нелишним повторить, что в любой момент времени общее количество $^{221}$Fr и $^{213}$Bi в контуре равно $N_2^0 e^{-\lambda_1 t_e}$ и $N_3^0 e^{-\lambda_1 t_e}$, т.е. система остается в состоянии "интегрального" подвижного равновесия, тогда как в ее элементах происходит перераспределение веществ.

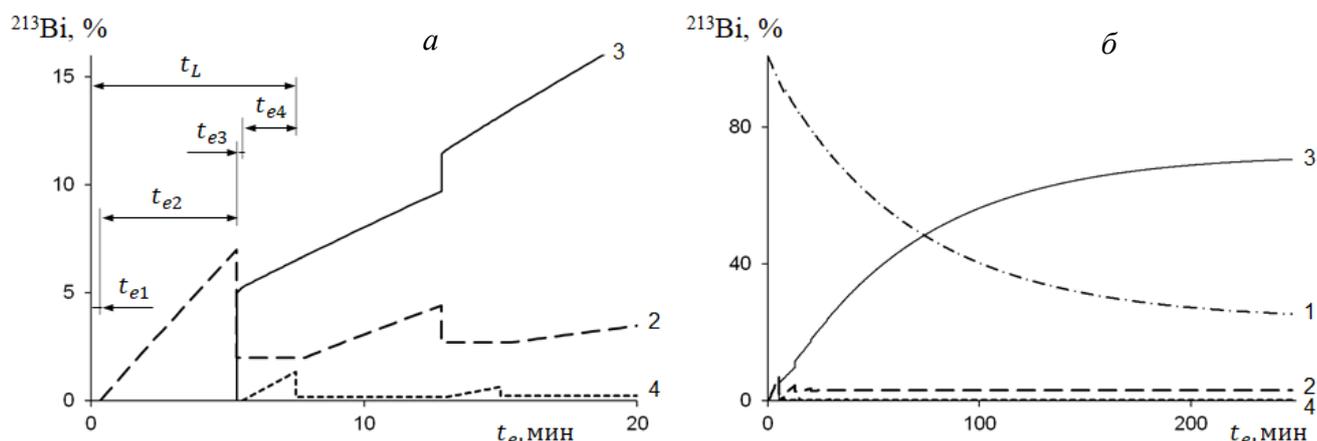

**Рис. 33**. Распределение $^{213}$Bi в элементах циркулирующего двух-колоночного $^{225}$Ac/$^{213}$Bi генератора: *а* – начальная стадия накопления $^{213}$Bi, на которой отмечены временные интервалы первого периода $t_L$ циркуляции: $0 \leq t_{e1} \leq \frac{V_{c1}}{Q}$; $\frac{V_{c1}}{Q} \leq t_{e2} \leq \frac{V_{cp}}{Q}$; $\frac{V_{cp}}{Q} \leq t_{e3} \leq \frac{V_{cpc}}{Q}$; $\frac{V_{cpc}}{Q} \leq t_{e4} \leq t_L$; *б* – достижение равновесного распределения (~ 4 часа). Обозначения: 1 – первая колонка $V_{c1}$ (диапазон $0 \div V_{c1}$); 2 – первая трубка $V_{p1}$ (диапазон $V_{c1} \div V_{cp}$); 3 – вторая колонка $V_{c2}$ (диапазон $V_{cp} \div V_{cpc}$); 4 – вторая трубка $V_{p2}$ (диапазон $V_{cpc} \div$ "0"). В расчете использованы значения: $Q = 1$ мл/мин; $V_{c1} = 0.34$ мл; $V_{c2} = 0.2$ мл; $V_{p1} = 5$ мл; $V_0 = 7.5$ мл.

На Рис. 33б показано, как замкнутая хроматографическая система достигает нового "дифференциального" подвижного равновесия по $^{213}$Bi, когда количества $^{213}$Bi в элементах системы становятся практически постоянными. Качественно картина почти не отличается от приведенной на Рис. 20б.

В этом состоянии выражение для концентрации $^{213}$Bi в колонках приходит к ур-ию (65рк). Равновесное количество $^{213}$Bi в первой колонке складывается из интеграла ур-ия (65рк) в пределах "0" $\div V_{c1}$ и количества $^{213}$Bi в точке "0":

$$N_3("0") = \frac{\lambda_2 \lambda_1 N_1^0}{(\lambda_3-\lambda_1)(\lambda_3-\lambda_2)(1-e^{-(\lambda_2-\lambda_1)t_L})} e^{-\lambda_1 t_e - \lambda_2 \frac{V_{cpc}}{Q}} \left( e^{-\lambda_2 \frac{V_{p2}}{Q}} - e^{-\lambda_3 \frac{V_{p2}}{Q}} \right) \qquad (82р)$$

Подобным образом, равновесное количество $^{213}$Bi во второй колонке складывается из интеграла ур-ия (65рк) в пределах $V_{cp} \div V_{cpc}$ и количества $^{213}$Bi в точке $V_{cp}$:

$$N_3(V_{cp}) = \frac{\lambda_2 \lambda_1 N_1^0}{(\lambda_3-\lambda_1)(\lambda_3-\lambda_2)(1-e^{-(\lambda_2-\lambda_1)t_L})} e^{-\lambda_1 t_e - \lambda_2 \frac{V_{c1}}{Q}} \left( e^{-\lambda_2 \frac{V_{p1}}{Q}} - e^{-\lambda_3 \frac{V_{p1}}{Q}} \right) \qquad (85р)$$

Равновесные профили концентрации $^{213}$Bi в трубках $V_{p1}$ и $V_{p2}$ равны:

$$\frac{dN_3}{dV} = \frac{\lambda_2 \lambda_1 N_1^0}{Q(\lambda_3-\lambda_2)(1-e^{-(\lambda_2-\lambda_1)t_L})} e^{-\lambda_1 t_e - (\lambda_2-\lambda_1)\frac{V_{c1}}{Q}} \left( e^{-(\lambda_2-\lambda_1)\frac{V-V_{c1}}{Q}} - e^{-(\lambda_3-\lambda_1)\frac{V-V_{c1}}{Q}} \right) \qquad (83р)$$

$$\frac{dN_3}{dV} = \frac{\lambda_2 \lambda_1 N_1^0}{Q(\lambda_3-\lambda_2)(1-e^{-(\lambda_2-\lambda_1)t_L})} e^{-\lambda_1 t_e - (\lambda_2-\lambda_1)\frac{V_{cpc}}{Q}} \left( e^{-(\lambda_2-\lambda_1)\frac{V-V_{cpc}}{Q}} - e^{-(\lambda_3-\lambda_1)\frac{V-V_{cpc}}{Q}} \right) \qquad (84р)$$

При скорости циркуляции $Q = 1$ мл/мин и объеме первой трубки $V_{p1} = 5$ мл (Рис. 33б) количество $^{213}$Bi, накопленного во второй колонке, достигает 70%, а суммарное содержание $^{213}$Bi во второй колонке и в первой трубке – около 73%. Для того, чтобы найти оптимальные значения $Q$ и $V_{p1}$, построим и проанализируем зависимость количества $^{213}$Bi, потенциально извлекаемого из генератора, от этих параметров (Рис. 34).

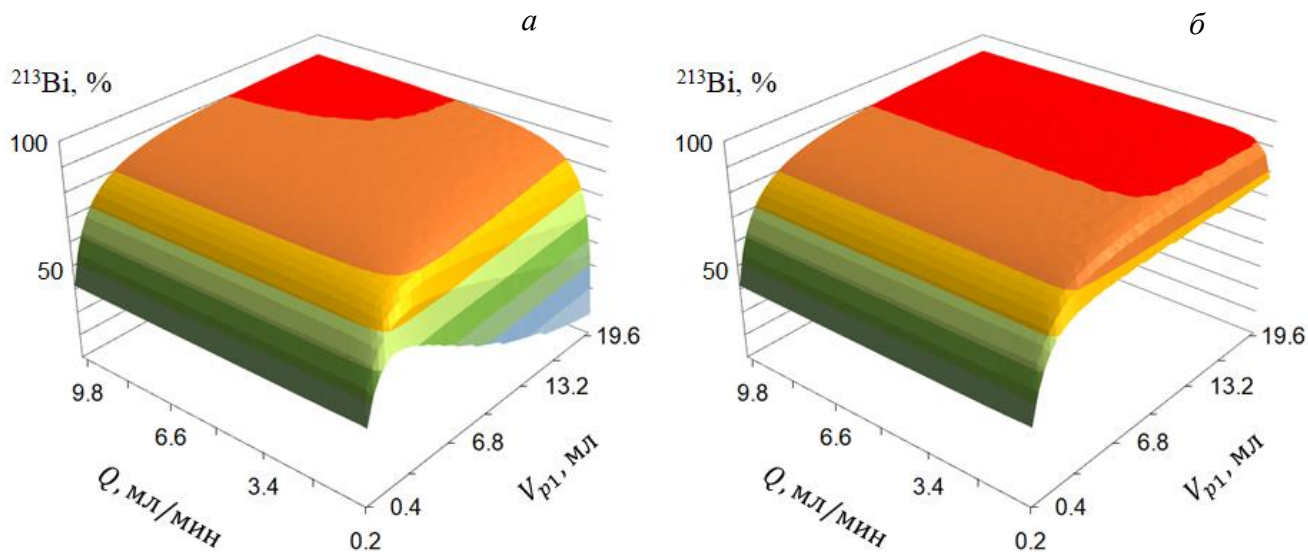

**Рис. 34**. Равновесное количество $^{213}$Bi во второй колонке (*а*) и суммарно во второй колонке и в первой трубке (*б*) циркулирующего двух-колоночного $^{225}$Ac/$^{213}$Bi генератора (время накопления ~ 4 часа). В расчете использованы значения: $V_{c1} = 0.34$ мл; $V_{c2} = 0.2$ мл; $V_{p2} = 0.46$ мл.

Мы видим, что обе зависимости не имеют максимума; с ростом переменных $Q$ и $V_{p1}$ функция количества $^{213}$Bi выходит на плато. Оценим равновесное распределение $^{213}$Bi в элементах генератора при достаточно большом увеличении $Q$.

Равновесное количество $^{213}$Bi в первой трубке определяется интегрированием ур-ия (83р) в пределах $V_{c1} \div V_{cp}$, а во второй – интегрированием ур-ия (84р) в пределах $V_{cpc} \div$ "0". С ростом $Q$ эти количества стремятся к нулю.

Первая колонка: ур-ие (82р) равновесного количества $^{213}$Bi в точке "0" с ростом $Q$ приходит к виду:

$$\lim_{Q\uparrow} N_3("0") = \frac{V_{p2}}{V_0} N_3^0 e^{-\lambda_1 t_e} \tag{86}$$

Интеграл ур-ия (65рк) в пределах "0" ÷ $V_{c1}$ с ростом $Q$ стремится к величине:

$$\lim_{Q\uparrow} N_3(V_{cp} \div V_{cpc}) = \frac{V_{c1}}{V_0} N_3^0 e^{-\lambda_1 t_e} \tag{87}$$

Тогда общее количество $^{213}$Bi в первой колонке:

$$\lim_{Q\uparrow} N_3(V_{c1}) = \frac{V_{p2}+V_{c1}}{V_0} N_3^0 e^{-\lambda_1 t_e} \tag{88}$$

Находим таким же образом общее количество $^{213}$Bi во второй колонке:

$$\lim_{Q\uparrow} N_3(V_{c2}) = \frac{V_{p1}+V_{c2}}{V_0} N_3^0 e^{-\lambda_1 t_e} \tag{89}$$

Из ур-ий (88) и (89) следует, что доля $^{213}$Bi, накапливаемого в колонке, в пределе равна сумме свободного объема сорбента самой колонки и объема предшествующей трубки, деленной на объем контура (н-р, доля $^{213}$Bi во второй колонке равна $\frac{V_{p1}+V_{c2}}{V_0}$). Кроме того, для достижения максимального выхода $^{213}$Bi из генератора объем $V_{p2}$ второй трубки, служащей для вспомогательных целей (н-р, подключение трехходового крана), должен быть минимальным.

Поскольку у функции количества $^{213}$Bi, изображенной на Рис. 34, нет физического максимума, оптимальные параметры $Q$ и $V_{p1}$ определяются технологическими соображениями такими, как надежность и компактность. Поясним с помощью графиков на Рис. 35, которые представляют собой двумерную версию Рис. 34.

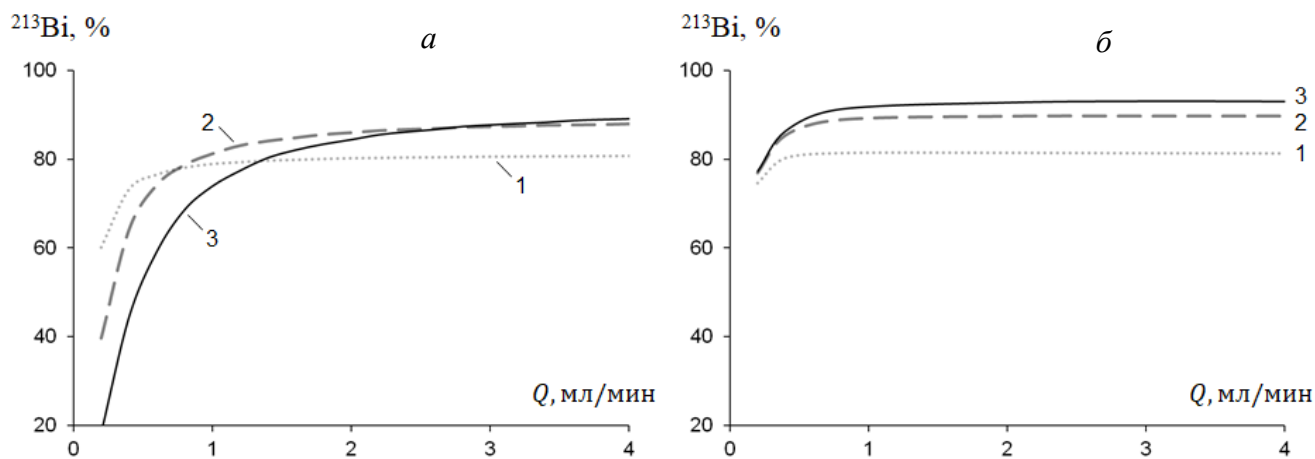

**Рис. 35**. Зависимость количества $^{213}$Bi во второй колонке (*а*) и суммарно во второй колонке и в первой трубке (*б*) от скорости циркуляции подвижной фазы (время накопления ~ 4 часа). Значения $V_{p1}$: 1 – 4 мл; 2 – 10 мл; 3 – 20 мл. В расчете использованы значения: $V_{c1} = 0.34$ мл; $V_{c2} = 0.2$ мл; $V_{p2} = 0.46$ мл.

Как видно из Рис. 35а, количество $^{213}$Bi во второй колонке быстрее выходит на плато при меньшем $V_{p1}$, но величина самого плато ниже. При $Q = 1$ мл/мин для диапазона значений $V_{p1}$ 4-10 мл количество $^{213}$Bi составляет ~ 80%. При увеличении скорости до $Q = 4$ мл/мин оно достигает 90%, а для $V_{p1} = 20$ мл – даже немного превышает. Однако рост скорости означает увеличение фактического объема раствора, пропущенного через колонки. Это может привести к размыванию тонкого слоя материнского $^{225}$Ac и росту его проскока, что в конечном счете ухудшит радионуклидную чистоту $^{213}$Bi, извлекаемого из генератора. Ответ на вопрос о допустимой скорости циркуляции дадут эксперименты.

Достичь практически максимального выхода без длительного использования высокой скорости позволяет описанный выше прием ее дискретного изменения. Из Рис. 35б следует, что суммарное содержание $^{213}$Bi в первой трубке и во второй колонке достигает 90% при $Q \sim 1$ мл/мин и $V_{p1} \sim 10$ мл, т.е. в первой трубке находится около 10% $^{213}$Bi. Увеличивая скорость циркуляции, скажем, до 10 мл/мин в течение одной минуты, мы вытесняем из трубки в колонку раствор, содержащий 10% $^{213}$Bi, гораздо менее концентрированным раствором. Используя рассуждения, приведенные в Разделах 2.1 и 3.2, читатель может самостоятельно получить необходимые уравнения.

Таким образом, на основе развитого подхода к кинетике последовательных реакций первого порядка в условиях хроматографического разделения построена модель циркулирующего двух-колоночного $^{225}$Ac/$^{213}$Bi генератора. Согласно расчетам и оценкам, требующим экспериментальной проверки, генератор обеспечивает выход $^{213}$Bi, сравнимый с выходом из наиболее распространенного сейчас генератора на основе AG MP-50 [15], и превосходит его по радионуклидной чистоте продукта, при этом конструкция генератора компактна, а его параметры, такие как размер колонок и скорость циркуляции раствора, технологичны и легко реализуемы.

## Заключение

Предложен подход, описывающий движение с различными скоростями и распределение в хроматографической среде веществ-участников последовательных реакций первого порядка. На примере цепочки ① → ② → ③ → построена одномерная модель движения веществ в колонке бесконечной длины, рассмотрены закономерности формирования профилей концентрации веществ при условии $\lambda_1 \ll \lambda_2, \lambda_3$, когда возможно достижение подвижного равновесия. Показано, что время движения дочернего вещества состоит из интервалов, а область его существования – из диапазонов, в каждом из которых действует свое уравнение для концентрации $\frac{dN_i}{dV} = f(t_e, V)$. Сформулирована математическая процедура нахождения концентрации дочернего вещества как функции времени и координаты в хроматографической колонке. Показано, что многообразие уравнений концентрации подчиняется единому правилу, которое можно выразить в форме, близкой к уравнению Бейтмана. Описан и раскрыт в примерах простой графический способ использования правила для нахождения $\frac{dN_i}{dV} = f(t_e, V)$ с помощью *V-t* диаграммы движения веществ. Предложенный подход к кинетике последовательных реакций применим и в случае, когда подвижное равновесие не достигается, и может быть расширен на большее число участников реакций, движение в многомерной среде, а также на реакции более высокого порядка.

Используя общий подход, проанализировано влияние на распределение дочернего вещества в хроматографической системе таких факторов, как дискретное изменение подвижной или неподвижной фазы и движение подвижной фазы в замкнутом контуре (циркуляция). Показано, что любое дискретное изменение в хроматографической системе приводит к образованию одного или нескольких вторичных фронтов, создающих дополнительные диапазоны в области существования дочернего вещества.

Развитие кинетики последовательных реакций, протекающих в хроматографической среде, позволило расширить возможности радионуклидных генераторов. В существующих сейчас генераторах стадия накопления дочернего вещества связана с отсутствием движения раствора, т.е. дочернее вещество накапливается в том же месте, где находится материнское вещество. Когда оно накопилось – включается движение раствора, и вещество извлекают из генератора. В настоящей работе предложен общий принцип генераторной схемы, основанный на непрерывном движении раствора на стадии накопления дочернего вещества. Со временем система приходит в состояние, когда радионуклид находится в подвижном равновесии с материнским, но пространственно от него отделен и затем выделен из генератора.

На примере модели радионуклидного $^{225}$Ac/$^{213}$Bi генератора, в котором $^{213}$Bi получают с помощью непрерывного отделения и распада промежуточного $^{221}$Fr, продемонстрировано практическое применение разработанного математического аппарата. При описании модели генератора приняты близкие к действительности допущения относительно скоростей движения веществ: $^{221}$Fr движется везде со скоростью подвижной фазы, скорости $^{225}$Ac и $^{213}$Bi в сорбенте равны нулю, а скорость $^{213}$Bi вне сорбента также равна скорости подвижной фазы. Модель позволяет проанализировать образующиеся распределения дочерних веществ и найти оптимальные параметры получения $^{213}$Bi. В результате предложена компактная и эффективная схема циркулирующего двух-колоночного генератора $^{213}$Bi высокой радионуклидной чистоты.

## Список литературы

1. H. Bateman, The solution of a system of differential equations occurring in the theory of radioactive transformations, Proc. Cambridge Philos. Soc 15 (1910) p. 423–427.
2. J. Cetnar, General solution of Bateman equations for nuclear transmutations, Ann. Nucl. Energy 33 (7) (2006) 640–645, http://dx.doi.org/10.1016/j.anucene.2006.02.004.
3. Rösch, F; Knapp, F F (2003). "Radionuclide Generators". In Vértes, Attila; Nagy, Sándor; Klencsár, Zoltan; Lovas, Rezső G. (eds.). Handbook of Nuclear Chemistry: Radiochemistry and radiopharmaceutical chemistry in life sciences.
4. V. M. Chudakov, B. L. Zhuikov, S. V. Ermolaev, V. M. Kokhanyuk, M. I. Mostova, V. V. Zaitsev, S. V. Shatik, N. A. Kostenikov, D. V. Ryzhkova, L. A. Tyutin. Characterization of a $^{82}$Rb Generator for Positron Emission Tomography. Radiochemistry, 2014, Vol. 56, No. 5, pp. 535–543.
5. CardioGen-82 rubidium Rb-82 generator for elution of rubidium chloride. $^{82}$Rb injection, *Bracco Diagnostics 43-8200*, USA, 2000.
6. D.A. Miller, S. Sun, J.H. Yi. Preparation of a $^{118}$Te/$^{118}$Sb radionuclide generator. J. Radioanal. Nucl. Chem. (1992) vol. 160, No. 2, pp. 467-476.
7. Atcher, R.W., Friedman, A.M., Huizenga, J.R., Spencer, R.P. A radionuclide generator for the production of $^{211}$Pb and its daughters. J. Radioanal. Nucl. Chem. Letters **135,** 215–221 (1989). https://doi.org/10.1007/BF02164974
8. S.V. Ermolaev, B.L. Zhuikov, V.M. Kokhanyuk, S.N. Kalmykov, R.A. Aliev. Production of radium-223 and its decay products from thorium irradiated with medium-energy protons. VII Russian Conference "Radiochemistry-2012", Dimitrovgrad, Russia, Oct. 15-19, 2012, Book of Abstracts, p. 370.
9. S. Hassfjell. A $^{212}$Pb generator based on a $^{228}$Th source. Appl. Radiat. Isot. 2001, vol. 55(4), pp. 433-439. https://doi.org/10.1016/S0969-8043(00)00372-9.
10. P.P. Boldyrev, B.V. Egorova, K.V. Kokov, Yu.A. Perminov, M.A. Proshin, D.Yu. Chuvilin. Physical and chemical processes on the $^{212}$Pb radionuclide production for nuclear medicine. J. Phys.: Conf. Ser., 2018, vol. 1099, p. 012003, https://doi.org/10.1088/1742-6596/1099/1/012003.
11. S.V. Ermolaev, A.K. Skasyrskaya. Motion of genetically related $^{221}$Fr and $^{213}$Bi radionuclides in a chromatographic medium. Radiochemistry, 2019, Vol. 61, No. 1, pp. 44–54.
12. S.V. Ermolaev, A.K. Skasyrskaya. Development of circulating generator systems $^{225}$Ac → $^{221}$Fr → $^{213}$Bi. III International Conference "Radiopharma-2019", Moscow, Russia, June 18-21, 2019, Book of Abstracts, p. 27.
13. A. Vasiliev, S. Ermolaev, E. Lapshina, N. Betenekov, E. Denisov, B. Zhuikov. Various chromatographic schemes for separation of $^{213}$Bi from $^{225}$Ac. J. Medical Imaging and Radiation Sciences, 2019, v. 50, No. 1, p. S21. DOI: 10.1016/j.jmir.2019.03.067.
14. McDevitt, M. R., Finn, R. D., Sgouros, G., Ma, D., Scheinberg, D. A. An $^{225}$Ac/$^{213}$Bi generator system for therapeutic clinical applications:construction and operation. Appl. Radiat. Isot. **50**(5), 895 (1999).

15. C. Apostolidis, R. Molinet, G. Rasmussen, A. Morgenstern. Production of Ac-225 from Th-229 for Targeted α-Therapy. Anal. Chem. (2005), 77, 6288-6291.
16. Boll, R. A., Mirzadeh, S., Kennel, S. J., DePaoli, D. W., Webb, O. F. $^{213}$Bi for alpha-particle-mediated radioimmunotherapy. J. Labelled Compd. Radiopharm. 40, 341 (1997).
17. Bray, L. A., Tingey, J. M., DesChane, J. R., Egorov, O. B., Tenforde, T. S., Wilbur, D. S., Hamlin, D. K., Pathare, P. M. Development of a unique bismuth (Bi-213) automated generator for use in cancer therapy. Ind. Eng. Chem. Res. **39**(9), 3189 (2000).
18. Wu, C., Brechbiel, M. W., Gansow, O. A. An improved generator for the production of $^{213}$Bi from $^{225}$Ac. Radiochim. Acta. **79**(2), 141 (1997).
19. McAlister, D. R., Horwitz, E. P. Automated two column generator systems for medical radionuclides. Appl. Radiat. Isot. **67**(11), 1985 (2009).
20. Guseva, L. I., Dogadkin, N. N.: Development of a tandem generator system $^{229}$Th/$^{225}$Ac/$^{213}$Bi for repeated production of short-lived α-emitting radionuclides. Radiochemistry **51**(2), 169 (2009).
21. A.N. Vasiliev, S.V. Ermolaev, E.V. Lapshina, B.L. Zhuikov, N.D. Betenekov. $^{225}$Ac/$^{213}$Bi generator based on inorganic sorbents. Radiochimica Acta, 2019, v. 107, No. 12, pp. 1203-1211. DOI: 10.1515/ract-2019-3137.
22. Morgenstern, A., Apostolidis, C., Kratochwil, C., Sathekge, M., Krolicki, L., Bruchertseifer, F. An overview of targeted alpha therapy with actinium-225 and bismuth-213. Curr. Radiopharm. 11(3), 200 (2018).
23. Б. Л. Жуйков, С. Н. Калмыков, С. В. Ермолаев, Р. А. Алиев, В. М. Коханюк, В. Л. Матушко, И. Г. Тананаев, Б. Ф. Мясоедов. Получение актиния-225 и радия 223 при облучении тория ускоренными протонами. Радиохимия, 2011, Т. 53, N 1, С. 66-72.
24. Ermolaev, S. V., Zhuikov, B. L., Kokhanyuk, V. M., Matushko, V. L., Kalmykov, S. N., Aliev, R. A., Tananaev, I. G., Myasoedov, B. F. (2012). Production of actinium, thorium and radium isotopes from natural thorium irradiated with protons up to 141 MeV. Radiochimica Acta, 100(4), 223-229.
25. Aliev, R. A., Ermolaev, S. V., Vasiliev, A. N., Ostapenko, V. S., Lapshina, E. V., Zhuikov, B. L., Zakharov, N. V., Pozdeev, V. V., Kokhanyuk, V. M., Myasoedov, B. F., Kalmykov, S. N.: Isolation of medicine-applicable actinium-225 from thorium targets irradiated by medium-energy protons, Solv. Extr. Ion Exch. 32(5), 468 (2014).
26. John K: US DOE tri-lab research and production effort to provide accelerator- produced $^{225}$Ac for radiotherapy: 2019 update. Eur J Nucl Med Mol Imaging 46:S722, 2019.
27. Hoehr, C., Bénard, F., Buckley, K., Crawford, J., Gottberg, A., Hanemaayer, V., Kunz, P., Ladouceur, K., Radchenko, V., Ramogida, C., Robertson, A., Ruth, T., Zacchia, N., Zeisler, S., Schaffer, P.: Medical isotope production at TRIUMF – from imaging to treatment. Phys. Procedia. 90, 200 (2017).
28. Horwitz E. P., Chiarizia R., Dietz M. L. // Solvent Extr. Ion Exch., 1992, Vol. 10, P. 313-336.
29. Horwitz E. P., Dietz M. L., Chiarizia R., Diamond H. // Anal. Chim. Acta, 1992, Vol. 266, P. 25-37.
30. Horwitz E. P., Dietz M. L., Chiarizia R., at al. // Anal. Chim. Acta, 1995, Vol. 310, P. 63-78.
31. Horwitz E. P., Chiarizia R., Dietz M. L. // Reactive and Functional Polym., 1997, Vol. 33, P. 25-36.